\documentclass[11pt]{article}

\usepackage{amsmath,amssymb,latexsym}
\usepackage[T1]{fontenc}
\usepackage{epsfig}
\usepackage{fullpage}
\usepackage{graphicx}
\usepackage{makeidx}
\usepackage{slashed}
\usepackage{color}
\usepackage{subcaption}
\usepackage{feynmf}
\usepackage{cite}
\usepackage{hyperref}

\long\def\symbolfootnote[#1]#2{\begingroup%
\def\thefootnote{\fnsymbol{footnote}}\footnote[#1]{#2}\endgroup}

\begin{document}

\begin{flushright}
QMUL-PH-19-10\\
Nikhef/2019-018
\end{flushright}

\vspace*{1.5cm}

\begin{center}
{\Large\sc Next-to-leading power threshold effects for inclusive and exclusive processes with final state jets }\\[10ex] 
 { 
 Melissa van Beekveld$^{a,b}$,
 Wim Beenakker$^{a,c}$,
 Eric Laenen$^{b,c,d}$,
 Chris D. White$^{e}$}\\[1cm]
 {\it 
$^a$ {Theoretical High Energy Physics, Radboud University
Nijmegen, Heyendaalseweg 135, 6525~AJ Nijmegen, NL}\\
$^b${Nikhef, Science Park 105, 1098 XG Amsterdam, NL}\\
$^c$ {Institute of Physics, University of Amsterdam, Science Park 904, 1018 XE Amsterdam, NL}\\
 $^d$ {ITF, Utrecht University, Leuvenlaan 4, 3584 CE Utrecht, NL}\\
 $^e$ {Centre for Research in String Theory, School of Physics and Astronomy, Queen Mary
University of London, 327 Mile End Road, London E1 4NS, UK}\\}
\end{center}

\vspace{1.5cm}
\vspace{1cm}

\begin{abstract}
\noindent{It is well known that cross-sections in perturbative QCD receive large corrections from soft and collinear radiation, whose properties must be resummed to all orders in the coupling. Whether or not the universal properties of this radiation can be extended to next-to-leading power (NLP) in the threshold expansion has been the subject of much recent study. In this paper, we consider two types of NLP effects: the interplay of next-to-soft and collinear radiation in processes with final state jets and the NLP contributions stemming from soft quarks. We derive an NLP amplitude for soft gluons and quarks, valid for an arbitrary number of coloured or colourless massless final state particles. We show explicitly that this framework can be used to correctly obtain the dominant NLP effects in three different types of processes at next-to-leading order: deep-inelastic scattering, hadroproduction via electron-positron annihilation and prompt photon production. Our results provide an important ingredient for developing a universal resummation formalism for NLP effects.}
\end{abstract}

\vspace*{\fill}

\newpage
\reversemarginpar

\section{Introduction}
The calculation of ever more precise results for cross-sections in perturbative Quantum Chromodynamics (QCD) is a perennial
necessity for current and forthcoming collider experiments. Results typically proceed on two frontiers. Firstly, one must proceed to
higher fixed orders in the strong coupling 
$\alpha_s$. Secondly, one must supplement fixed order calculations with contributions which are enhanced in certain kinematic regions. One such region is that of production of particles near
threshold. In that case, the phase space for the emission of additional radiation is squeezed, leading to an incomplete cancellation between real and virtual singularities, and thus the
appearance of large contributions at all orders in perturbation theory. 
More specifically, if $\xi$ is a dimensionless kinematic ratio
such that $\xi\rightarrow 0$ near threshold, the corresponding differential cross-section has the following form
\begin{eqnarray}
\label{eq:defthreshold}
\frac{{\rm d}\sigma}{{\rm d}\xi} \propto \sum_{n=0}^{\infty}\left(\frac{\alpha_s}{\pi}\right)^n\sum_{m=0}^{2n-1}\left[c_{nm}^{R}\left(\frac{\ln^m \xi}{\xi}\right)_++c_n^{V}\delta(\xi)+c_{nm}^{\rm NLP}\ln^m\xi+\dots\right].
\end{eqnarray}
The first two sets of terms originate from soft and collinear radiation (real or virtual). They make up the {\it leading power} (LP) contributions in the threshold variable $\xi$ and are localized at $\xi = 0$. The contributions have a universal form, which allows for their all-order resummation  \cite{Parisi:1980xd,Curci:1979am,Sterman:1987aj,Catani:1989ne,Catani:1990rp,Gatheral:1983cz,Frenkel:1984pz,Sterman:1981jc,Korchemsky:1993xv,Korchemsky:1993uz,Forte:2002ni,Contopanagos:1997nh,Becher:2006nr,Schwartz:2007ib,Bauer:2008dt,Chiu:2009mg}. 

The third set of terms in Eq.~(\ref{eq:defthreshold}) make up the {\it next-to-leading power} (NLP) contributions to the differential cross-section, as they are suppressed by a single power of the threshold variable. Although subleading, the increasing precision both of LP resummation and of experimental data makes such terms numerically relevant~\cite{Kramer:1996iq,Herzog:2014wja}. As at LP, the highest power of the NLP log at a given order is referred to as {\it leading-logarithmic} (LL). One may then worry about next-to-leading
logarithmic (NLL) contributions, and so on. It may well be the case for collider processes of interest that the LL NLP terms (or beyond) must be resummed for an adequate comparison of theory with data. Even if not, elucidation of NLP contributions at a fixed order in $\alpha_s$ may have a key role to play in estimating higher order cross-sections. It can furthermore aid in the development of subtraction schemes for the efficient numerical cancellation of infrared singularities~\cite{Boughezal:2016zws,Boughezal:2018mvf}.

Both LP and NLP threshold contributions arise from radiation that is (next-to-)soft and/or collinear. Next-to-soft radiation in gauge theory was first studied in the classic works of Refs.~\cite{Low:1958sn,Burnett:1967km}, and more recently in Ref.~\cite{DelDuca:1990gz}. Since then, a variety of approaches have been used to try to systematically elucidate the structure of
next-to-soft corrections~\cite{Laenen:2008gt,Laenen:2010uz,Soar:2009yh,Moch:2009hr,Moch:2009mu,deFlorian:2014vta,Presti:2014lqa,Laenen:2008ux,Grunberg:2007nc,Grunberg:2009yi,Grunberg:2009vs}. There
has recently been a revival of interest in this topic, partially motivated by more formal work on so-called {\it next-to-soft theorems} of Refs.~\cite{Cachazo:2014fwa,Casali:2014xpa} (see also Ref.~\cite{White:2011yy}), which related soft physics to asymptotic symmetries in gauge theory and gravity. This has led to a great deal of activity aiming to systematically classify NLP contributions to cross-sections, using either diagrammatic factorisation formulas
that generalise their LP counterparts~\cite{Gervais:2017yxv,Gervais:2017zky,Gervais:2017zdb,Bonocore:2015esa,Bonocore:2016awd},
or the framework of soft-collinear effective theory
(SCET)~\cite{Beneke:2004in,Larkoski:2014bxa,Kolodrubetz:2016uim,Moult:2016fqy,Moult:2017rpl,Feige:2017zci,Chang:2017atu}. A resummation of LL NLP effects in Drell-Yan production has recently
been presented using the SCET approach~\cite{Beneke:2018gvs},
confirming earlier expectations from
Refs.~\cite{Soar:2009yh,Moch:2009hr,Moch:2009mu,deFlorian:2014vta,Presti:2014lqa},
and results using the diagrammatic approach are in progress \cite{resum:1}. The various approaches have also been used to examine NLP effects at fixed
order in the coupling~\cite{DelDuca:2017twk,Bonocore:2014wua,Bahjat-Abbas:2018hpv,Boughezal:2016zws,Boughezal:2018mvf}. Of particular relevance for the present study is Ref.~\cite{DelDuca:2017twk}, which derived a universal form of the cross-section for the production of an arbitrary number of colour
singlet particles at NLO, up to NLP level, in either the $q\bar{q}$ or $gg$ channel. An especially elegant result was that the NLP cross-section could be expressed in terms of a simple kinematic shift of the LO result. This both illustrates the phenomenological use of
next-to-soft factorisation formulas, and provides analytic information where this was previously absent (such as in di-Higgs production). 

To date, most studies of NLP effects (with the exception of the conjectural but well-motivated resummation proposal of Refs.~\cite{Soar:2009yh,Moch:2009hr,Moch:2009mu,deFlorian:2014vta,Presti:2014lqa}) have focused on processes in which all real radiation is manifestly
(next-to) soft, such as Drell-Yan production. This is less complicated theoretically than the completely general case of a process containing coloured final state particles at LO. In such processes, both soft and collinear real emissions lead to threshold enhanced contributions, such that one
must carefully disentangle them in aiming to classify all NLP effects. It is then natural to ponder what the recently developed next-to-soft formalisms are able to capture. In particular it is interesting to find out whether the simple kinematic shift observed in Ref.~\cite{DelDuca:2017twk} remains relevant and to what degree. Here we will significantly extend this previous result. We will study processes containing final state massless coloured particles, in which we will {\it indeed} have to worry about collinear effects associated with real radiation. Furthermore, we will consider the effect of different partonic channels to the LO process, which open up for the first time at NLO. Such corrections are known to contain NLP contributions, and we will construct a universal operator that can include them. In particular, we will be able to derive an NLP amplitude for the emission of soft (anti-)quarks, up to NLO in perturbation theory.

As a case study, we will use three examples of processes with final state jets: deep-inelastic scattering, quark-antiquark pair production in electron-positron annihilation, and the production of a photon in association with a hard coloured particle ({\it prompt photon production}). There are a number of motivations for our investigation. Firstly, elucidating the structure of NLP effects is a crucial prerequisite to being able to resum
them, and it is important in particular to work out what does and does not contribute to the LL NLP terms, which would be resummed first. Secondly, the great interest in next-to-soft theorems has yet to be supplemented with a systematic investigation of subleading collinear behaviour (although
Ref.~\cite{Nandan:2016ohb} is a notable exception), and we hope that our results provide a useful springboard for further work.

This paper is structured as follows. In section \ref{sec:NLP} we derive an explicit expression for the NLO amplitude of a coloured final state in quark-antiquark, gluon-gluon or quark-gluon scattering, valid up to NLP level in the soft expansion. This NLP amplitude can be subdivided into two separate contributions: a gluonic contribution (section \ref{sec:NLPgluon}) and a quark contribution (section \ref{sec:NLPquark}). As in Ref.~\cite{DelDuca:2017twk}, the amplitudes we obtain are fully general, and the results thus provide universal corrections to any Born process with massless coloured particles in the final state. We then illustrate how to apply our formalism in a number of examples of increasing complexity. In section~\ref{sec:DIS} we consider deep-inelastic scattering, whose Born amplitude contains a single final-state parton, and in section~\ref{sec:epem} we examine hadroproduction in electron-positron annihilation, where two final state partons are present at LO. In section~\ref{sec:results} we look at prompt photon production, which adds the complication of a final state which is not fully inclusive. In all cases, we find that leading logarithmic effects up to NLP order are completely captured by performing a similar kinematic shift to that observed for colour singlet production processes~\cite{DelDuca:2017twk}, in addition to inclusion of soft quark radiation. Finally, in section~\ref{sec:discuss}, we discuss the implications of our results before concluding.

\section{Universal NLO amplitudes for (next-to-)soft radiation}
\label{sec:NLP}

As stated above, Ref.~\cite{DelDuca:2017twk} examined dressing the amplitude for production of $N$ massive colour singlet particles with an additional gluon, and derived a universal form for the NLO cross-section for any such process, valid up to next-to-soft order in the momentum of the emitted radiation. For the emission of gluons, we will recover the results of Refs.~\cite{Low:1958sn,Burnett:1967km,DelDuca:1990gz,Casali:2014xpa} where overlap exists, and the derivation presented here will allow us to set up careful notation needed for what follows. A significant extension of previous results, however, is a universal next-to-soft amplitude for the emission of soft quarks, which we present here for the first time. The latter effect is known to be absent at LP in the threshold expansion, but must be included at NLP level and beyond. 

As is well-known, whenever a massless external line of an amplitude gets dressed with a tree level vertex, there is a potential for an infrared (IR) divergence, with associated threshold logarithms in the final cross-section. This potential divergence originates from two different momentum limits, which may overlap: the emission carrying momentum $k$ can become either soft and/or collinear to its emitter. Whether or not and in what form the divergence will develop depends upon the kinematics of the leading-order process. For the production of massive colour singlet particles considered in Ref.~\cite{DelDuca:2017twk}, the only IR divergences that appear are manifestly associated with (next-to-)soft radiation. When final state massless particles are present, however, one must also worry about (hard-)collinear effects, even at LP in the threshold expansion. In this section, we restrict ourselves to (next-to-)soft effects only, where our aim is to generalise the universal NLP amplitude formula of Ref.~\cite{DelDuca:2017twk}. We will return to the issue of hard collinear contributions when looking at specific processes in the sections that follow.

\subsection{Radiation of (next-to-)soft gluons} 
\label{sec:NLPgluon}
\begin{figure}
\centering
\begin{subfigure}{0.31\textwidth}
\centering

\begin{fmffile}{diagramgen1}
\begin{fmfchar*}(100,100)
  \fmfleft{i1,i2}
  \fmfright{f1,f2}
  \fmfforce{(0,0)}{i2} 
  \fmfforce{(0,h)}{i1} 
  \fmfforce{(.45w,.5h)}{v4}
  \fmfforce{(.5w,.5h)}{v1}
  \fmfforce{(.95w,.05h)}{v2}
  \fmfforce{(.95w,.95h)}{v3}
  \fmfforce{(w,h)}{f1} 
  \fmfforce{(w,0)}{f2} 
  \fmf{double}{i1,v1}
  \fmf{double}{v1,i2}
  \fmflabel{$p_{1}$}{i1}
  \fmflabel{$p_{2}$}{i2}
  \fmflabel{$p_{3}$}{f1}
  \fmflabel{$p_{n+2}$}{f2}
  \fmf{dots,right=0.25}{v2,v3}
  \fmf{double}{v2,v1}
  \fmf{double}{v3,v1}
  \fmf{double}{f1,v3}
  \fmf{double}{f2,v2}
  \fmfv{decor.shape=circle,decor.filled=empty, d.si=.45w,label=$\mathcal{M}_{\rm H}$,label.dist=-11}{v1}
\end{fmfchar*}
\end{fmffile} 
\vspace{0.5cm}
\caption{} \label{fig:1a}
\end{subfigure}
\hspace{0.3cm}
\begin{subfigure}{0.31\textwidth}
\centering
\begin{fmffile}{diagramgen2}
\begin{fmfchar*}(100,100)
  \fmfleft{i1,i2}
  \fmfright{f1,f2}
  \fmfforce{(0,0)}{i2} 
  \fmfforce{(0,h)}{i1} 
  \fmfforce{(.3w,h)}{f3} 
  \fmfforce{(.15w,.85h)}{g1}
  \fmfforce{(.5w,.5h)}{v1}
  \fmfforce{(.95w,.05h)}{v2}
  \fmfforce{(.95w,.95h)}{v3}
  \fmfforce{(w,h)}{f1} 
  \fmfforce{(w,0)}{f2} 
  \fmf{fermion}{i1,g1} 
  \fmf{fermion,tension=1,label=$p_{1,,c_j}'$,label.side=right,label.dist=-0.5}{g1,v1}
  \fmf{gluon}{g1,f3}
  \fmf{double}{v1,i2}
  \fmflabel{$p_{1,c_i}$}{i1}
  \fmflabel{$p_{2}$}{i2}
  \fmflabel{$p_{3}$}{f1}
  \fmflabel{$k_{\sigma,c}$}{f3}
  \fmflabel{$p_{n+2}$}{f2}
  \fmf{dots,right=0.25}{v2,v3}
  \fmf{double}{v2,v1}
  \fmf{double}{v3,v1}
  \fmf{double}{f1,v3}
  \fmf{double}{f2,v2}
  \fmfv{decor.shape=circle,decor.filled=empty, d.si=.45w,label=$\mathcal{M}_{c_j}(p_1')$,label.dist=-18}{v1}
\end{fmfchar*}
\end{fmffile} 
\vspace{0.5cm}
\caption{}\label{fig:1b}
\end{subfigure}
\hspace{0.3cm}
\begin{subfigure}{0.31\textwidth}
\centering
\begin{fmffile}{diagramgen3c}
\begin{fmfchar*}(100,100)
  \fmfleft{i1,i2}
  \fmfright{f1,f2,f3}
  \fmfforce{(0,0)}{i2} 
  \fmfforce{(0,h)}{i1} 
  \fmfforce{(.3w,h)}{f3} 
  \fmfforce{(.15w,.85h)}{g1}
  \fmfforce{(.5w,.5h)}{v1}
  \fmfforce{(.95w,.05h)}{v2}
  \fmfforce{(.95w,.95h)}{v3}
  \fmfforce{(w,h)}{f1} 
  \fmfforce{(w,0)}{f2} 
  \fmf{gluon}{i1,g1} 
  \fmf{gluon}{g1,f3}
     \fmf{gluon,tension=1,label=$p_{1,,\rho,,b}'$,label.side=right,label.dist=1.2}{g1,v1}
   \fmf{double}{v1,i2}
  \fmflabel{$p_{1,\mu,a}$}{i1}
  \fmflabel{$p_{2}$}{i2}
  \fmflabel{$p_{3}$}{f1}
  \fmflabel{$k_{\sigma,c}$}{f3}
  \fmflabel{$p_{n+2}$}{f2}
  \fmf{dots,right=0.25}{v2,v3}
  \fmf{double}{v2,v1}
  \fmf{double}{v3,v1}
  \fmf{double}{f1,v3}
  \fmf{double}{f2,v2}
  \fmfv{decor.shape=circle,decor.filled=empty, d.si=.45w,label=$\mathcal{M}_{\rho,,b}(p_1')$,label.dist=-20.5}{v1}
  \fmfv{decor.shape=circle,decor.filled=full, d.si=.02w}{g1}
\end{fmfchar*}
\end{fmffile} 
\vspace{0.5cm}
\caption{}\label{fig:1c}
\end{subfigure}

\vspace{1.5cm}
\begin{subfigure}{0.31\textwidth}
\centering
\begin{fmffile}{diagramgen4}
\begin{fmfchar*}(100,100)
  \fmfleft{i1,i2}
  \fmfright{f1,f2}
  \fmfforce{(0,0)}{i2} 
  \fmfforce{(0,h)}{i1} 
  \fmfforce{(.w,.7h)}{f3} 
  \fmfforce{(.w,.5h)}{f4} 
  \fmfforce{(.85w,.5h)}{g1}
  \fmfforce{(.5w,.5h)}{v1}
  \fmfforce{(.8w,.2h)}{v2}
  \fmfforce{(.8w,.8h)}{v3}
  \fmfforce{(w,h)}{f1} 
  \fmfforce{(w,0)}{f2} 
  \fmf{double}{i1,v1} 
  \fmf{double}{v1,i2}
  \fmflabel{$p_{1}$}{i1}
  \fmflabel{$p_{2}$}{i2}
  \fmflabel{$p_{3}$}{f1}
  \fmflabel{$p_{i,{c_i}}$}{f4}
  \fmflabel{$k_{\sigma,c}$}{f3}
  \fmflabel{$p_{n+2}$}{f2}
  \fmf{dots,right=0.25}{v2,v3}
  \fmf{double}{v2,v1}
  \fmf{double}{v3,v1}
  \fmf{fermion}{v1,g1}
  \fmf{fermion}{g1,f4}
  \fmf{gluon}{f3,g1}
  \fmf{double}{f1,v3}
  \fmf{double}{f2,v2} 
  \fmfv{decor.shape=circle,decor.filled=empty, d.si=.45w,label=$\mathcal{M}_{c_j}(p_i')$,label.dist=-18}{v1}
\end{fmfchar*}
\end{fmffile} 
\vspace{0.5cm}
\caption{}\label{fig:1d}
\end{subfigure}
\hspace{0.3cm}
\begin{subfigure}{0.31\textwidth}
\centering
\begin{fmffile}{diagramgen4b}
\begin{fmfchar*}(100,100)
  \fmfleft{i1,i2}
  \fmfright{f1,f2}
  \fmfforce{(0,0)}{i2} 
  \fmfforce{(0,h)}{i1} 
  \fmfforce{(.w,.65h)}{f3} 
  \fmfforce{(.w,.35h)}{f4} 
  \fmfforce{(.9w,.5h)}{g1}
  \fmfforce{(.5w,.5h)}{v1}
  \fmfforce{(.8w,.2h)}{v2}
  \fmfforce{(.8w,.8h)}{v3}
  \fmfforce{(w,h)}{f1} 
  \fmfforce{(w,0)}{f2} 
  \fmf{double}{i1,v1} 
  \fmf{double}{v1,i2}
  \fmflabel{$p_{1}$}{i1}
  \fmflabel{$p_{2}$}{i2}
  \fmflabel{$p_{3}$}{f1}
  \fmflabel{$p_{i,\mu,a}$}{f4}
  \fmflabel{$k_{\sigma,c}$}{f3}
  \fmflabel{$p_{n+2}$}{f2}
  \fmf{dots,right=0.25}{v2,v3}
  \fmf{double}{v2,v1}
  \fmf{double}{v3,v1}
  \fmf{gluon}{g1,v1}
  \fmf{gluon}{f4,g1}
  \fmf{gluon}{f3,g1}
  \fmf{double}{f1,v3}
  \fmf{double}{f2,v2} 
  \fmfv{decor.shape=circle,decor.filled=empty, d.si=.45w,label=$\mathcal{M}_{\rho,,b}(p_i')$,label.dist=-20}{v1}
  \fmfv{decor.shape=circle,decor.filled=full, d.si=.02w}{g1}
\end{fmfchar*}
\end{fmffile} 
\vspace{0.5cm}
\caption{}\label{fig:1e}
\end{subfigure}
\hspace{0.3cm}
\begin{subfigure}{0.31\textwidth}
\centering
\begin{fmffile}{diagramgen5}
\begin{fmfchar*}(100,100)
  \fmfleft{i1,i2}
  \fmfright{f1,f2}
  \fmfforce{(0,0)}{i2} 
  \fmfforce{(0,h)}{i1} 
  \fmfforce{(.w,.5h)}{f4} 
  \fmfforce{(.5w,.5h)}{v1}
  \fmfforce{(.8w,.2h)}{v2}
  \fmfforce{(.8w,.8h)}{v3}
  \fmfforce{(w,h)}{f1} 
  \fmfforce{(w,0)}{f2} 
  \fmf{double}{i1,v1} 
  \fmf{double}{v1,i2}
  \fmflabel{$p_{1}$}{i1}
  \fmflabel{$p_{2}$}{i2}
  \fmflabel{$p_{3}$}{f1}
  \fmflabel{$k_{\sigma}$}{f4}
  \fmflabel{$p_{n+2}$}{f2}
  \fmf{dots,right=0.25}{v2,v3}
  \fmf{double}{v2,v1}
  \fmf{double}{v3,v1}
  \fmf{gluon}{f4,v1}
  \fmf{double}{f1,v3}
  \fmf{double}{f2,v2}
  \fmfv{decor.shape=circle,decor.filled=empty, d.si=.45w,l=$\mathcal{M}_{{\rm int},,\sigma}$,label.dist=-15}{v1}
\end{fmfchar*}
\end{fmffile} 
\vspace{0.5cm}
\caption{}\label{fig:1f}
\end{subfigure}
\vspace{0.5cm}
\label{fig:diagrams1}
\caption{(a) Feynman diagram for a generic $2\rightarrow n$ scattering process where all external particles are colour charged and either in the adjoint, fundamental or anti-fundamental representation, which are all indicated by a double line. By $\mathcal{M}_{\rm H}$ we denote the matrix element without any additional radiation, containing all asymptotic states (spinors and polarisation vectors). (b) Feynman diagram with the emission of one additional gluon carrying momentum $k$ and colour $c$ from an initial state quark carrying momentum $p_1$ and colour $c_i$. The notation is such that $\mathcal{M}$ contains all colour generators, spinors and/or polarisation vectors, except for the ones stemming from the line that is emitting. Since the gluon carries away momentum $k$ and changes the colour of the quark line, this matrix element will depend on $p_1'= p_1 - k$ and carries a colour label $c_j$. (c) Feynman diagram with the emission of one additional gluon from an initial state gluon. (d) Feynman diagram with an emission off a final state quark. Here the hard scattering matrix element depends on $p_i' = p_i + k$. (e) Feynman diagram with an emission off a final state gluon. (f) Feynman diagram where the gluon is emitted from an internal line. The matrix element $\mathcal{M}_{{\rm int},\sigma}$ contains all colour generators and external states, except for $\epsilon^*_{\sigma}(k)$.}
\end{figure}
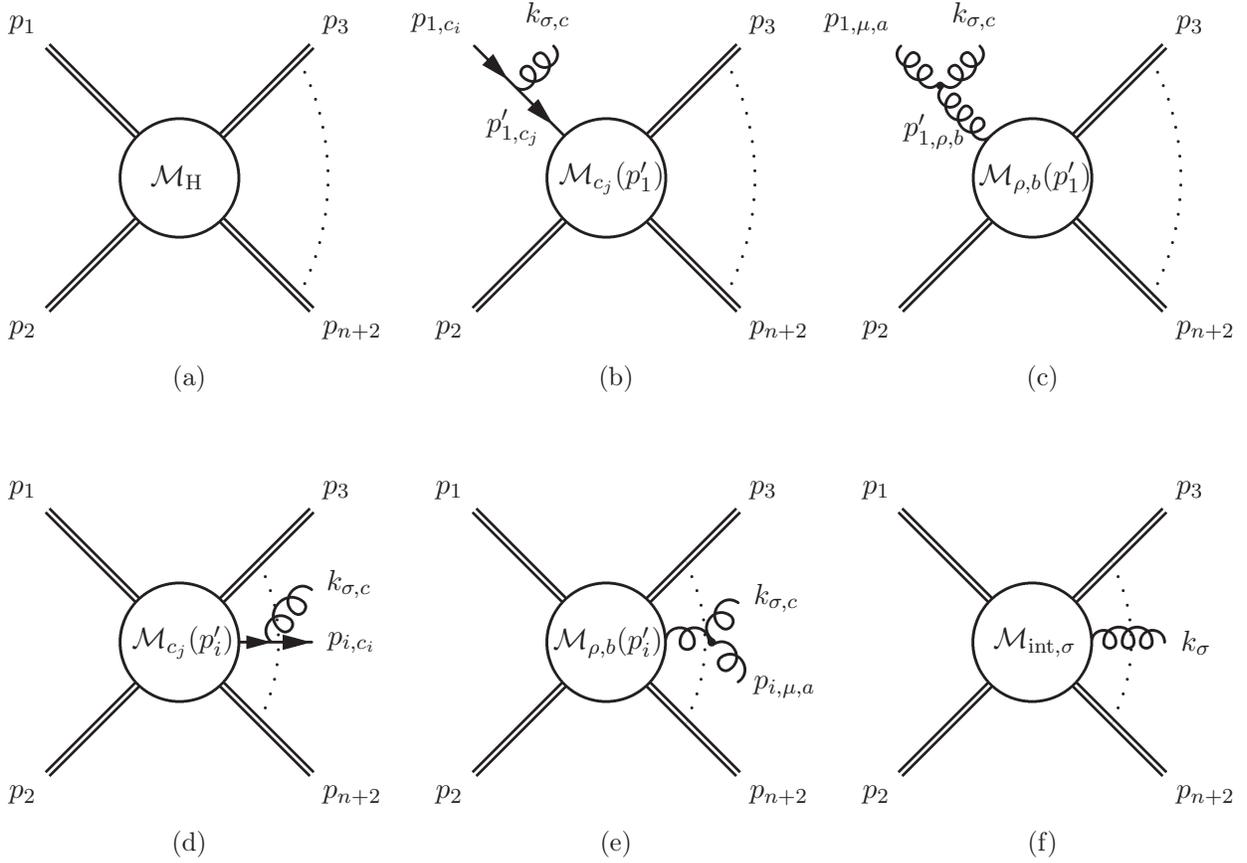
Let us first consider the emission of (next-to-)soft gluons. We will do this for a generic Born level process with 2 initial state coloured particles and $n$ final state coloured particles (see Fig. \ref{fig:1a}). As mentioned above, this extends the work of Ref.~\cite{DelDuca:2017twk} for colour-singlet production, itself based on the earlier work of Refs.~\cite{Bonocore:2015esa,Bonocore:2016awd} (see also Refs.~\cite{White:2014qia,Larkoski:2014bxa,Gervais:2017yxv,Gervais:2017zky,Gervais:2017zdb}). All particles are considered massless, which will be the case for all processes considered throughout the paper~\footnote{We note in passing that our formalism would also apply for massive coloured particles.}. We must then consider all possible ways in which a gluon can be emitted, namely the contributions of Fig.~\ref{fig:1b}--\ref{fig:1f}, where momenta and colour indices are defined as shown. The first of these contributions, Fig.~\ref{fig:1b}, yields the matrix element
\begin{eqnarray}
i\mathcal{M}_{1,q} &=& -\frac{i g_s t^{c}_{c_jc_i}}{(p_1-k)^2+i\varepsilon} \mathcal{M}_{c_j}(p_1-k,p_2,\dots,p_{n+2})(\slashed{p}_1-\slashed{k})\gamma^{\sigma}u(p_1)\epsilon^*_{\sigma}(k) \nonumber\\
 &=& -\frac{i g_s t^{c}_{c_jc_i}}{(p_1-k)^2+i\varepsilon}\mathcal{M}_{c_j}(p_1,p_2,\dots,p_{n+2})\nonumber\\
 && \hspace{3cm} \times \left((2p_1^{\sigma}-k^{\sigma}) - 2i S^{\sigma\mu}k_{\mu} - 2k^{\mu}\frac{\overleftarrow{\partial}}{\partial p_1^{\mu}}p_1^{\sigma}\right)u(p_1)\epsilon^*_{\sigma}(k),
\label{Mqdef}
\end{eqnarray}
where $\{t^a_{ij}\}$ are elements of the colour generator in the fundamental representation, and $\mathcal{M}_{c_j}$ is the hard scattering matrix element carrying colour label $c_j$ for the incoming quark. The notation is such that $\mathcal{M}_{c_j}$ contains all colour generators, spinors and/or polarisation vectors, except for the ones stemming from the line that is emitting. We have also introduced the generator of Lorentz transformations for fermionic fields:
\begin{equation}
S^{\sigma\mu} \equiv \frac{i}{4}\left[\gamma^{\sigma},\gamma^{\mu}
\right].
\label{Sspindef}
\end{equation}
In the second line of Eq.~(\ref{Mqdef}), the first two terms in the bracket come from rewriting the Dirac propagator in a suitable form. The derivative term stems from Taylor expanding the hard scattering matrix element $\mathcal{M}_{c_j}(p_1-k)$ to first order in $k$, where we have assumed this to be Taylor expandable. This assumption will fail at higher loop orders due to the presence of virtual collinear singularities, as first studied by Ref.\cite{DelDuca:1990gz}, and further developed by subsequent works~\cite{Bonocore:2015esa,Bonocore:2016awd,White:2014qia,Larkoski:2014bxa,Gervais:2017yxv,Gervais:2017zky,Gervais:2017zdb}. However, the assumption is valid for tree- and loop-induced Born processes that are free of virtual collinear effects, which applies to all examples studied in this paper. Carrying out a similar exercise for the case where the initial state is an anti-quark, we find
\begin{eqnarray}
i\mathcal{M}_{1,\bar{q}} = \frac{i g_s t^{c}_{c_ic_j}}{(p_1-k)^2+i\varepsilon}\bar{v}(p_1)\left((2p_1^{\sigma}-k^{\sigma}) + 2i S^{\sigma\mu}k_{\mu} - 2p_1^{\sigma}k^{\mu}\frac{\partial}{\partial p_1^{\mu}}\right)\mathcal{M}_{c_j}(p_1,p_2,\dots,p_{n+2})\epsilon^*_{\sigma}(k).
\end{eqnarray}
In the case of an emitting gluon (Fig. \ref{fig:1c}) the resulting matrix element is:
\begin{align}
i\mathcal{M}_{1,g} &= \frac{ g_s f^{cab}}{(p_1-k)^2+i\varepsilon}\epsilon_{\mu}(p_1)\mathcal{M}_{\rho,b}(p_1-k,p_2,\dots,p_{n+2})\left(-g^{\sigma\mu}(p_1+k)^{\rho}+g^{\mu\rho}(2p_1-k)^{\sigma}\right.\notag\\
&\left.\quad+2g^{\sigma\rho}k^{\mu}\right)\epsilon^*_{\sigma}(k)\notag\\
&= \frac{g_s f^{cab}}{(p_1-k)^2+i\varepsilon}\epsilon_{\mu}(p_1)\left((2p_1-k)^{\sigma}g^{\rho\mu}-2ik_{\alpha}M^{\sigma\alpha,\rho\mu}- 2 p_1^\sigma k^{\alpha}\frac{\partial}{\partial p_1^{\alpha}}g^{\rho\mu}\right)
\notag\\&\quad\times \mathcal{M}_{\rho,b}(p_1,p_2,\dots,p_{n+2})\epsilon^*_{\sigma}(k),
\label{Mgdef}
\end{align}
where 
\begin{equation}
M^{\sigma\alpha,\rho\mu} = i(g^{\sigma\rho}g^{\mu\alpha}-g^{\sigma\mu}g^{\rho\alpha})
\label{Mspindef}
\end{equation}
denotes the Lorentz generator for spin 1 particles and $\mathcal{M}_{\rho,b}$ the hard scattering matrix element with adjoint colour index $b$ and Minkowski index $\rho$ for the incoming gluon. To go from the first to the second line in Eq.~(\ref{Mgdef}) we have used the physical polarisation condition for the incoming gluon to write~\cite{White:2014qia}
\begin{eqnarray}
p_1^{\rho}\mathcal{M}_{\rho}(p_1,p_2,\dots,p_{n+2}) \equiv 0 \hspace{0.3cm} \rightarrow \hspace{0.3cm} p_1^{\rho}k^{\alpha}\frac{\partial}{\partial p_1^{\alpha}}\mathcal{M}_{\rho}(p_1,p_2,\dots,p_{n+2}) = - k^{\rho}\mathcal{M}_{\rho}(p_1,p_2,\dots,p_{n+2}).
\end{eqnarray}
We thus observe that for all species of incoming parton, the next-to-soft  matrix element with an additional gluon emission from the incoming leg consists of three terms: a universal scalar term (which is proportional to $2p_i^{\sigma} - k^{\sigma}$), a term that is sensitive to the spin of the emitter (which is proportional to either $S$ or $M$) and a universal derivative term acting on the nonradiative amplitude. 

We may carry out a similar analysis for hard emitting particles in the final state. However, the fact that the gluon is emitted {\it after} the hard scattering results in a sign difference for the derivative term. More specifically, in considering the emission of a gluon of momentum $k$ from a final state hard particle of momentum $p_i+k$ leads to a momentum-shifted amplitude
\begin{eqnarray}
\mathcal{M}(p_1,p_2,\dots,p_{i}+k,\dots,p_{n+2}) &=& \mathcal{M}(p_1,p_2,\dots,p_{i},\dots,p_{n+2}) \nonumber \\
&&\hspace{1cm}+   k^{\alpha}\frac{\partial}{\partial p_i^{\alpha}} \mathcal{M}(p_1,p_2,\dots,p_{i},\dots,p_{n+2}).
\end{eqnarray}
The next-to-soft matrix element for a final state quark emitter (Fig. \ref{fig:1d}) is then found to be
\begin{eqnarray}
i\mathcal{M}_{i,q} = - \frac{ig_s t^{c}_{c_ic_j}}{(p_i+k)^2+i\varepsilon}\bar{u}(p_i)\left(2p_i^{\sigma}+k^{\sigma}+2iS^{\alpha\sigma}k_{\alpha}+2p_i^{\sigma}k^{\alpha}\frac{\partial}{\partial p_i^{\alpha}}\right)\otimes \mathcal{M}_{c_j}(p_1,p_2,\dots,p_{n+2})\epsilon^*_{\sigma}(k),
\end{eqnarray}
and for a final state anti-quark emitter:
\begin{eqnarray}
i\mathcal{M}_{i,\bar{q}} = \frac{ig_s t^{c}_{c_jc_i}}{(p_i+k)^2+i\varepsilon}\mathcal{M}_{c_j}(p_1,p_2,\dots,p_{n+2})\left(2p_i^{\sigma}+k^{\sigma}-2iS^{\alpha\sigma}k_{\alpha}+2k^{\alpha}\frac{\overleftarrow{\partial}}{\partial p_i^{\alpha}}p_i^{\sigma}\right)v(p_i)\epsilon^*_{\sigma}(k).
\end{eqnarray}
For a final state gluon emitter the next-to-soft matrix element is (Fig. \ref{fig:1e}):
\begin{align}
i\mathcal{M}_{i,g} &= \frac{g_s f^{cba}}{(p_i+k)^2+i\varepsilon}\epsilon^*_{\mu}(p_i)\left(g^{\mu\rho}(2p_i^{\sigma}+k^{\sigma})+2iM_{\mathcal{L}}^{\sigma\alpha,\rho\mu}k_{\alpha}+2g^{\mu\rho}p_i^{\sigma}k^{\alpha}\frac{{\partial}}{{\partial} p_i^{\alpha}}\right)\notag\\
&\quad\times
\mathcal{M}_{\rho,b}(p_1,p_2,\dots,p_{n+2})\epsilon^*_{\sigma}(k).
\end{align}
As for an initial state emitter, the NLP amplitude for a final state emitter also consists of a universal scalar term, a term that is sensitive to the spin of the emitter and a universal derivative term. 

So far, we have considered only emissions from the external legs of the non-radiative amplitude. We must also consider the emission of a gluon from {\it inside} the hard interaction, as shown in Fig.~\ref{fig:1f}. To this end, we may consider the Ward identity for the emitted gluon, which takes the form
\begin{eqnarray}
i\mathcal{M}_{{\rm NLP},\sigma}k^{\sigma} = \sum_{j=1}^{n+2}i\mathcal{M}_{j,\sigma}k^{\sigma} + i\mathcal{M}_{{\rm int},\sigma}k^{\sigma}  =  0 \hspace{0.5cm}
\rightarrow  \hspace{0.5cm} i\mathcal{M}_{{\rm int},\sigma}k^{\sigma} = -\sum_{j=1}^{n+2}i\mathcal{M}_{j,\sigma}k^{\sigma},
\label{derivterm}
\end{eqnarray}
where $\mathcal{M}_{j,\sigma}$ is the contribution to the {\it total} matrix element arising from gluon emission from an external line $j$, each consisting of a scalar, spin and derivative contribution as shown above. It is straightforwardly verified that the scalar and spin contributions vanish automatically upon contracting with $k^\sigma$, leaving only the derivative contribution, so that upon removing the gluon 4-momentum from both sides one obtains~\footnote{In principle, one may add a contribution $C_\sigma$ to the right-hand side of Eq.~(\ref{derivterm}), that is transverse by itself i.e. $k\cdot C=0$. Such contributions, however, can be ruled out based on gauge invariance and locality (see e.g. Refs.~\cite{Bern:2014vva,Broedel:2014fsa} for a recent discussion). } 
\begin{eqnarray}
\label{eq:NLPsoftgluonint}
i\mathcal{M}_{{\rm int},\sigma} =  \sum_{j}\eta_j g_s\mathbf{T}_j \otimes \frac{\partial}{\partial p_j^{\sigma}}\left[i\mathcal{M}_{\rm{H}}\right],
\end{eqnarray}
where $\eta_j = +1 (-1)$ for a hard emitting particle in the initial (final) state respectively. We use the symbol $\otimes$ to denote the fact that the action of the colour generator for each external leg should be interpreted with appropriate coupling of colour indices to the hard interaction. The derivative does not act on the asymptotic states of the hard scattering matrix element $\mathcal{M}_{\rm{H}}$. Combining this expression with the other contributions above, we can now write down a general formula for the emission of a soft gluon from an arbitrary amplitude up to next-to-soft level:
\begin{eqnarray}
\nonumber
\mathcal{A}_{\rm NLP} &=& \mathcal{A}_{\rm scal} + \mathcal{A}_{\rm spin} + \mathcal{A}_{\rm orb} \\
\label{eq:NLPgluon}
&=& \sum_{j=1}^{n+2} \frac{g_s \mathbf{T}_j}{2p_j \cdot k}\left(\mathcal{O}_{{\rm scal},j}^{\sigma}+\mathcal{O}_{{\rm spin},j}^{\sigma}+\mathcal{O}_{{\rm orb},j}^{\sigma}\right)\otimes i\mathcal{M}_{\rm H}(p_1,\dots,p_i,
\dots,p_{n+2})\epsilon^*_{\sigma}(k),
\end{eqnarray}
where $\mathcal{M}_{\rm H}$ again denotes the hard scattering matrix element, and the first two terms on the right-hand side constitute the scalar-like and spin contributions respectively. Furthermore, the third term is the orbital angular momentum operator associated with each external leg, and ${\bf T}_j$ a colour generator in the appropriate representation. We use the symbol $\otimes$ in the same way as before, with the extension that now also the spin generator should be interpreted with the appropriate coupling of the spinor and/or vector indices to the hard interaction. We define each of these actions carefully, for all possible types of external leg, in appendix~\ref{app:definitions}. Note that the scalar contribution commences at leading power (LP) in the soft expansion, whereas both of the angular momentum contributions are NLP  only.

In Eq.~(\ref{eq:NLPgluon}), the hard scattering matrix element contains only those momenta that are also present at LO. These do not obey momentum conservation once the extra radiation is present, and there appears to be an ambiguity in how one shares the momentum of the additional radiation between these existing momenta (see e.g. ref.~\cite{Gervais:2017yxv} for a particularly complete discussion of this point). We will see in Sections~\ref{sec:DIS}--\ref{sec:results} that actually there is no such ambiguity, as the form of the momentum shift created by the angular momentum operators of Eq.~(\ref{eq:NLPgluon}) is completely fixed. Furthermore, exact momentum conservation is enforced by integrating over the complete phase space, which is not included in the amplitude itself.  

The result of Eq.~(\ref{eq:NLPgluon}) has previously been derived in a more formal context~\cite{Casali:2014xpa}, where it is known as the {\it next-to-soft theorem}. It was motivated by a similar result in gravity~\cite{Cachazo:2014fwa,White:2011yy}, that generalises the leading soft results of Ref.~\cite{Weinberg:1965nx}. Our reason for carefully rederiving this result here is twofold. Firstly, we may contrast this derivation with a similar analysis for the emission of soft quarks, to be carried out in the following section. Secondly, in applying Eq.~(\ref{eq:NLPgluon}) to example scattering processes in the remainder of the paper, it is useful to have a precise record of how to keep track of colour and spinor/vector indices. The above derivation (and the results of appendix~\ref{app:definitions}) are particularly useful in this regard. Before moving on, note that the next-to-soft formalism of Eq.~(\ref{eq:NLPgluon}) applies for arbitrary tree-level induced processes with external momenta in the non-radiative amplitude held fixed. As such, we are certainly entitled to apply it to the processes considered in this paper, but must bear in mind that divergent threshold contributions which are {\it not} associated with (next-to)-soft behaviour may not be correctly described. We will see this explicitly in what follows. 

\subsection{Radiation of soft quarks}\label{sec:NLPquark}

\begin{figure}
\centering
\begin{subfigure}{0.31\textwidth}
\centering
\begin{fmffile}{diagramsoftquark1}
\begin{fmfchar*}(100,100)
  \fmfleft{i1,i2}
  \fmfright{f1,f2}
  \fmfforce{(0,0)}{i2} 
  \fmfforce{(0,h)}{i1} 
  \fmfforce{(.3w,h)}{f3} 
  \fmfforce{(.15w,.85h)}{g1}
  \fmfforce{(.5w,.5h)}{v1}
  \fmfforce{(.95w,.05h)}{v2}
  \fmfforce{(.95w,.95h)}{v3}
  \fmfforce{(w,h)}{f1} 
  \fmfforce{(w,0)}{f2} 
  \fmf{gluon}{i1,g1} 
  \fmf{fermion,tension=1,label=$p_{1,,c_j}'$,label.side=left,label.dist=-0.5}{v1,g1}
  \fmf{fermion}{g1,f3}
  \fmf{double}{v1,i2}
  \fmflabel{$p_{1,\mu,a}$}{i1}
  \fmflabel{$p_{2}$}{i2}
  \fmflabel{$p_{3}$}{f1}
  \fmflabel{$k_{,c_m}$}{f3}
  \fmflabel{$p_{n+2}$}{f2}
  \fmf{dots,right=0.25}{v2,v3}
  \fmf{double}{v2,v1}
  \fmf{double}{v3,v1}
  \fmf{double}{f1,v3}
  \fmf{double}{f2,v2}
  \fmfv{decor.shape=circle,decor.filled=empty, d.si=.45w,label=$\mathcal{M}_{c_j}(p_1')$,label.dist=-18}{v1}
\end{fmfchar*}
\end{fmffile} 
\vspace{0.5cm}
\caption{ } \label{fig:2aa}
\end{subfigure}
\hspace{0.5cm}
\begin{subfigure}{0.31\textwidth}
\centering
\begin{fmffile}{diagramgen3}
\begin{fmfchar*}(100,100)
  \fmfleft{i1,i2}
  \fmfright{f1,f2,f3}
  \fmfforce{(0,0)}{i2} 
  \fmfforce{(0,h)}{i1} 
  \fmfforce{(.3w,h)}{f3} 
  \fmfforce{(.15w,.85h)}{g1}
  \fmfforce{(.5w,.5h)}{v1}
  \fmfforce{(.95w,.05h)}{v2}
  \fmfforce{(.95w,.95h)}{v3}
  \fmfforce{(w,h)}{f1} 
  \fmfforce{(w,0)}{f2} 
  \fmf{fermion}{i1,g1} 
  \fmf{fermion}{g1,f3}
     \fmf{gluon,tension=1,label=$p_{1,,\rho,,b}'$,label.side=right,label.dist=1.2}{g1,v1}
   \fmf{double}{v1,i2}
  \fmflabel{$p_{1,c_i}$}{i1}
  \fmflabel{$p_{2}$}{i2}
  \fmflabel{$p_{3}$}{f1}
  \fmflabel{$k_{,c_m}$}{f3}
  \fmflabel{$p_{n+2}$}{f2}
  \fmf{dots,right=0.25}{v2,v3}
  \fmf{double}{v2,v1}
  \fmf{double}{v3,v1}
  \fmf{double}{f1,v3}
  \fmf{double}{f2,v2}
  \fmfv{decor.shape=circle,decor.filled=empty, d.si=.45w,label=$\mathcal{M}_{\rho,,b}(p_1')$,label.dist=-20.5}{v1}
\end{fmfchar*}
\end{fmffile} 
\vspace{0.5cm}
\caption{ } \label{fig:2bb}
\end{subfigure}
\vspace{0.5cm}
    \caption{(a) Feynman diagram for the emission of one additional quark carrying momentum $k$ and colour $c_m$ from an initial state gluon carrying momentum $p_1$ and colour $a$. The momenta of the particles is defined to flow from left to right for all external lines. The hard scattering matrix element $\mathcal{M}_{c_j}(p_1')$ is defined to contain all external states, except for the polarisation vector $\epsilon^{\mu}(p_1)$ and the spinor $\bar{u}(k)$. By the emission of a quark, the identity and the colour of the external gluon changes. (b) Feynman diagram for the emission of one additional quark carrying momentum $k$ and colour $c_m$ from an initial state quark carrying momentum $p_1$ and colour $c_i$. The hard scattering matrix element $\mathcal{M}_{\rho,b}(p_1')$ is defined to contain all external states, except for the spinors $u(p_1)$ and $\bar{u}(k)$.}
    \label{fig:diagrams}
\end{figure}
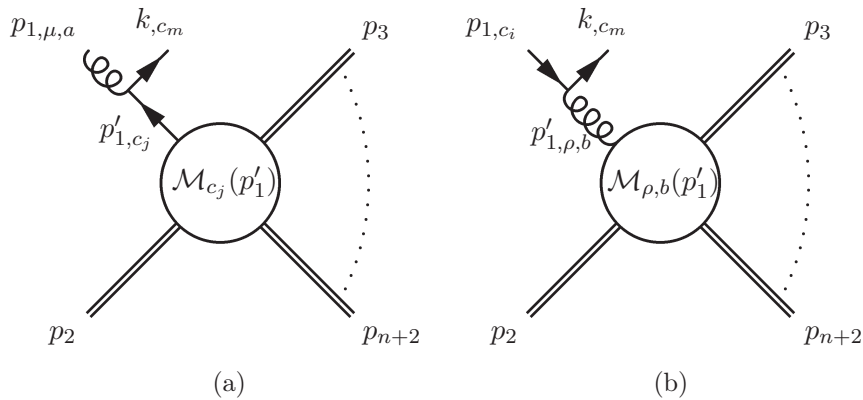

Having reviewed the universal NLO amplitude for the emission of a (next-to-)soft gluon, we now turn to the emission of one additional soft quark. One must then consider all possible partonic splittings that can lead to such an emission, which we show for the case of emission from the initial state in Fig.~\ref{fig:diagrams}. Let us first consider an initial state gluon splitting into a quark-antiquark pair, where the antiquark participates in the hard interaction (Fig. \ref{fig:2aa}). The resulting matrix element is
\begin{eqnarray}
i\mathcal{M}_{\mathcal{Q},1,g} = \frac{ig_st^{a}_{c_mc_j}}{(p_1-k)^2+i\varepsilon} \epsilon^{\mu}(p_1)\bar{u}(k)\gamma_{\mu}( \slashed{p_1}-\slashed{k})\mathcal{M}_{c_j}(p_1 - k,p_2,\dots,p_{n+2}),
\end{eqnarray}
where momenta and colour/Lorentz indices are labelled in the figure. The subscript $\mathcal{Q}$ is used to indicate the emission of a soft quark. From the fermion completeness relation for the emitted soft quark
\begin{eqnarray}
\sum_{\rm spins} u(k)\bar{u}(k)=\slashed{k},
\end{eqnarray}
we see that the spinor for the emitted quark scales with soft momentum as ${\cal O}(k^{1/2})$. Thus, the leading power of divergence for $k^{\sigma} \rightarrow 0$ in the matrix element is $\mathcal{O}\left(k^{-1/2}\right)$ (as opposed to ${\cal O}(k^{-1})$ for the soft gluon case). It will therefore not give rise to a leading power threshold contribution, but will instead contribute at NLP accuracy. Furthermore, as the leading contribution from soft quark emission is already ${\cal O}(k^{-1/2})$, there will be no additional contribution from the $\mathcal{O}(k)$ terms in the hard scattering matrix element or the Dirac propagator. The matrix element for soft quark emission becomes
\begin{eqnarray}
i\mathcal{M}_{\mathcal{Q},1,g} =  \frac{ig_st^{a}_{c_mc_j}}{(p_1-k)^2+i\varepsilon}\epsilon^{\mu}(p_1)\bar{u}(k)\gamma_{\mu}\slashed{p}_1\mathcal{M}_{c_j}(p_1,p_2,\dots,p_{n+2}).
\end{eqnarray}
A similar exercise can be performed if the initial state involves a quark splitting into a quark-gluon pair (Fig.~\ref{fig:2bb}), and one obtains
\begin{eqnarray}
i\mathcal{M}_{\mathcal{Q},1,g} =  \frac{ig_st^{b}_{c_mc_i}}{(p_1-k)^2+i\varepsilon}\bar{u}(k)\gamma^{\rho}u(p_1)\mathcal{M}_{\rho,b}(p_1,p_2,\dots,p_{n+2}).
\end{eqnarray}

The analysis for a final state particle emitting soft quarks is similar, as is the case of antiquark emission. Thus, we do not explicitly report the intermediate steps here. In the previous analysis of gluon emission, we needed to consider the possibility that a gluon was emitted from inside the hard scattering process, i.e. Fig.~\ref{fig:1f}. There is no such possibility here, which follows from the fact that the soft quark emission is by itself already a subleading effect in the momentum expansion, and also that any internal line is by definition far off-shell.  Emission of an internal soft quark is then sub-sub-leading in the momentum expansion, thus irrelevant at NLP order.

\begin{figure}
\centering 
\begin{fmffile}{action1}
\begin{eqnarray*}
\mathcal{Q}_j\left\{\hspace{0.1cm}\begin{gathered}
\begin{fmfchar*}(50,50)
  \fmfleft{i1}
  \fmfforce{(0,.5h)}{i1} 
  \fmfforce{(.w,.5h)}{v1}
  \fmf{gluon,tension=1,label=$p_{j,,\mu,,a}$,label.side=left}{i1,v1}
  \fmfv{decor.shape=circle,decor.filled=shaded, d.si=.4w}{v1}
\end{fmfchar*}
\hspace{0.5cm}
\end{gathered}\right\}&=&\hspace{0.1cm}\begin{gathered}
\begin{fmfchar*}(80,50)
  \fmfleft{i1}
  \fmfright{f1}
  \fmfforce{(0,.5h)}{i1} 
  \fmfforce{(.8w,0)}{f1} 
  \fmfforce{(.w,.5h)}{v1}
  \fmfforce{(.4w,.5h)}{v2}
  \fmf{gluon,tension=1,label=$p_{j,,\mu,,a}$,label.side=left}{v2,v1}
  \fmf{fermion,tension=1,label=$p_{j,,c_j}$,label.side=left}{i1,v2}
  \fmf{fermion,tension=1,label=$k_{,,c_m}$,label.side=right}{v2,f1}
  \fmfv{decor.shape=circle,decor.filled=shaded, d.si=.4h}{v1}
  \fmfv{decor.shape=circle,decor.filled=full, d.si=.02h}{v2}
\end{fmfchar*} 
\hspace{0.5cm}
\end{gathered}+
\hspace{0.1cm}\begin{gathered}
\begin{fmfchar*}(80,50)
  \fmfleft{i1}
  \fmfright{f1}
  \fmfforce{(0,.5h)}{i1} 
  \fmfforce{(.8w,0)}{f1} 
  \fmfforce{(.w,.5h)}{v1}
  \fmfforce{(.4w,.5h)}{v2}
  \fmf{gluon,tension=1,label=$p_{j,,\mu,,a}$,label.side=left}{v2,v1}
  \fmf{fermion,tension=1,label=$p_{j,,c_j}$,label.side=right}{v2,i1}
  \fmf{fermion,tension=1,label=$k_{,,c_m}$,label.side=left}{f1,v2}
  \fmfv{decor.shape=circle,decor.filled=shaded, d.si=.4h}{v1}
  \fmfv{decor.shape=circle,decor.filled=full, d.si=.02h}{v2}
\end{fmfchar*}
\end{gathered}\\%%%%%%%%%%%%%%%%%%%%%%%%%%%%%%%%%%%%%%%%%%%%%%%
\mathcal{Q}_j\left\{\hspace{0.1cm}\begin{gathered}
\begin{fmfchar*}(50,50)
  \fmfleft{i1}
  \fmfforce{(0,.5h)}{i1} 
  \fmfforce{(.w,.5h)}{v1}
  \fmf{fermion,tension=1,label=$p_{j,,c_j}$,label.side=left}{i1,v1}
  \fmfv{decor.shape=circle,decor.filled=shaded, d.si=.4w}{v1}
\end{fmfchar*}
\hspace{0.5cm}
\end{gathered}\right\}&=&\hspace{0.1cm}\begin{gathered}
\begin{fmfchar*}(80,50)
  \fmfleft{i1}
  \fmfright{f1}
  \fmfforce{(0,.5h)}{i1} 
  \fmfforce{(.8w,0)}{f1} 
  \fmfforce{(.w,.5h)}{v1}
  \fmfforce{(.4w,.5h)}{v2}
  \fmf{fermion,tension=1,label=$p_{j,,c_j}$,label.side=left}{v2,v1}
  \fmf{gluon,tension=1,label=$p_{j,,\mu,,a}$,label.side=left}{i1,v2}
  \fmf{fermion,tension=1,label=$k_{,,c_m}$,label.side=left}{f1,v2}
  \fmfv{decor.shape=circle,decor.filled=shaded, d.si=.4h}{v1}
  \fmfv{decor.shape=circle,decor.filled=full, d.si=.02h}{v2}
\end{fmfchar*} 
\end{gathered}\\%%%%%%%%%%%%%%%%%%%%%%%%%%%%%%%%%%%
\mathcal{Q}_j\left\{\hspace{0.1cm}\begin{gathered}
\begin{fmfchar*}(50,50)
  \fmfleft{i1}
  \fmfforce{(0,.5h)}{i1} 
  \fmfforce{(.w,.5h)}{v1}
  \fmf{fermion,tension=1,label=$p_{j,,c_j}$,label.side=right}{v1,i1}
  \fmfv{decor.shape=circle,decor.filled=shaded, d.si=.4w}{v1}
\end{fmfchar*}
\hspace{0.5cm}
\end{gathered}\right\}&=&\hspace{0.1cm}\begin{gathered}
\begin{fmfchar*}(80,50)
  \fmfleft{i1}
  \fmfright{f1}
  \fmfforce{(0,.5h)}{i1} 
  \fmfforce{(.8w,0)}{f1} 
  \fmfforce{(.w,.5h)}{v1}
  \fmfforce{(.4w,.5h)}{v2}
  \fmf{fermion,tension=1,label=$p_{j,,c_j}$,label.side=right}{v1,v2}
  \fmf{gluon,tension=1,label=$p_{j,,\mu,,a}$,label.side=left}{i1,v2}
  \fmf{fermion,tension=1,label=$k_{,,c_m}$,label.side=right}{v2,f1}
  \fmfv{decor.shape=circle,decor.filled=shaded, d.si=.4h}{v1}
  \fmfv{decor.shape=circle,decor.filled=full, d.si=.02h}{v2}
\end{fmfchar*} 
\end{gathered}\\%%%%%%%%%%%%%%%%%%%%%%%%%%%%%%%%%%%
\mathcal{Q}_j\left\{\hspace{0.5cm}\begin{gathered}
\begin{fmfchar*}(50,50)
  \fmfright{f1}
  \fmfleft{v1}
  \fmfforce{(.w,.5h)}{f1} 
  \fmfforce{(0,.5h)}{v1}
  \fmf{fermion,tension=1,label=$p_{j,,c_j}$,label.side=left}{v1,f1}
  \fmfv{decor.shape=circle,decor.filled=shaded, d.si=.4w}{v1}
\end{fmfchar*}
\hspace{.1cm}
\end{gathered}\right\}&=&\hspace{.5cm}\begin{gathered}
\begin{fmfchar*}(80,50)
  \fmfright{f1,f2}
  \fmfleft{v1}
  \fmfforce{(.w,.5h)}{f1} 
  \fmfforce{(.w,0)}{f2} 
  \fmfforce{(0,.5h)}{v1}
  \fmfforce{(.6w,.5h)}{v2}
  \fmf{fermion,tension=1,label=$p_{j,,c_j}$,label.side=left}{v1,v2}
  \fmf{gluon,tension=1,label=$p_{j,,\mu,,a}$,label.side=left}{v2,f1}
  \fmf{fermion,tension=1,label=$k_{,,c_m}$,label.side=right}{v2,f2}
  \fmfv{decor.shape=circle,decor.filled=shaded, d.si=.4h}{v1}
  \fmfv{decor.shape=circle,decor.filled=full, d.si=.02h}{v2}
\end{fmfchar*} 
\end{gathered}\\%%%%%%%%%%%%%%%%%%%%%%%%%%%%%%%%%%%
\mathcal{Q}_j\left\{\hspace{.5cm}\begin{gathered}
\begin{fmfchar*}(50,50)
  \fmfright{f1}
  \fmfleft{v1}
  \fmfforce{(.w,.5h)}{f1} 
  \fmfforce{(0,.5h)}{v1}
  \fmf{fermion,tension=1,label=$p_{j,,c_j}$,label.side=right}{f1,v1}
  \fmfv{decor.shape=circle,decor.filled=shaded, d.si=.4w}{v1}
\end{fmfchar*}
\hspace{.1cm}
\end{gathered}\right\}&=&\hspace{.5cm}\begin{gathered}
\begin{fmfchar*}(80,50)
  \fmfright{f1,f2}
  \fmfleft{v1}
  \fmfforce{(.w,.5h)}{f1} 
  \fmfforce{(.w,0)}{f2} 
  \fmfforce{(0,.5h)}{v1}
  \fmfforce{(.6w,.5h)}{v2}
  \fmf{fermion,tension=1,label=$p_{j,,c_j}$,label.side=right}{v2,v1}
  \fmf{gluon,tension=1,label=$p_{j,,\mu,,a}$,label.side=left}{v2,f1}
  \fmf{fermion,tension=1,label=$k_{,,c_m}$,label.side=left}{f2,v2}
  \fmfv{decor.shape=circle,decor.filled=shaded, d.si=.4h}{v1}
  \fmfv{decor.shape=circle,decor.filled=full, d.si=.02h}{v2}
\end{fmfchar*} 
\end{gathered}\\%%%%%%%%%%%%%%%%%%%%%%%%%%%%%%%%%%%
\mathcal{Q}_j\left\{\hspace{.5cm}\begin{gathered}
\begin{fmfchar*}(50,50)
  \fmfright{f1}
  \fmfleft{v1}
  \fmfforce{(.w,.5h)}{f1} 
  \fmfforce{(0,.5h)}{v1}
  \fmf{gluon,tension=1,label=$p_{j,,\mu,,a}$,label.side=left}{v1,f1}
  \fmfv{decor.shape=circle,decor.filled=shaded, d.si=.4w}{v1}
\end{fmfchar*}
\hspace{.1cm}
\end{gathered}\right\}&=&\hspace{.5cm}\begin{gathered}
\begin{fmfchar*}(80,50)
  \fmfright{f1,f2}
  \fmfleft{v1}
  \fmfforce{(.w,.5h)}{f1} 
  \fmfforce{(.w,0)}{f2} 
  \fmfforce{(0,.5h)}{v1}
  \fmfforce{(.6w,.5h)}{v2}
  \fmf{gluon,tension=1,label=$p_{j,,\mu,,a}$,label.side=left}{v1,v2}
  \fmf{fermion,tension=1,label=$p_{j,,c_j}$,label.side=left}{v2,f1}
  \fmf{fermion,tension=1,label=$k_{,,c_m}$,label.side=left}{f2,v2}
  \fmfv{decor.shape=circle,decor.filled=shaded, d.si=.4h}{v1}
  \fmfv{decor.shape=circle,decor.filled=full, d.si=.02h}{v2}
\end{fmfchar*} 
\hspace{.1cm}
\end{gathered}+\hspace{.5cm}\begin{gathered}
\begin{fmfchar*}(80,50)
  \fmfright{f1,f2}
  \fmfleft{v1}
  \fmfforce{(.w,.5h)}{f1} 
  \fmfforce{(.w,0)}{f2} 
  \fmfforce{(0,.5h)}{v1}
  \fmfforce{(.6w,.5h)}{v2}
  \fmf{gluon,tension=1,label=$p_{j,,\mu,,a}$,label.side=left}{v1,v2}
  \fmf{fermion,tension=1,label=$p_{j,,c_j}$,label.side=right}{f1,v2}
  \fmf{fermion,tension=1,label=$k_{,,c_m}$,label.side=right}{v2,f2}
  \fmfv{decor.shape=circle,decor.filled=shaded, d.si=.4h}{v1}
  \fmfv{decor.shape=circle,decor.filled=full, d.si=.02h}{v2}
\end{fmfchar*} 
\end{gathered}\\%%%%%%%%%%%%%%%%%%%%%%%%%%%%%%%%%%%
\end{eqnarray*}
\end{fmffile}
\caption{Action of the quark emission operator ${\cal Q}_j$ on an external parton line $j$, where all possible cases of incoming or outgoing line, and all parton species are considered. All momenta are defined to flow from left to right. The explicit contributions to the amplitude from each possibility are collected in appendix~\ref{app:definitions}. }
\label{fig:Soperator}
\end{figure}

As for the gluon case, we can write a compact universal formula for soft quark emission. In order to do this, we introduce a {\it quark emission operator} ${\cal Q}_i$, which acts on a given external parton line $i$ to produce the emission of a quark or antiquark. The action of this operator on every species of incoming/outgoing parton leg is shown diagrammatically in Fig.~\ref{fig:Soperator}, and we collect the explicit rules for the amplitude from each possibility in appendix~\ref{app:definitions}. Armed with the quark emission operator, we may write the following general formula for the next-to-soft amplitude arising from soft (anti-)quark emission:
\begin{eqnarray}
\label{eq:NLPquark}
\mathcal{A}_{\rm NLP,\mathcal{Q}} &=& \sum_{j=1}^{n+2} \frac{g_s}{2p_j\cdot k}\mathcal{Q}_j \otimes i\mathcal{M}_{H,j}(p_1,p_2,\dots,p_j,\dots,p_{n+2}).
\end{eqnarray}
This is very different to the next-to-soft gluon formalism of Eq.~(\ref{eq:NLPgluon}), in that there is no equivalent of the scalar and orbital angular momentum contributions. The quark emission operator generates a single ``external emission'' from the non-radiative amplitude, that commences at NLP order in the momentum expansion. Unlike the gluon case, it must change the identity of the parton that enters the hard scattering process. 

In this and the previous section, we have reviewed the derivation of a universal next-to-soft amplitude for the emission of a single additional gluon from a general Born process, up to next-to-soft order in its momentum. We have also derived a similar result for the emission of soft (anti-) quarks, which involved introducing the quark emission operator of Fig.~\ref{fig:Soperator}. We will now illustrate how to use these expressions in a series of example scattering processes, of increasing complication. 

\section{NLP contributions in DIS at NLO}
\label{sec:DIS}
\begin{figure}
\centering
\centering
\begin{subfigure}{0.18\textwidth}
\centering
\begin{fmffile}{DISkinqq1}
\begin{fmfchar*}(60,60)
  \fmfleft{i1,i2}
  \fmfright{f1,f2}
  \fmfforce{(0,0)}{i2} 
  \fmfforce{(0,h)}{i1} 
  \fmfforce{(.5w,.5h)}{v1}
  \fmfforce{(w,.5h)}{f1} 
  \fmfforce{(.w,.h)}{f2} 
  \fmfforce{(.9w,.5h)}{v2}
  \fmf{photon}{i1,v1}
  \fmf{fermion}{i2,v1}
  \fmf{fermion}{v1,f1}
  \fmflabel{$q_{\mu}$}{i1}
  \fmflabel{$p_{,c_i}$}{i2}
  \fmflabel{$k_{2,c_j}$}{f1}
  \fmfv{decor.shape=circle,decor.filled=shaded, d.si=.4w}{v1}
  \fmfv{decor.shape=circle,decor.filled=full, d.si=.02w}{v2}
\end{fmfchar*}
\end{fmffile}
\vspace{0.5cm}
\caption{ } \label{fig:8a}
\end{subfigure}
\hspace{1cm}
\begin{subfigure}{0.18\textwidth}
\centering
\begin{fmffile}{DISkinqq2}
\begin{fmfchar*}(60,60)
    \fmfleft{i1,i2}
  \fmfright{f1,f2}
  \fmfforce{(0,0)}{i2} 
  \fmfforce{(0,h)}{i1} 
  \fmfforce{(.5w,.5h)}{v1}
  \fmfforce{(w,.5h)}{f1} 
  \fmfforce{(.4w,0)}{f2} 
  \fmfforce{(.2w,.2h)}{v2}
  \fmf{photon}{i1,v1}
  \fmf{fermion}{i2,v1}
  \fmf{fermion}{v1,f1}
  \fmf{gluon}{v2,f2}
  \fmflabel{$q$}{i1}
  \fmflabel{$p$}{i2}
  \fmflabel{$k_2$}{f1}
  \fmflabel{$k_{\sigma,a}$}{f2}
  \fmfv{decor.shape=circle,decor.filled=shaded, d.si=.4w}{v1}
  \fmfv{decor.shape=circle,decor.filled=full, d.si=.02w}{v2}
\end{fmfchar*}
\end{fmffile}
\vspace{0.5cm}
\caption{ } \label{fig:8b}
\end{subfigure}
\hspace{0.8cm}
\begin{subfigure}{0.18\textwidth}

\centering
\begin{fmffile}{DISkinqq3}
\begin{fmfchar*}(60,60)
    \fmfleft{i1,i2}
  \fmfright{f1,f2}
  \fmfforce{(0,0)}{i2} 
  \fmfforce{(0,h)}{i1} 
  \fmfforce{(.5w,.5h)}{v1}
  \fmfforce{(w,.5h)}{f1} 
  \fmfforce{(.w,.3h)}{f2} 
  \fmfforce{(.8w,.5h)}{v2}
  \fmf{photon}{i1,v1}
  \fmf{fermion}{i2,v1}
  \fmf{fermion}{v1,f1}
  \fmf{gluon}{v2,f2}
  \fmflabel{$q$}{i1}
  \fmflabel{$p$}{i2}
  \fmflabel{$k_2$}{f1}
  \fmflabel{$k$}{f2}
  \fmfv{decor.shape=circle,decor.filled=shaded, d.si=.4w}{v1}
  \fmfv{decor.shape=circle,decor.filled=full, d.si=.02w}{v2}
\end{fmfchar*}
\end{fmffile}
\vspace{0.5cm}
\caption{ } \label{fig:8c}
\end{subfigure}
\hspace{0.8cm}
\begin{subfigure}{0.18\textwidth}
\centering
\begin{fmffile}{DISqq4}
\begin{fmfchar*}(60,60)
   \fmfleft{i1,i2}
  \fmfright{f1,f2}
  \fmfforce{(0,0)}{i2} 
  \fmfforce{(0,h)}{i1} 
  \fmfforce{(.5w,.5h)}{v1}
  \fmfforce{(w,.5h)}{f1} 
  \fmfforce{(.w,0)}{f2} 
  \fmfforce{(.8w,.5h)}{v2}
  \fmf{photon}{i1,v1}
  \fmf{fermion}{i2,v1}
  \fmf{fermion}{v1,f1}
  \fmf{gluon}{v1,f2}
  \fmflabel{$q$}{i1}
  \fmflabel{$p$}{i2}
  \fmflabel{$k_2$}{f1}
  \fmflabel{$k$}{f2}
  \fmfv{decor.shape=circle,decor.filled=shaded, d.si=.4w}{v1}
\end{fmfchar*}
\end{fmffile}
\vspace{0.5cm}
\caption{ } \label{fig:8d}
\end{subfigure}
\caption{Diagrams for the DIS process: (a) shows the LO contribution, whilst (b)--(d) show all possible gluon emissions at NLO for the quark channel. Here $(p,q,k_2)$ denote the 4-momenta, $\mu$ and $\sigma$ are Lorentz indices, and $c_i$ ($a$) denotes a colour index in the fundamental (adjoint) representation.}
\label{fig:DIS}.
\end{figure}
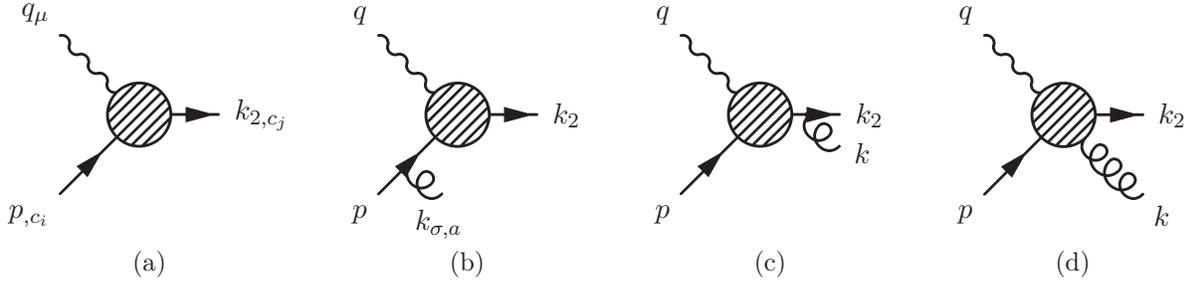
In this section, we will consider arguably the simplest process that has an unobserved parton in the final state, namely deep inelastic scattering (DIS):
\begin{equation}
e^-(k)+q(p)\rightarrow e^-(k')+q(k_2).\nonumber
\label{DISdef}
\end{equation}
We can then use the formulae derived in section \ref{sec:NLPgluon} and \ref{sec:NLPquark} to compute the NLO cross-section, up to next-to-soft order in the momentum of the emitted radiation. For the calculation we will utilise dimensional regularisation in $d=4-2\varepsilon$ dimensions, and use $\mu$ to indicate the renormalisation scale. The strong coupling is denoted as usual by $\alpha_s \equiv \alpha_s(\mu^2) =g_s^2 / (4\pi) $.  As is customary (see e.g. Refs.~\cite{Zijlstra:1992bb,Zijlstra:1992qd}), the complete squared amplitude before summing/averaging over spins, polarisations and colours can be written as
\begin{equation}
|\mathcal{A}|^2 \sim L^{\mu\nu}H_{\mu\nu},
\end{equation}
where $H^{\mu\nu}$ ($L^{\mu\nu}$) is the {\it hadronic} ({\it leptonic}) tensor respectively. To calculate the proton structure functions, it is sufficient to consider the hadronic tensor only. Thus, we may assume an initial state consisting of a quark and a spacelike off-shell photon, as shown in Fig.~\ref{fig:DIS}. Let us now consider the NLO cross-section up to NLP order, which we calculate using Eq.~(\ref{eq:NLPgluon}). For the hadronic tensor, we need the squared amplitude with different space-time indices for the off-shell photon, which reads
\begin{equation}
{\cal A}^\mu {\cal A}^{\dag\,\nu}=
{\cal A}^\mu_{\rm scal}{\cal A}^{\dag\,\nu}_{\rm scal}
+2{\rm Re}\left[{\cal A}^\mu_{\rm scal}{\cal A}^{\dag\,\nu}_{\rm spin}
+{\cal A}^\mu_{\rm scal}{\cal A}^{\dag\,\nu}_{\rm orb}\right]+\ldots,
\label{ampsq}
\end{equation}
where we have included only those terms in the squared amplitude that are up to NLP order in the next-to-soft expansion. By explicit calculation (similar to those performed in Ref.~\cite{DelDuca:2017twk}), we find that the first term on the right-hand side of Eq.~(\ref{ampsq}) is given by
\begin{eqnarray}
\langle{\cal A}_{{\rm scal},\mu}{\cal A}^{\dag}_{{\rm scal},\nu}\rangle = g_s^2 C_F \frac{p\cdot k_2}{(p\cdot k) (k_2\cdot k)}{\rm Tr}\left[\slashed{k_2}\mathcal{M}_{\mu}(p,k_2)\slashed{p}\mathcal{M}^{\dagger}_{\nu}(p,k_2)\right],
\end{eqnarray}
where the bracket notation indicates that we have averaged over the initial state color and spin of the quark (resulting in a factor of $\frac{1}{2N_C}$), and summed over final state spins and gluon polarisations. Furthermore, we have defined ${\cal M}^\mu(p,k_2)\equiv{\cal M}^\mu$ to be the LO amplitude, with external fermion wavefunctions removed, which we have allowed at present to be fully general. The scalar-spin contribution is found to be
\begin{eqnarray}
\label{eq:spinscal} 
\langle 2{\rm Re}\left[{\cal A}_{{\rm scal},\mu}{\cal A}^{\dag}_{{\rm spin},\nu}\right]\rangle &=& - g_s^2 C_F \left(\frac{1}{p\cdot k}-\frac{1}{k\cdot k_2}\right){\rm Tr}\left[\slashed{k_2}\mathcal{M}_{\mu}\slashed{p}\mathcal{M}^{\dagger}_{\nu}\right].
\end{eqnarray}
Finally, the scalar-orbital squared amplitude is given by
\begin{eqnarray}
\langle 2{\rm Re}\left[{\cal A}_{{\rm scal},\mu}{\cal A}^{\dag}_{{\rm orb},\nu}\right] \rangle&=& g_s^2C_F\frac{p\cdot k_2}{(p\cdot k)(k_2\cdot k)}\Bigg[{\rm Tr}\left[\slashed{k_2}\mathcal{M}_{\mu}\slashed{p}\left(\delta p \cdot\frac{\partial}{\partial p}-\delta k_2 \cdot\frac{\partial}{\delta k_2}\right)\mathcal{M}_{\nu}^{\dagger}\right]\nonumber \\
&& \hspace{3.1cm} + {\rm Tr}\left[\slashed{p}\mathcal{M}_{\nu}^{\dagger}\slashed{k_2}\left(\delta p \cdot\frac{\partial}{\partial p}-\delta k_2 \cdot\frac{\partial}{\partial k_2}\right)\mathcal{M}_{\mu}\right] \Bigg],
\end{eqnarray}
where we have defined the momentum shifts
\begin{eqnarray}
\delta p^{\alpha} &\equiv& -\frac{1}{2}\left(k^{\alpha}+\frac{k_2\cdot k}{p\cdot k_2}p^{\alpha} - \frac{p\cdot k}{p\cdot k_2}k_2^{\alpha}\right)\\
\delta k_2^{\alpha} &\equiv& -\frac{1}{2}\left(k^{\alpha}+\frac{p\cdot k}{p\cdot k_2}k_2^{\alpha} - \frac{k_2\cdot k}{p\cdot k_2}p^{\alpha}\right).
\label{momshifts}
\end{eqnarray}
We are now in a position to calculate the full NLP squared amplitude for DIS. First we will make use of the chain rule to write
\begin{eqnarray}
\langle 2{\rm Re}\left[{\cal A}_{{\rm scal},\mu}{\cal A}^{\dag}_{{\rm orb},\nu}\right]\rangle &=&
g_s^2C_F\frac{p\cdot k_2}{(p\cdot k)(k_2\cdot k)}\Bigg[\left(\delta p \cdot\frac{\partial}{\partial p}-\delta k_2 \cdot\frac{\partial}{\partial k_2}\right){\rm Tr}\left[\slashed{k_2}\mathcal{M}_{\mu}\slashed{p}\mathcal{M}_{\nu}^{\dagger}\right]\nonumber \\ 
&&\hspace{3.1cm}  + {\rm Tr}\left[\slashed{\delta k_2}\mathcal{M}_{\mu}\slashed{p}\mathcal{M}_{\nu}^{\dagger}\right] - {\rm Tr}\left[\slashed{k_2}\mathcal{M}_{\mu}\slashed{\delta p}\mathcal{M}_{\nu}^{\dagger}\right]  \Bigg].
\label{scalorbDIS}
\end{eqnarray}
The first term generates a momentum shift on the entire trace. The other two terms can be rewritten using a Sudakov decomposition for the emitted gluon momentum:
\begin{eqnarray}
\label{eq:sudakov}
k^{\mu}= \frac{p\cdot k}{p \cdot k_2}k_2^{\mu}+\frac{k_2\cdot k}{p \cdot k_2}p^{\mu}+k_T^{\mu},
\end{eqnarray}
so that one finds
\begin{eqnarray}
\langle 2{\rm Re}\left[{\cal A}_{{\rm scal},\mu}{\cal A}^{\dag}_{{\rm orb},\nu}\right] \rangle &=&
g_s^2C_F\frac{p\cdot k_2}{(p\cdot k)(k_2\cdot k)}\Bigg[\left(\delta p \cdot\frac{\partial}{\partial p}-\delta k_2 \cdot\frac{\partial}{\partial k_2}\right){\rm Tr}\left[\slashed{k_2}\mathcal{M}_{\mu}\slashed{p}\mathcal{M}_{\nu}^{\dagger}\right] \\
&&\hspace{3cm}  + \frac{k_2\cdot k}{p \cdot k_2}{\rm Tr}\left[\slashed{k_2}\mathcal{M}_{\mu}\slashed{p}\mathcal{M}_{\nu}^{\dagger}\right] - \frac{p\cdot k}{p \cdot k_2}{\rm Tr}\left[\slashed{k_2}\mathcal{M}_{\mu}\slashed{ p}\mathcal{M}_{\nu}^{\dagger}\right]  \Bigg]. \nonumber
\end{eqnarray}
Here we have ignored terms linear in $k_T$, as they will ultimately vanish upon integration over the final state phase space. The latter two terms can be combined with the scalar-spin contribution in Eq.~(\ref{eq:spinscal}), after which they also vanish. Putting everything together, the complete NLP squared amplitude can be written in terms of the LO hadronic tensor
\begin{equation}
H_{\mu\nu}(p,k_2)= \langle {\cal A}^{(0)}_\mu(p,k_2)\,
{\cal A}^{(0)^\dag}_\nu(p,k_2)\rangle ,
\label{Hmunudef}
\end{equation}
but with momenta shifted according to Eq.~(\ref{momshifts}):
\begin{eqnarray}
\langle {\cal A}_{\mu}{\cal A}^{\dag}_{\nu}\rangle \Big{|}_{{\rm LP+NLP}}= g_s^2 C_F \frac{p\cdot k_2}{(p\cdot k)(k_2\cdot k)}H_{\mu\nu}(p + \delta p, k_2 - \delta k_2).
\label{Asqres}
\end{eqnarray}
This is directly analogous to the case of colour singlet production examined in Ref.~\cite{DelDuca:2017twk}, which also found that the squared amplitude for the one real emission contribution could be written in terms of the momentum-shifted non-radiative amplitude. The forms of the shifts found here differ only in that the shift in $k_2$ has an opposite sign, owing to the fact that it is a final-, rather than initial-state momentum. Up to now we have allowed the LO stripped amplitude to be fully general, but we now use the explicit result for DIS: \footnote{Note that we have not included a factor of $iQ_{q}g_{\rm EM}$ here, which we define to be part of the leptonic tensor.} 
\begin{equation}
{\cal M}^\mu=\gamma^\mu,
\label{MmuDIS}
\end{equation}
before projecting the squared amplitude of Eq.~(\ref{Asqres}) with:
\begin{equation}
T_{2}^{\mu\nu} = -\frac{1}{4\pi}\frac{1}{2-2\varepsilon}\left(g^{\mu\nu}+(3-2\varepsilon)\frac{q^2}{(p\cdot q)^2}p^{\mu}p^{\nu}\right)
\end{equation}
to obtain the proton structure function $F_2^\gamma(x,Q^2)$ (see e.g. Refs.~\cite{Zijlstra:1992bb,Zijlstra:1992qd})\footnote{One may also consider the structure function $F_L$. However, this does not exhibit any logarithmic terms at NLO. See Ref.~\cite{Moch:2009mu} for a detailed discussion of threshold contributions at higher orders.}. To calculate the structure function, we use the following momentum parameterisation~\cite{Zijlstra:1992bb,Zijlstra:1992qd}:
\begin{eqnarray*}
p &=& \frac{s+Q^2}{2\sqrt{s}}\left(1,0,\dots,0,1\right)\\
q &=& \left(\frac{s-Q^2}{2\sqrt{s}},0,\dots,0,-\frac{s+Q^2}{2\sqrt{s}}\right)\\
k &=& \frac{\sqrt{s}}{2}\left(1,0,\dots,0,\sin{\theta},\cos{\theta}\right)\\
k_2 &=& \frac{\sqrt{s}}{2}\left(1,0,\dots,0,-\sin{\theta},-\cos{\theta}\right).
\end{eqnarray*}
Next defining $\cos \theta = 2y -1$ and $s = \frac{Q^2(1-x)}{x}$, one has
\begin{equation}
-q^2 = Q^2,\quad p\cdot k = \frac{Q^2(1-y)}{2x},\quad
p \cdot q = \frac{Q^2}{2x},\quad q \cdot k = \frac{Q^2 (y - x)}{2 x},
\end{equation}
such that the two-body final state phase space may be written as
\begin{eqnarray}
\int{\rm d}\Phi_2 = \frac{2\pi}{(4\pi)^{d/2}\Gamma\left(\frac{d-4}{2}\right)}\left(\frac{Q^2}{\mu^2}\right)^{d/2-2}\left(\frac{1-x}{x}\right)^{\frac{d-4}{2}}\int_0^1{\rm d}y \left(y(1-y)\right)^{\frac{d-4}{2}}.
\label{PS2}
\end{eqnarray}
Note that there is an overall factor of $x^{-(d-4)/2}\equiv x^{\varepsilon}$, which when expanded about $x=1$ contributes to NLP terms in the final result suppressed by a power of $\varepsilon$. Thus, NLP corrections to the phase space will not affect leading logarithmic behaviour. Using these ingredients, the result for the structure function, valid up to NLP order, is
\begin{eqnarray}
F_{2,{\rm LP+NLP}}^{\gamma}(x,Q^2) &=& \int{\rm d}\Phi_2 T_2^{\mu\nu} \langle {\cal A}_{\mu}{\cal A}^{\dag}_{\nu}\rangle \Big{|}_{{\rm LP+NLP}} \label{F2NLP}
\\ &=& \frac{\alpha_s}{4\pi}\left(-\frac{4}{\varepsilon}\frac{1}{1-x}+\frac{4}{\varepsilon}-\frac{4-4\ln(1-x)}{1-x} + 8 - 4\ln(1-x)+\mathcal{O}(1-x)\right),\nonumber
\end{eqnarray}
where we have set
\begin{equation}
\bar{\mu}^2 \equiv 4\pi{\rm e}^{-\gamma_E}\mu^2 = Q^2.
\end{equation}
This is the result obtained for the structure function in the next-to-soft approximation, as opposed to the full NLO result
\begin{eqnarray}
F_{2,{\rm NLO}}^{\gamma}(x,Q^2) = \frac{\alpha_s}{4\pi}\left(-\frac{4}{\varepsilon}\frac{1}{1-x}+\frac{4}{\varepsilon}-\frac{3-4\ln(1-x)}{1-x} + 14 - 4\ln(1-x)+\mathcal{O}(1-x)\right).
\label{F2full}
\end{eqnarray}
Comparison of Eqs.~(\ref{F2NLP}) and~(\ref{F2full}) shows that the next-to-soft expansion misses a LP term, and a constant at NLP order:
\begin{eqnarray}
F_{2,{\rm NLO}}^{\gamma}(x,Q^2) - F_{2,{\rm LP+NLP}}^{\gamma}(x,Q^2) = \frac{\alpha_s}{4\pi}\left(\frac{1}{1-x} + 6 +\mathcal{O}(1-x)\right).
\label{discrepancy}
\end{eqnarray}
The first term on the right-hand side forms a plus distribution when combined with virtual corrections. As is well-known, LP contributions in DIS correspond to the emitted gluon being either soft and/or collinear with the final state parton $k_2$, as shown in Fig.~\ref{fig:massivejet}. However, we see explicitly that the corrections to the next-to-soft approximation (i.e. collinear effects which are next-to-next-to-soft level and beyond) do not contribute to the {\it leading logarithms} at LP or NLP order in the threshold expansion. Thus, to capture NLP effects at LL order, we find that it is sufficient to employ the momentum shift of Eq.~(\ref{Asqres}). For the LL term, one may also approximate the phase space in Eq.~(\ref{PS2}), ignoring the factor $x^{\varepsilon}$. We therefore find that the LL contributions at LP and NLP order arise solely from the next-to-soft matrix element, combined with a LP-like phase space. This is similar to the conclusions of Ref.~\cite{DelDuca:2017twk}. 

It is in fact possible to describe in more detail the discrepancy of Eq.~(\ref{discrepancy}). Using our framework of soft quark emission, there is a way of obtaining the missing LP NLL contribution. If we let the quark emission operator $\mathcal{Q}_i$ act on the outgoing quark, we generate the diagram shown in Fig.~\ref{fig:2a}, whose contribution reads
\begin{equation}
F^\gamma_{2,a}=\frac{\alpha_s}{4\pi}\left(\frac{1}{1-x}+{\cal O}(1-x)\right).
\end{equation}
This shows that the LP NLL effects that do not arise from the next-to-soft gluon formalism can in this case be correctly obtained using the soft-quark formalism. The missing LP NLL term corresponds to the situation where the gluon becomes hard-collinear to the final state unobserved quark, in which case the quark is allowed to get soft. It is known at LP that this hard-collinear gluon information can also be captured by including jet functions \cite{Sterman:1987aj}, but here we explicitly see that this contribution is {\it exactly} generated by the soft quark emission operator. Our soft quark framework also allows us to include the other partonic DIS channel in a natural way, namely the one where the hard scattering is induced by a gluon that splits into a quark-anti-quark pair, shown for convenience in Fig. \ref{fig:2b}. This contribution turns out to be
\begin{eqnarray}
F^\gamma_{2,b}=\frac{\alpha_s}{4\pi}\left(-\frac{2}{\varepsilon}+2\ln(1-x)+\mathcal{O}(1-x)\right).
\end{eqnarray}
One observes the presence of a collinear pole associated with the quark-anti-quark pair. This in turn generates an NLP logarithmic term in the finite part, due to the interplay of the collinear pole with the factor $(1-x)^{-\varepsilon}$ in the 2-body phase space of Eq.~(\ref{PS2}). 

The constant in Eq.~(\ref{discrepancy}) is generated by the diagram shown in Fig. \ref{fig:2c} where all final state partons carry a hard momentum. It therefore cannot be obtained in our soft-quark and gluon formalism. This momentum configuration contributes as a constant and reads
\begin{equation}
F^\gamma_{2,c}=\frac{\alpha_s}{4\pi}\Big(6+{\cal O}(1-x)\Big).
\end{equation}
This is a non-leading NLP contribution, that is not captured by the soft expansion. It stems from squaring the contribution of the emission of a gluon from an initial state line, which makes it a {\it next-to-collinear} effect. The study of such contributions for arbitrary processes and at higher orders in perturbation theory deserves further investigation (although see Ref.~\cite{Nandan:2016ohb} for a discussion in a more formal context).

In this section, we have examined a first process with a final state parton (DIS), and found the next-to-soft formalism as derived in section \ref{sec:NLPgluon} can be used to derive a similar result to that obtained for colour singlet particle production in Ref.~\cite{DelDuca:2017twk}. That is, the NLO amplitude up to LP + NLP LL order in the next-to-soft expansion can be written in terms of the LO amplitude with shifted external momenta (Eq.~(\ref{Asqres})). Contrary to Ref.~\cite{DelDuca:2017twk}, in the present case, the next-to-soft gluon amplitude only leads to LP and NLP LL accuracy in the final result. However, the missing LP NLL information can actually be captured by the soft quark amplitude. This is because the missing LP NLL information actually originates from a momentum configuration where the quark and gluon are collinear to one another. The soft quark amplitude contains this missing collinear information exactly at NLO. To investigate how general this situation is, it is instructive to consider a second inclusive process with two final state jets, which we do in the following section.

\begin{figure}
\centering
\begin{subfigure}{0.25\textwidth}
\begin{fmffile}{mass}
\begin{fmfchar*}(120,60)
  \fmfleft{i1,i2}
  \fmfright{f1,f2}
  \fmfforce{(0,0)}{i2} 
  \fmfforce{(0,h)}{i1} 
  \fmfforce{(.2w,.5h)}{v1}
  \fmfforce{(.8w,.5h)}{v2}
  \fmfforce{(.5w,.5h)}{v2a}
  \fmfforce{(.35w,.5h)}{v3}
  \fmfforce{(.65w,.5h)}{v4}
  \fmfforce{(.5w,1.1h)}{v5}
  \fmfforce{(.5w,-0.1h)}{v6}
  \fmfforce{(.w,0)}{f1} 
  \fmfforce{(.w,h)}{f2} 
  \fmf{photon}{i1,v1}
  \fmf{fermion}{i2,v1}
  \fmf{fermion}{v1,v3}
  \fmf{gluon}{v4,v3}
  \fmf{fermion}{v4,v2}
  \fmf{photon}{f2,v2}
  \fmf{fermion}{v2,f1}
  \fmf{phantom, tension=1}{v3,v4}
  \fmf{fermion, right=1}{v3,v4}
  \fmf{dashes}{v5,v6}
  \fmflabel{$q$}{i1}
  \fmflabel{$p$}{i2}
  \fmflabel{$q$}{f2}
  \fmflabel{$p$}{f1}
  \fmfv{decor.shape=circle,decor.filled=shaded, d.si=.15w}{v1}
  \fmfv{decor.shape=circle,decor.filled=shaded, d.si=.15w}{v2}
\end{fmfchar*}
\end{fmffile}
\caption{ } \label{fig:2a}
\end{subfigure}
\hspace{1.5cm}
\begin{subfigure}{0.25\textwidth}
\begin{fmffile}{mass2IS}
\begin{fmfchar*}(120,60)
  \fmfleft{i1,i2}
  \fmfright{f1,f2}
  \fmfforce{(0,0)}{i2} 
  \fmfforce{(0,h)}{i1} 
  \fmfforce{(.2w,.5h)}{v1}
  \fmfforce{(.8w,.5h)}{v2}
  \fmfforce{(.5w,.5h)}{v2a}
  \fmfforce{(.1w,.2h)}{v3}
  \fmfforce{(.9w,.2h)}{v4}
  \fmfforce{(.5w,1.1h)}{v5}
  \fmfforce{(.5w,-0.1h)}{v6}
  \fmfforce{(.w,0)}{f1} 
  \fmfforce{(.w,h)}{f2} 
  \fmf{photon}{i1,v1}
  \fmf{gluon}{i2,v3}
  \fmf{fermion}{v3,v1}
  \fmf{fermion}{v1,v2a}
  \fmf{fermion}{v2a,v2}
  \fmf{photon}{f2,v2}
  \fmf{gluon}{v4,f1}
  \fmf{fermion}{v2,v4}
  \fmf{phantom, tension=1}{v3,v4}
  \fmf{fermion}{v4,v3}
  \fmf{dashes}{v5,v6}
  \fmflabel{$q$}{i1}
  \fmflabel{$p$}{i2}
  \fmflabel{$q$}{f2}
  \fmflabel{$p$}{f1}
  \fmfv{decor.shape=circle,decor.filled=shaded, d.si=.15w}{v1}
  \fmfv{decor.shape=circle,decor.filled=shaded, d.si=.15w}{v2}
\end{fmfchar*}
\end{fmffile}
\caption{ } \label{fig:2b}
\end{subfigure}
\hspace{1.5cm}
\begin{subfigure}{0.25\textwidth}
\begin{fmffile}{massIS}
\begin{fmfchar*}(120,60)
  \fmfleft{i1,i2}
  \fmfright{f1,f2}
  \fmfforce{(0,0)}{i2} 
  \fmfforce{(0,h)}{i1} 
  \fmfforce{(.2w,.5h)}{v1}
  \fmfforce{(.8w,.5h)}{v2}
  \fmfforce{(.5w,.5h)}{v2a}
  \fmfforce{(.07w,.12h)}{v3}
  \fmfforce{(.95w,.12h)}{v4}
  \fmfforce{(.5w,1.1h)}{v5}
  \fmfforce{(.5w,-0.1h)}{v6}
  \fmfforce{(.w,0)}{f1} 
  \fmfforce{(.w,h)}{f2} 
  \fmf{photon}{i1,v1}
  \fmf{fermion}{i2,v1}
  \fmf{fermion}{v1,v2a}
  \fmf{fermion}{v2a,v2}
  \fmf{photon}{f2,v2}
  \fmf{fermion}{v2,f1}
  \fmf{phantom, tension=1}{v3,v4}
  \fmf{gluon}{v4,v3}
  \fmf{dashes}{v5,v6}
  \fmflabel{$q$}{i1}
  \fmflabel{$p$}{i2}
  \fmflabel{$q$}{f2}
  \fmflabel{$p$}{f1}
  \fmfv{decor.shape=circle,decor.filled=shaded, d.si=.15w}{v1}
  \fmfv{decor.shape=circle,decor.filled=shaded, d.si=.15w}{v2}
\end{fmfchar*}
\end{fmffile}
\caption{ } \label{fig:2c}
\end{subfigure}
\vspace{0.3cm}
\caption{Contributions for the DIS $F_{2,{\rm NLO}}(x,Q^2)$ structure function, where the gluon is next-to-next-to-soft or beyond. In the first two diagrams, the soft momentum is carried by the (anti-)quark and in the final diagram none of the partons become soft.}
\label{fig:massivejet}
\end{figure}
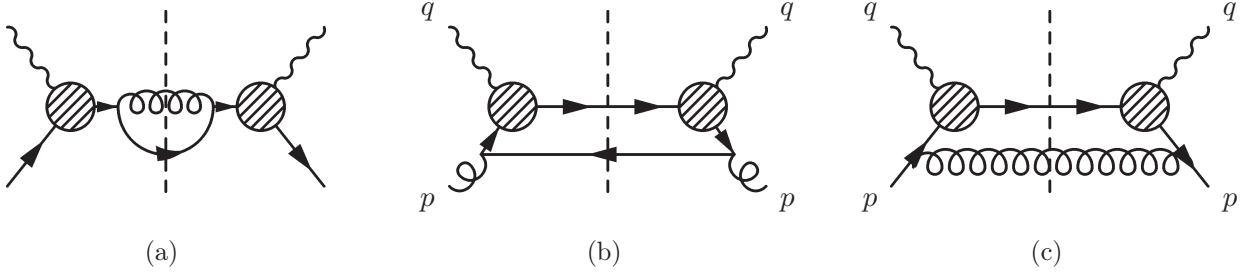

\section{NLP contributions in $e^+ e^-\rightarrow$ jets at NLO}
\label{sec:epem}
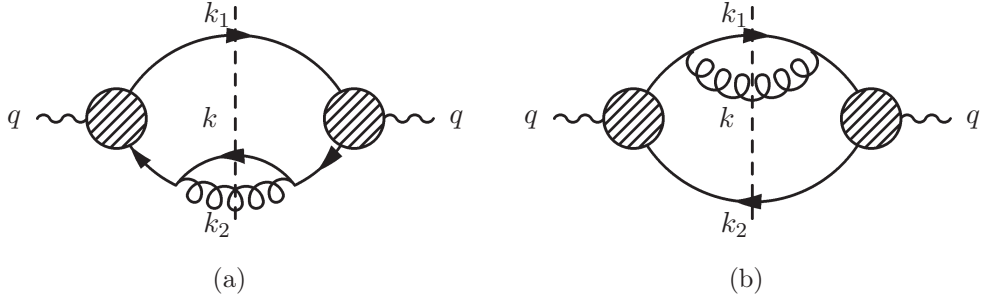
\begin{figure}[t]
\centering
\centering

\begin{subfigure}{0.31\textwidth}
\begin{fmffile}{mass2}
\begin{fmfchar*}(150,60)
  \fmfleft{i1,i2}
  \fmfright{f1,f2}
  \fmfforce{(0,.5h)}{i1} 
  \fmfforce{(.2w,.5h)}{v1}
  \fmfforce{(.8w,.5h)}{v2}
  \fmfforce{(.35w,.08h)}{v3}
  \fmfforce{(.8w,.h)}{v2a}
  \fmfforce{(.8w,0)}{v2b}
  \fmfforce{(.65w,.2h)}{v3a}
  \fmfforce{(.65w,.08h)}{v4}
  \fmfforce{(.5w,1.2h)}{v5}
  \fmfforce{(.5w,-0.2h)}{v6}
  \fmfforce{(.w,.5h)}{f1} 
  \fmf{photon}{i1,v1}
  \fmf{phantom,left=0.32,label=$k_1$,label.side=left}{v1,v2a}
  \fmf{phantom,right=0.32,label=$k_2$}{v1,v2b}
  \fmf{phantom,right=0.01,label=$k$}{v3a,v1}
  \fmf{fermion,left=.7}{v1,v2}
  \fmf{fermion,left=.2}{v3,v1}
  \fmf{fermion,left=.2}{v2,v4}
  \fmf{gluon,left=.3}{v4,v3}
  \fmf{fermion,right=.5}{v4,v3}
  \fmf{photon}{f1,v2}
  \fmf{dashes}{v5,v6}
  \fmflabel{$q$}{i1}
  \fmflabel{$q$}{f1}
  \fmfv{decor.shape=circle,decor.filled=shaded, d.si=.15w}{v1}
  \fmfv{decor.shape=circle,decor.filled=shaded, d.si=.15w}{v2}
\end{fmfchar*}
\end{fmffile}

\vspace{2mm}
\caption{ } \label{fig:3a}
\end{subfigure}
\hspace{1.5cm}
\begin{subfigure}{0.31\textwidth}
\begin{fmffile}{mass1}
\begin{fmfchar*}(150,60)
  \fmfleft{i1,i2}
  \fmfright{f1,f2}
  \fmfforce{(0,.5h)}{i1} 
  \fmfforce{(.2w,.5h)}{v1}
  \fmfforce{(.8w,.5h)}{v2}
  \fmfforce{(.35w,.92h)}{v3}
  \fmfforce{(.8w,.h)}{v2a}
  \fmfforce{(.8w,0)}{v2b}
  \fmfforce{(.65w,.2h)}{v3a}
  \fmfforce{(.65w,.92h)}{v4}
  \fmfforce{(.5w,1.2h)}{v5}
  \fmfforce{(.5w,-0.2h)}{v6}
  \fmfforce{(.w,.5h)}{f1} 
  \fmf{photon}{i1,v1}
  \fmf{fermion,left=.7}{v2,v1}
  \fmf{phantom,left=0.32,label=$k_1$,label.side=left}{v1,v2a}
  \fmf{phantom,right=0.32,label=$k_2$}{v1,v2b}
  \fmf{phantom,right=0.01,label=$k$}{v3a,v1}
  \fmf{fermion,left=.7}{v1,v2}
  \fmf{gluon,right=.5}{v3,v4}
  \fmf{photon}{f1,v2}
  \fmf{dashes}{v5,v6}
  \fmflabel{$q$}{i1}
  \fmflabel{$q$}{f1}
  \fmfv{decor.shape=circle,decor.filled=shaded, d.si=.15w}{v1}
  \fmfv{decor.shape=circle,decor.filled=shaded, d.si=.15w}{v2}
\end{fmfchar*}
\end{fmffile}
\vspace{2mm}
\caption{ } \label{fig:3b}
\end{subfigure}

\vspace{0.3cm}
\caption{Missing contributions for the parton fragmentation function $\mathcal{F}_{T,{\rm NLO}}(z,Q^2)$.}
\label{fig:massiveepemjet}
\end{figure}

The next process we will consider is that of hadroproduction in electron-positron annihilation:
\begin{eqnarray*}
e^+(p_1) + e^-(p_2) \rightarrow \gamma(q) \rightarrow q(k_1) + \bar{q}(k_2) + g(k),
\end{eqnarray*}
where we will choose $q(k_1)$ to fragment into the observed hadron, and be inclusive for the other final state particles. We are interested in the transverse parton fragmentation function (see e.g.~\cite{Rijken:1996ns})\footnote{We could also consider the longitudinal fragmentation function. However, as for the longitudinal structure function $F_L$ in DIS, this does not contain logarithmic contributions at NLO.}
\begin{eqnarray}
\mathcal{F}_{T}(z,Q^2) = \frac{1}{d-2}\left(-\frac{2k_1\cdot q}{q^2}W^{\mu}_{\mu}-\frac{2}{k_1\cdot q}k_1^{\mu}k_2^{\nu}W_{\mu\nu}\right).
\end{eqnarray}
Here $W_{\mu\nu}$ is the parton structure tensor
\begin{eqnarray}
W_{\mu\nu}(p,q)= \frac{z^{d-3}}{4\pi}\int {\rm d}\Phi_{2} \langle \mathcal{A}_{\mu}\mathcal{A}_{\nu}^{\dagger}\rangle ,
\label{Wmunudef}
\end{eqnarray}
where $\mathcal{A}_{\mu}$ now denotes the matrix element of the sub-process $\gamma^*(q) \rightarrow q(k_1)\bar{q}(k_2)g(k)$, and we have also defined the partonic scaling variable
\begin{eqnarray}
z=\frac{2k_1\cdot q}{Q^2}.
\end{eqnarray}
The phase space for the two unobserved final state partons takes the form
\begin{eqnarray}
\int {\rm d}\Phi_2 = (\mu)^{4-d} \int \frac{{\rm d}^d k}{(2\pi)^{d-1}}\delta^+(k^2)\frac{{\rm d}^d k_2}{(2\pi)^{d-1}}\delta^+(k_2^2)(2\pi)^d\delta^{(d)}\left(q-k_1-k_2-k\right),
\end{eqnarray}
where the momenta of the photon and the outgoing partons can be parameterised as follows~\cite{Rijken:1996ns}:
\begin{eqnarray*}
q &=& \sqrt{s}(1,0,\dots,0)\\
k_1 &=& \frac{s-s_{12}}{2\sqrt{s}}(1,0,\dots,0,1)\\
k_2 &=& \frac{s-s_2}{2\sqrt{s}}(1,0,\dots,0,\sin\theta,\cos\theta)\\
k& =& q - k_1 - k_2.
\end{eqnarray*}
We have introduced the invariants
\begin{eqnarray*}
s = Q^2,\hspace{1cm} s_1 = (k_1 + k_2)^2, \hspace{1cm} s_2 = (k_1+k)^2, \hspace{1cm} s_{12} = (k_2+k)^2,
\end{eqnarray*}
satisfying $s = s_1 + s_2 + s_{12}$. Using momentum conservation and the on-shell conditions for the anti-quark and the gluon, we can parameterise the phase space in a convenient way using
\begin{eqnarray*}
\cos\theta = \frac{s_2 s_{12} - s_1 s}{(s-s_{12})(s-s_2)}, \hspace{1cm} s_1 = z(1-y)s, \hspace{1cm} s_{12} = (1-z)s, \hspace{1cm} s_2 = yzs,
\end{eqnarray*}
such that the 2-body phase space reads
\begin{eqnarray}
\int {\rm d}\Phi_2 = \frac{1}{8\pi}\frac{1}{\Gamma(1-\varepsilon)}\left(\frac{4\pi\mu^2}{s}\right)^{\varepsilon}(1-z)^{-\varepsilon}\int_0^1{\rm d}y (y(1-y))^{-\varepsilon}.
\end{eqnarray}
As in the case of DIS, we may now calculate the one real emission correction to the $\gamma^* \rightarrow q + X$ amplitude up to NLP order, by applying the next-to-soft formalism of Eq.~(\ref{eq:NLPgluon}). This proceeds directly analogously to the previous calculation, and we find that the squared amplitude may be written as
\begin{eqnarray}
\langle \mathcal{A}_{\mu}\mathcal{A}_{\nu}^{\dagger}\rangle \Big{|}_{{\rm LP+NLP}}  = g_s^2 C_F \frac{k_1\cdot k_2}{(k_1\cdot k)(k_2 \cdot k)}H_{\mu\nu}^{\gamma^* \rightarrow q \bar{q}}(k_1-\delta k_1, k_2 - \delta k_2),
\label{epemam}
\end{eqnarray}
where the squared Born process is denoted by $H_{\mu\nu}^{\gamma^*\rightarrow q\bar{q}}$ and the momentum shifts are defined via
\begin{eqnarray}
\label{epemres1}
\delta k_1^{\alpha} &=& -\frac{1}{2}\left(k^{\alpha}+\frac{k_2\cdot k}{k_1\cdot k_2}k_1^{\alpha}-\frac{k_1\cdot k}{k_1\cdot k_2}k_2^{\alpha}\right) \\
\delta k_2^{\alpha} &=& -\frac{1}{2}\left(k^{\alpha}+\frac{k_1\cdot k}{k_1\cdot k_2}k_2^{\alpha}-\frac{k_2\cdot k}{k_1\cdot k_2}k_1^{\alpha}\right).
\label{epemres}
\end{eqnarray}
Thus, as in the DIS and colour singlet production cases~\cite{DelDuca:2017twk}, we again find that all NLP effects can be written in terms of the momentum-shifted non-radiative amplitude. The momentum shifts are similar in form to the previous cases, but both have a negative sign in Eq.~(\ref{epemres1}) and Eq.~(\ref{epemres}) owing to the fact that both hard partons are now in the final state. Putting together all of the above ingredients, the next-to-soft result for the parton fragmentation function $\mathcal{F}_{T,{\rm LP+NLP}}(z,Q^2) $ is~\footnote{Following Eq.~(\ref{Wmunudef}), we keep an overall factor of $z$ unexpanded, which would cancel with a similar factor in forming the hadronic fragmentation function.}
\begin{eqnarray}
\mathcal{F}_{T,{\rm LP+NLP}}(z,Q^2) &=& \frac{\alpha_s z}{\pi}\Bigg(\frac{1}{\varepsilon}\frac{1}{1-z}-\frac{1}{\varepsilon}-\frac{\ln(1-z)}{1-z} +\ln(1-z) \nonumber \\
&& \hspace{4cm} -\frac{z}{1-z}\left(2\ln{z}-1+\ln\frac{s}{\bar{\mu}^2}\right)\Bigg).
\end{eqnarray}
This may be compared with the full NLO expression, which is
\begin{eqnarray}
\mathcal{F}_{T,{\rm NLO}}(z,Q^2)& =& \frac{\alpha_s z}{4\pi}\Bigg(\frac{2}{\varepsilon}\frac{1+z^2}{1-z}-2\frac{\ln(1-z)}{1-z}(1+z^2) \nonumber\\
&& \hspace{1cm} -\frac{1}{1-z}\left(4(1+z^2)\ln{z}+3(z-2)z + 2(1+z^2)\ln\frac{s}{\bar{\mu}^2} \right)\Bigg).
\end{eqnarray}
Similar to the DIS case, the next-to-soft formalism predicts the LL behaviour, but fails to capture a LP term and a constant piece at NLP order: 
\begin{eqnarray}
\label{eq:missingepem}
\mathcal{F}_{T,{\rm NLO}}(z,Q^2)-\mathcal{F}_{T,{\rm LP+NLP}}(z,Q^2)  &=&  \\
&& \hspace{-3cm} \frac{\alpha_s z}{4\pi}\Bigg(-\frac{1}{1-z}+4 +(1-z)\Big[\frac{2}{\varepsilon} -3-2\ln(1-z)-2\ln\frac{s}{\bar{\mu}^2}-4\ln z\Big]\Bigg).\nonumber
\end{eqnarray}
Again, the soft quark formalism allows us to generate the missing LP NLL contribution by using the $\mathcal{Q}_i$-operator on the LO matrix element, generating the diagram in Fig. \ref{fig:3a}. Its contribution reads
\begin{eqnarray}
\mathcal{F}_{T,a}  = \frac{\alpha_s z}{4\pi}\left(-\frac{1}{1-z}+2+{\cal O}(1-z)\right),
\label{FTab}
\end{eqnarray}
which exactly corresponds to the needed LP NLL term, as in the DIS case. Since the quark carrying momentum $k_1$ is observed and therefore non-soft, the other contribution $\mathcal{F}_{T,b}$ (shown in Fig.~\ref{fig:3b}) cannot be obtained using our soft-quark or next-to-soft-gluon formalism. This contribution will however contribute at NLP NLL and beyond. It reads
\begin{eqnarray}
\mathcal{F}_{T,b}  = \frac{\alpha_s z}{4\pi}\left(2+\frac{2(1-z)}{\varepsilon}+\mathcal{O}(1-z)\right),
\label{FTab}
\end{eqnarray}
where we have also shown some ${\cal O}(1-z)$ terms, given that the latter contribution itself includes a pole in $\varepsilon$. This can be traced back to the singularity associated with $k$ being hard and collinear with the parton of 4-momentum $k_1$, which is observed. As in the DIS case, we thus note the presence of an NLP term of collinear origin, and therefore a next-to-collinear contribution.

In this section we have examined a second example with a final state parton that is unobserved. We again find that we can successfully use Eq.~(\ref{eq:NLPgluon}) to obtain an NLO result for the amplitude that is accurate up to NLP order in the next-to-soft expansion. Similar as to the previous case, we find that the result is accurate only to LL. Surprisingly, we find in both the DIS case and the present case that the missing NLL LP information is fully accounted for by adding the soft quark contribution. 

\section{NLP cross-section for NLO prompt photon production}
\label{sec:promptphoton}
In this section we will consider the production of a single photon that recoils against a hard parton at NLO, where the latter is unobserved. This process has more partonic sub-channels than the ones previously considered and it has more than one colour structure to consider at NLO. This makes the prompt photon production process an interesting testing ground for our next-to-soft gluon and soft quark formalisms.  

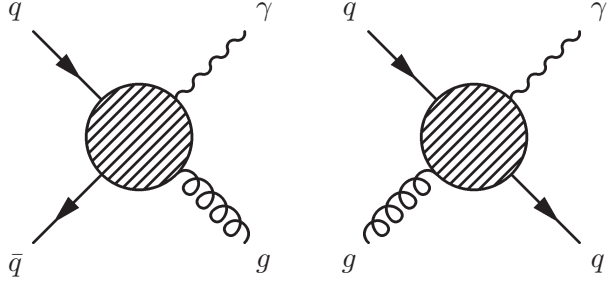
\begin{figure}
    \centering
\begin{fmffile}{kin}
\begin{fmfchar*}(80,80)
  \fmfleft{i1,i2}
  \fmfright{f1,f2}
  \fmfforce{(0,0)}{i2} 
  \fmfforce{(0,h)}{i1} 
  \fmfforce{(.5w,.5h)}{v1}
  \fmfforce{(w,h)}{f2} 
  \fmfforce{(w,0)}{f1} 
  \fmf{fermion}{i1,v1}
  \fmflabel{$q$}{i1}
  \fmflabel{$\bar{q}$}{i2}
  \fmflabel{$g$}{f1}
  \fmflabel{$\gamma$}{f2}
  \fmf{fermion}{v1,i2}
  \fmf{gluon}{f1,v1}
  \fmf{photon}{f2,v1}
  \fmfv{decor.shape=circle,decor.filled=shaded, d.si=.5w}{v1}
\end{fmfchar*}
\end{fmffile}
\hspace{1cm}
\begin{fmffile}{kin2}
\begin{fmfchar*}(80,80)
  \fmfleft{i1,i2}
  \fmfright{f1,f2}
  \fmfforce{(0,0)}{i2} 
  \fmfforce{(0,h)}{i1} 
  \fmfforce{(.5w,.5h)}{v1}
  \fmfforce{(w,h)}{f2} 
  \fmfforce{(w,0)}{f1} 
  \fmf{fermion}{i1,v1}
  \fmflabel{$q$}{i1}
  \fmflabel{$g$}{i2}
  \fmflabel{$q$}{f1}
  \fmflabel{$\gamma$}{f2}
  \fmf{gluon}{v1,i2}
  \fmf{fermion}{v1,f1}
  \fmf{photon}{f2,v1}
  \fmfv{decor.shape=circle,decor.filled=shaded, d.si=.5w}{v1}
\end{fmfchar*}
\end{fmffile}
\vspace{0.5cm}
    \caption{Feynman diagrams for the LO processes $q\bar{q} \rightarrow g \gamma$ (left) and $q g \rightarrow \gamma q$ (right). }
    \label{fig:LOfeynman}
\end{figure}

At leading order, the prompt photon production process ($p p \rightarrow \gamma + X$) consists of two subprocesses: $q\bar{q}\rightarrow \gamma g$ and $q g \rightarrow q \gamma $, as shown in Fig.~\ref{fig:LOfeynman}. 

At next-to-leading order in the coupling, one additional particle can be radiated. This creates a variety of new diagrams that one has to consider. Firstly, one can add an additional gluon to the two Born processes: $q\bar{q}\rightarrow \gamma g g$ and $q g \rightarrow q g \gamma $. The presence of this additional gluon will create the leading power logarithmic contributions to the NLO cross-section. Apart from that, there will be diagrams that appear for the first time at NLO and contain an extra quark in the final state: $q\bar{q}\rightarrow \gamma q^{(')}\bar{q}^{(')}$,  $g g\rightarrow \gamma q\bar{q}$ and  $q q^{(')}\rightarrow \gamma q q^{(')}$. Although only the two Born processes need to be considered for a  leading power analysis, all of the other additional subprocesses will also contribute at NLP only due to the possibility of either a quark or gluon becoming soft and/or collinear. 

Due to the presence of different subprocesses, we have divided this section into three subsections. In subsection \ref{sec:qqbargamma} we will consider the channel that only has gluons in the final state: $q\bar{q}\rightarrow g g \gamma$. Then, in subsection \ref{sec:qqbartoqqbargamma}, we instead consider the quark radiative cross-section ($q\bar{q}\rightarrow q\bar{q}\gamma$), which does not contribute at LP and only commences at NLP order. In the last subsection, we will consider the most involved example, namely one where both gluons and quarks are present in the final state. We will find there that the next-to-soft gluon effects are factorised from the soft quark effects, therefore their contributions can be calculated separately, as was suggested in section \ref{sec:NLP}. For our calculations we will use $\mu_F$ to indicate the factorisation scale and the electromagnetic finestructure constant is given by $\alpha = g_{\rm EM}^2/(4\pi)$.

\label{sec:results}
\subsection{$q\bar{q}\rightarrow g g \gamma$ channel}
\label{sec:qqbargamma}

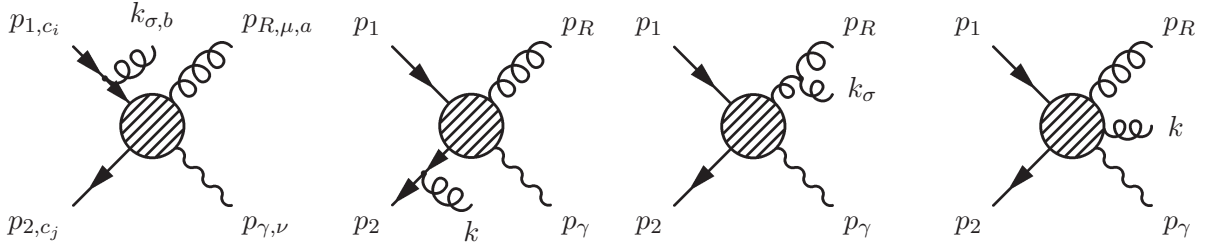
\begin{figure}
\centering
\begin{fmffile}{kin5}
\begin{fmfchar*}(60,60)
  \fmfleft{i1,i2}
  \fmfright{f1,f2,f3}
  \fmfforce{(0,0)}{i2} 
  \fmfforce{(0,h)}{i1} 
  \fmfforce{(.5w,.5h)}{v1}
  \fmfforce{(w,h)}{f1} 
  \fmfforce{(w,0)}{f2} 
  \fmfforce{(.5w,h)}{f3} 
  \fmfforce{(.2w,.8h)}{v2}
  \fmf{fermion}{i1,v2}
  \fmf{fermion}{v2,v1}
  \fmf{fermion}{v1,i2}
  \fmf{gluon}{v2,f3}
  \fmf{photon}{f2,v1}
  \fmflabel{$p_{1,c_i}$}{i1}
  \fmflabel{$p_{2,c_j}$}{i2}
  \fmflabel{$p_{R,\mu,a}$}{f1}
  \fmflabel{$k_{\sigma,b}$}{f3}
  \fmflabel{$p_{\gamma,\nu}$}{f2}
  \fmf{gluon}{f1,v1}
  \fmfv{decor.shape=circle,decor.filled=shaded, d.si=.4w}{v1}
  \fmfv{decor.shape=circle,decor.filled=full, d.si=.02w}{v2}
\end{fmfchar*}
\end{fmffile}
\hspace{1.5cm}
\begin{fmffile}{kin4}
\begin{fmfchar*}(60,60)
  \fmfleft{i1,i2}
  \fmfright{f1,f2,f3}
  \fmfforce{(0,0)}{i2} 
  \fmfforce{(0,h)}{i1} 
  \fmfforce{(.5w,.5h)}{v1}
  \fmfforce{(w,h)}{f1} 
  \fmfforce{(w,0)}{f2} 
  \fmfforce{(.5w,0)}{f3} 
  \fmfforce{(.2w,.2h)}{v2}
  \fmf{fermion}{i1,v1}
  \fmf{fermion}{v1,v2}
  \fmf{fermion}{v2,i2}
  \fmf{gluon}{v2,f3}
  \fmf{photon}{f2,v1}
  \fmflabel{$p_1$}{i1}
  \fmflabel{$p_2$}{i2}
  \fmflabel{$p_R$}{f1}
  \fmflabel{$k$}{f3}
  \fmflabel{$p_{\gamma}$}{f2}
  \fmf{gluon}{f1,v1}
  \fmfv{decor.shape=circle,decor.filled=shaded, d.si=.4w}{v1}
  \fmfv{decor.shape=circle,decor.filled=full, d.si=.02w}{v2}
\end{fmfchar*}
\end{fmffile}
\hspace{1cm}
\begin{fmffile}{kin3}
\begin{fmfchar*}(60,60)
  \fmfleft{i1,i2}
  \fmfright{f1,f2,f3}
  \fmfforce{(0,0)}{i2} 
  \fmfforce{(0,h)}{i1} 
  \fmfforce{(.5w,.5h)}{v1}
  \fmfforce{(w,h)}{f1} 
  \fmfforce{(w,0)}{f2} 
  \fmfforce{(w,.7h)}{f3} 
  \fmfforce{(.8w,.8h)}{v2}
  \fmf{fermion}{i1,v1}
  \fmf{fermion}{v1,i2}
  \fmf{photon}{f2,v1}
  \fmflabel{$p_{1}$}{i1}
  \fmflabel{$p_{2}$}{i2}
  \fmflabel{$p_{R}$}{f1}
  \fmflabel{$k_{\sigma}$}{f3}
  \fmflabel{$p_{\gamma}$}{f2}
  \fmf{gluon}{f1,v2}
  \fmf{gluon}{v2,v1}
  \fmf{gluon}{v2,f3}
  \fmfv{decor.shape=circle,decor.filled=shaded, d.si=.4w}{v1}
  \fmfv{decor.shape=circle,decor.filled=full, d.si=.02w}{v2}
\end{fmfchar*}
\end{fmffile}
\hspace{1.5cm}
\begin{fmffile}{kin6}
\begin{fmfchar*}(60,60)
  \fmfleft{i1,i2}
  \fmfright{f1,f2,f3}
  \fmfforce{(0,0)}{i2} 
  \fmfforce{(0,h)}{i1} 
  \fmfforce{(.5w,.5h)}{v1}
  \fmfforce{(w,h)}{f1} 
  \fmfforce{(w,0)}{f2} 
  \fmfforce{(.w,.5h)}{f3} 
  \fmf{fermion}{i1,v1}
  \fmf{fermion}{v1,i2}
  \fmf{gluon}{v1,f3}
  \fmf{photon}{f2,v1}
  \fmflabel{$p_1$}{i1}
  \fmflabel{$p_2$}{i2}
  \fmflabel{$p_R$}{f1}
  \fmflabel{$k$}{f3}
  \fmflabel{$p_{\gamma}$}{f2}
  \fmf{gluon}{f1,v1}
  \fmfv{decor.shape=circle,decor.filled=shaded, d.si=.4w}{v1}
\end{fmfchar*}
\end{fmffile}
\vspace{0.5cm}
\caption{NLO Feynman diagrams for the NLP contributions of the process $q(p_1)\bar{q}(p_2)\rightarrow g(p_R) g(k) \gamma(p_{\gamma})$.}
\label{fig:NLPqq}
\end{figure}

We first consider the process obtained by dressing $q\bar{q}\rightarrow g  \gamma$ with one additional gluon, where the relevant Feynman diagrams for this are shown in Fig.~\ref{fig:NLPqq}. Given that gluons are the only partons in the final state, it should be sufficient to describe this process, up to next-to-soft level, with the amplitude of Eq.~(\ref{eq:NLPgluon}). To compute the cross-section, we must calculate the squared amplitude, summed and averaged over final/initial state colours and spins. This involves summing over all polarisations for the emitted gluon, and one may restrict this to be over physical degrees of freedom by defining an arbitrary lightlike vector $l$ such that $l\cdot\epsilon(k)=0$, where $\epsilon_\mu(k)$ is the polarisation vector of the gluon, thus obtaining
\begin{equation}
\sum_{\rm phys.}\epsilon^\dag_\alpha (k)\epsilon_\beta(k)
=-\eta_{\alpha\beta}+\frac{l_\alpha\,k_\beta+l_\beta\,k_\alpha}
{l\cdot k}.
\label{polsum1}
\end{equation}
Alternatively, one may sum over {\it all} polarisations (including an unphysical longitudinal degree of freedom), provided one introduces external ghost particles to remove the spurious contributions. As argued in Ref.~\cite{DelDuca:2017twk}, however, soft ghosts do not contribute up to NLP order in the momentum expansion, and thus one may replace Eq.~(\ref{polsum1}) with~\footnote{A similar distinction was not needed for the DIS and $e^+e^-$ processes, due to the fact that ghosts cannot couple directly to quarks.}
\begin{equation}
\sum_{\rm pols.}\epsilon^\dag_\alpha (k)\epsilon_\beta(k)
=-\eta_{\alpha\beta}.
\end{equation}
Since the current process is more involved than the DIS and $e^+e^-$ cases due to the presence of more than two NLO diagrams, we will again provide explicit details of how to construct the next-to-soft squared amplitude, thus illustrating the use of Eq.~(\ref{eq:NLPgluon}) in a more complicated scenario. As before, we will start with the scalar amplitude 
\begin{eqnarray}
\mathcal{A}_{\rm scal} &=& \bar{v}(p_2)\Bigg(\frac{ig_s t^{b}_{c_kc_i}(2p_1^{\sigma}-k^{\sigma})}{2p_1\cdot  k}\mathcal{M}^{\mu\nu,a}_{c_jc_k}-\frac{ig_s t^{b}_{c_jc_k}(2p_2^{\sigma}-k^{\sigma})}{2p_2\cdot  k}\mathcal{M}^{\mu\nu,a}_{c_kc_i} \nonumber \\
&& \hspace{4cm} - \frac{g_s f^{bac}(2p_R^{\sigma}+k^{\sigma})}{ 2 p_R\cdot k}g^{\mu}_{\rho}\mathcal{M}^{\rho\nu,c}_{c_jc_i} \Bigg) u(p_1)\epsilon^*_{\mu}(p_R)\epsilon^*_{\nu}(p_{\gamma})\epsilon^*_{\sigma}(k),
\end{eqnarray}
where the matrices $\mathcal{M}^{\mu\nu,a}_{c_jc_k}$, $\mathcal{M}^{\mu\nu,a}_{c_kc_i}$ and $\mathcal{M}^{\rho\nu,c}_{c_jc_i}$ correspond to the shaded circles in the first three diagrams in Fig.~\ref{fig:NLPqq}, whose dependence on the momenta $p_1$, $p_2$, $p_R$ and $p_{\gamma}$ is implicitly understood. For clarity we have included explicit colour labels on the matrix element, indicating that it still depends on the colour structure via the SU(3) generators. The double scalar contribution to the matrix element squared, inclusive of spin/colour averaging factors, is easily computed and results in 
\begin{eqnarray}
\nonumber
\langle |\mathcal{A}_{{\rm scal}}|^2 \rangle &=& \frac{Q_q^2g_{\rm EM}^2g_s^4C_F}{4C_A}{\rm Tr}\left[\slashed{p}_2 \Gamma^{\mu\nu} \slashed{p}_1 \Gamma^*_{\mu\nu}\right] \times \Bigg[C_F \frac{2 p_1\cdot p_2}{(p_1 \cdot k)(p_2\cdot k)} \\
&& \hspace{1cm}  +\frac{1}{2}C_A\left(\frac{2 p_1\cdot p_R}{(p_1 \cdot k)(p_R\cdot k)}-\frac{2 p_2\cdot p_R}{(p_2 \cdot k)(p_R\cdot k)}+\frac{2 p_1\cdot p_2}{(p_1 \cdot k)(p_2\cdot k)}\right)\Bigg].
\end{eqnarray}
Here $\Gamma^{\mu\nu} \equiv \Gamma^{\mu\nu}(p_1,p_2,p_R,p_{\gamma}) $ denotes the (non-radiative) hard scattering matrix element for the process $q(p_1)\bar{q}(p_2) \rightarrow g(p_R) \gamma(p_{\gamma}) $ stripped of its polarisation vectors, spinors, colour factors and charges, and where we suppress the momenta labels for brevity.   \\
We now move on to the scalar-spin interference term. The spin amplitude for the $q\bar{q}\rightarrow gg \gamma$ process evaluates to 
\begin{eqnarray}
\mathcal{A}_{\rm spin} &=& \bar{v}(p_2)\left(\frac{ig_s t^{b}_{c_kc_i}}{2p_1\cdot  k}\mathcal{M}^{\mu\nu,a}_{c_jc_k}\gamma^{\sigma}\slashed{k}-\frac{ig_s t^{b}_{c_jc_k}}{2p_2\cdot  k}\slashed{k}\gamma^{\sigma}\mathcal{M}^{\mu\nu,a}_{c_kc_i} + \frac{g_s f^{bac}}{p_R\cdot k}\mathcal{M}^{\rho\nu,c}_{c_jc_i} (g^{\sigma}_{\rho}k^{\mu}-g^{\sigma\mu}k_{\rho}) \right)\\
&& \hspace{10cm}u(p_1)\epsilon^*_{\mu}(p_R)\epsilon^*_{\nu}(p_{\gamma})\epsilon^*_{\sigma}(k). \nonumber
\end{eqnarray}
Contracting the spin amplitude with the scalar amplitude then results in 
\begin{eqnarray}
2 {\rm Re}\left[\mathcal{A}_{\rm scal}\mathcal{A}_{\rm spin}^*\right] = -\frac{Q_q^2g_{\rm EM}^2g_s^4C_F^2}{4C_A}\frac{2p_1\cdot p_2}{(p_1\cdot k)(p_2 \cdot k)}\frac{(p_1+p_2)\cdot k}{p_1 \cdot p_2}{\rm Tr}\left[\slashed{p}_2 \Gamma^{\mu\nu} \slashed{p}_1 \Gamma_{\mu\nu}^* \right].
\label{eq:scalspin}
\end{eqnarray}
Finally, we must evaluate the scalar-orbital interference term. To this end, we may first write down the orbital amplitude
\begin{eqnarray}
\mathcal{A}_{\rm orb} &=& ig_s k_{\alpha} \bar{v}(p_2)\Bigg(\frac{t^{b}_{c_kc_i}}{p_1\cdot  k}\left(p_1^{\alpha}\frac{\partial}{\partial p_{1,\sigma}}-p_1^{\sigma}\frac{\partial}{\partial p_{1,\alpha}}\right)\mathcal{M}^{\mu\nu,a}_{c_jc_k}-\frac{t^{b}_{c_jc_k}}{p_2\cdot  k}\left(p_2^{\alpha}\frac{\partial}{\partial p_{2,\sigma}}-p_2^{\sigma}\frac{\partial}{\partial p_{2,\alpha}}\right)\mathcal{M}^{\mu\nu,a}_{c_kc_i} \nonumber \\
&&\hspace{3cm} - \frac{i f^{bac}}{p_R\cdot k}\left(p_R^{\alpha}\frac{\partial}{\partial p_{R,\sigma}}-p_R^{\sigma}\frac{\partial}{\partial p_{R,\alpha}}\right)\mathcal{M}^{\mu\nu,c}_{c_jc_i} \Bigg)u(p_1)\epsilon^*_{\mu}(p_R)\epsilon^*_{\nu}(p_{\gamma})\epsilon^*_{\sigma}(k),
\end{eqnarray}
to be contracted with the scalar amplitude. Adopting the abbreviation
\begin{equation}
\label{eq:deltamom}
\delta p_{i;j}^{\alpha} \equiv -\frac{1}{2}\left(k^{\alpha}+\frac{p_j\cdot k}{p_i \cdot p_j}p_i^{\alpha} - \frac{p_i \cdot k}{p_i \cdot p_j}p_j^{\alpha}\right),
\end{equation}
we get
\begin{eqnarray}
\langle \mathcal{A}_{\rm orb }\mathcal{A}_{\rm scal}^*\rangle  &=& \frac{Q_q^2g_{\rm EM}^2g_s^4C_F}{4C_A} \Bigg[\left(C_F-\frac{1}{2}C_A\right) \frac{2 p_1\cdot p_2}{(p_1 \cdot k)(p_2 \cdot k)}\nonumber\\
&&\hspace{3.5cm} \times {\rm Tr}\left[\slashed{p}_2 \left(\delta p_{1;2}^{\alpha}\frac{\partial}{\partial p_1^{\alpha}} + \delta p_{2;1}^{\alpha} \frac{\partial}{\partial p_2^{\alpha}}\right)\Gamma^{\mu\nu} \slashed{p}_1 \Gamma_{\mu\nu}^* \right] \nonumber \\
&& +\frac{1}{2}C_A \frac{2 p_1\cdot p_R}{(p_1 \cdot k)(p_R \cdot k)}{\rm Tr}\left[\slashed{p}_2 \left(\delta p_{1;R}^{\alpha}\frac{\partial}{\partial p_1^{\alpha}} - \delta p_{R;1}^{\alpha} \frac{\partial}{\partial p_R^{\alpha}}\right)\Gamma^{\mu\nu} \slashed{p}_1 \Gamma_{\mu\nu}^* \right]\nonumber \\
&&+\frac{1}{2}C_A \frac{2 p_2\cdot p_R}{(p_2 \cdot k)(p_R \cdot k)}{\rm Tr}\left[\slashed{p}_2 \left(\delta p_{2;R}^{\alpha}\frac{\partial}{\partial p_2^{\alpha}} - \delta p_{R;2}^{\alpha} \frac{\partial}{\partial p_R^{\alpha}}\right)\Gamma^{\mu\nu} \slashed{p}_1 \Gamma_{\mu\nu}^* \right]\Bigg].\label{eq:scalorb}
\end{eqnarray}
The expression for $\langle \mathcal{A}_{\rm scal }\mathcal{A}_{\rm orb}^* \rangle $ looks similar, but with the derivatives acting on $\Gamma_{\mu\nu}^*$. As in section~\ref{sec:DIS}, we may transform the derivatives in these expressions into total derivatives acting on the complete trace using the chain rule, which results in
\begin{eqnarray}
\langle  2{\rm Re}\left[\mathcal{A}_{\rm orb }\mathcal{A}_{\rm scal}^*\right] \rangle &=& \frac{Q_q^2 g_{{\rm EM}}^2g_s^4C_F}{4C_A}\Bigg[C_F\frac{2 p_1\cdot p_2}{(p_1 \cdot k)(p_2 \cdot k)}\Bigg\{\left(\delta p_{1;2}^{\alpha}\frac{\partial}{\partial p_1^{\alpha}} + \delta p_{2;1}^{\alpha} \frac{\partial}{\partial p_2^{\alpha}}\right){\rm Tr}\left[\slashed{p}_2 \Gamma^{\mu\nu} \slashed{p}_1 \Gamma_{\mu\nu}^* \right]\nonumber \\
 && \hspace{2cm} - {\rm Tr}\left[\delta\slashed{p}_{2;1} \Gamma^{\mu\nu} \slashed{p}_1 \Gamma_{\mu\nu}^* \right] - {\rm Tr}\left[\slashed{p}_2 \Gamma^{\mu\nu} \delta\slashed{p}_{1;2}\Gamma_{\mu\nu}^* \right]\Bigg\}\nonumber\\
&&+ \frac{1}{2}C_A \frac{2 p_1\cdot p_R}{(p_1 \cdot k)(p_R \cdot k)}\left(\delta p_{1;R}^{\alpha}\frac{\partial}{\partial p_1^{\alpha}} - \delta p_{R;1}^{\alpha} \frac{\partial}{\partial p_R^{\alpha}}\right){\rm Tr}\left[\slashed{p}_2 \Gamma^{\mu\nu} \slashed{p}_1 \Gamma_{\mu\nu}^* \right]\nonumber \\
&&+ \frac{1}{2}C_A \frac{2 p_2\cdot p_R}{(p_2 \cdot k)(p_R \cdot k)}\left(\delta p_{2;R}^{\alpha}\frac{\partial}{\partial p_2^{\alpha}} - \delta p_{R;2}^{\alpha} \frac{\partial}{\partial p_R^{\alpha}}\right){\rm Tr}\left[\slashed{p}_2 \Gamma^{\mu\nu}\slashed{p}_1 \Gamma_{\mu\nu}^* \right]\nonumber \\
&&-\frac{1}{2}C_A \frac{2 p_1\cdot p_2}{(p_1 \cdot k)(p_2 \cdot k)}\left(\delta p_{1;2}^{\alpha}\frac{\partial}{\partial p_1^{\alpha}} + \delta p_{2;1}^{\alpha} \frac{\partial}{\partial p_2^{\alpha}}\right){\rm Tr}\left[\slashed{p}_2 \Gamma^{\mu\nu} \slashed{p}_1 \Gamma_{\mu\nu}^* \right]\Bigg]. \label{eq:scalorb2}
\end{eqnarray}
Here we see explicitly the presence of two different colour structures, namely terms in the square bracket which are proportional to $C_F$ and $C_A$ respectively. For the $C_F$ contribution, the derivative term is accompanied by additional contributions, involving the shifts $\delta\slashed{p}_{2;1}$ and $\delta\slashed{p}_{1;2}$. A similar situation occurred in Eq.~(\ref{scalorbDIS}) for the DIS process, where the additional contributions ended up being cancelled by the spin-scalar interference term. The same turns out to happen here if one introduces a Sudakov decomposition for $k$, defined here such that
\begin{equation}
\delta p_{i;j}^{\alpha} =  -\frac{1}{2}\left(k_{T(i,j)}^{\alpha}+2\frac{p_j\cdot k}{p_i \cdot p_j}p_i^{\alpha}\right),
\end{equation}
where $k_{T(i,j)}$ is defined to be orthogonal to both $p_i$ and $p_j$. As in the DIS and $e^+e^-$ case, terms that are proportional to $k_T$ will ultimately vanish upon integration over the final state phase space, so we will ignore them in what follows. Interestingly, for the $C_A$ term there is no need for a cancellation originating from a spin-scalar contribution, as the terms proportional to $C_A$ vanish directly up to $\mathcal{O}(k_T)$. Putting all pieces together, the complete LP + NLP squared amplitude can then be written as
\begin{eqnarray}
\label{eq:NLPqq}
\langle |\mathcal{A}_{{\rm LP+NLP},q\bar{q}\rightarrow \gamma g g}|^2 \rangle  &=& \frac{Q_q^2g_{{\rm EM}}^2g_s^4C_F}{4C_A} \Bigg[\left(C_F-\frac{1}{2}C_A\right)\frac{2p_1\cdot p_2}{(p_1\cdot k)(p_2\cdot k)} \\
&&\hspace{4cm} \times H_{q\bar{q}\rightarrow \gamma g}(p_1+\delta p_{1;2},p_2+\delta p_{2;1},p_{\gamma},p_R) \nonumber \\
&&\hspace{0.8cm} + \frac{1}{2}C_A\frac{2p_1\cdot p_R}{(p_1\cdot k)(p_R\cdot k)}H_{q\bar{q}\rightarrow \gamma g}(p_1+\delta p_{1;R},p_2,p_{\gamma},p_R-\delta p_{R;1}) \nonumber \\
&& \hspace{0.8cm} + \frac{1}{2}C_A\frac{2p_2\cdot p_R}{(p_2\cdot k)(p_R\cdot k)}H_{q\bar{q}\rightarrow \gamma g}(p_1,p_2+\delta p_{2;R},p_{\gamma},p_R-\delta p_{R;2}) \Bigg],\nonumber
\end{eqnarray}
where $H_{q\bar{q}\rightarrow \gamma g}(p_1+\delta p_{1;2},p_2+\delta p_{2;1},p_{\gamma},p_R)$ denotes the trace appearing in e.g. Eq.~\eqref{eq:scalorb2}, but where the momenta $p_1$ and $p_2$ are shifted by $\delta p_{1;2}$ and $\delta p_{2;1}$ respectively. This result is directly analogous to the previous cases, which also found that the squared amplitude for the one real emission contribution could be written in terms of the momentum-shifted non-radiative amplitude. There is a notable difference with respect to our previous results, however. Both the DIS and $e^+e^-$ cases had only two parton legs in the Born interaction, and the final result for the squared amplitude consisted of a dipole-like eikonal factor dressing the momentum shifted hard interaction (Eqs.~(\ref{Asqres}) and~(\ref{epemam}) respectively). In the present case, we see multiple dipole-like terms, each consisting of an eikonal factor involving two hard momenta $p_i$ and $p_j$ dressing a hard interaction where the {\it same} momenta are shifted. Furthermore, different dipole terms have correspondingly different colour structures. 
As remarked above, the momentum shifts in Eq.~(\ref{eq:NLPqq}) are generated by a combination of the spin and orbital term for the $C_F$ colour structure and only by the orbital term for the $C_A$ colour structure. The orbital terms act as a momentum shift operator on the hard scattering of the matrix element, while the spin term takes care of the same shift on the asymptotic states. 

We are now in a position to integrate over the final state momenta $p_R$ and $k$ and compute the differential cross-section.  We will separate the three-body phase space into two two-body phase spaces, one containing the unobserved partons, the other describing the photon and the collective effect of the unobserved partons. To this end, we first factor the three-body phase space as
\begin{eqnarray}
{\rm d}\Phi_3 = \frac{1}{2\pi} \int {\rm d}P^2 \,\,{\rm d}\Phi_2 (p_1 + p_2; p_{\gamma}, P)\,\,{\rm d}\Phi_2 (P; p_R, k),
\end{eqnarray}
where the second two-body phase space factor on the right-hand side is in the center of mass frame of the two unobserved partons, and takes the form
\begin{eqnarray}
{\rm d}\Phi_2(P;p_R, k) = \left(\frac{16\pi\mu^2}{P^2}\right)^{\varepsilon}\frac{1}{16\pi \Gamma\left(\frac{1}{2}\right)\Gamma\left(\frac{1}{2}-\varepsilon\right)}\int_0^{\pi}{\rm d}\theta_1 \left(\sin\theta_1\right)^{1-2\varepsilon}\int_0^{\pi}{\rm d}\theta_2 \left(\sin\theta_2\right)^{-2\varepsilon}
\end{eqnarray}
after having parameterised the momenta as follows:
\begin{eqnarray*}
p_R = \frac{\sqrt{P^2}}{2}\left(1,0,\dots,0,\sin\theta_1\sin\theta_2,\sin\theta_1\cos\theta_2,\cos\theta_1\right)\\
k = \frac{\sqrt{P^2}}{2}\left(1,0,\dots,0,-\sin\theta_1\sin\theta_2,-\sin\theta_1\cos\theta_2,-\cos\theta_1\right).
\end{eqnarray*}
In terms of the invariants $s = (p_1 + p_2)^2$, $u_1 = (p_1 - p_{\gamma})^2$, $t_1 = (p_2-p_{\gamma})^2$ and $s_4 = (p_1+p_2-p_{\gamma})^2 = (k+p_R)^2$, the other phase space is given by 
\begin{eqnarray}
{\rm d}\Phi_2(p_1+p_2;p_{\gamma}, P) = \left(\frac{4 \pi s \mu^2}{t_1 u_1}\right)^{\varepsilon}\frac{1}{8\pi s\Gamma(1-\varepsilon)} \delta^+\left(P^2-s_4\right) \int {\rm d} t_1 \int {\rm d} u_1.
\end{eqnarray}
To compare our results with the NLO calculation presented in Ref. \cite{Gordon:1993qc}, we will make a change of variables:
\begin{eqnarray}
u_1  &\equiv& -svw \nonumber\\
t_1  &\equiv& s(v-1) \nonumber\\ 
s_4 &=& s+t_1+u_1 = sv(1-w),
\end{eqnarray}
where $(1-w)$ plays the role of the threshold variable $\xi$ in Eq.~(\ref{eq:defthreshold}) (i.e. $w\rightarrow 1$ at threshold). In terms of these invariants the complete three body phase space now reads 
\begin{eqnarray}
\label{eq:3bodyphase}
{\rm d}\Phi_3 = s\left(\frac{4\pi\mu^2}{s}\right)^{2\varepsilon}\frac{v\left(v^2(1-v)w(1-w)\right)^{-\varepsilon}}{(4\pi)^{4}\Gamma(1-2\varepsilon)}\int {\rm d} v\int {\rm d}w \int_0^{\pi}{\rm d}\theta_1 \left(\sin\theta_1\right)^{1-2\varepsilon}\int_0^{\pi}{\rm d}\theta_2 \left(\sin\theta_2\right)^{-2\varepsilon}.
\end{eqnarray}
Furthermore, we will extract a common factor of $vw(1-v)s$ from the differential cross-section as was done in Ref. \cite{Gordon:1993qc}, and we obtain
\begin{eqnarray}
vw(1-v)s\frac{{\rm d}\sigma^{\rm LP+NLP}_{q\bar{q}\rightarrow \gamma g g }}{{\rm d}{v}{\rm d}w} &=&  s \left(\frac{4\pi\mu^2}{s}\right)^{2\varepsilon}\frac{v^2 w (1-v)\left(v^2(1-v)w(1-w)\right)^{-\varepsilon}}{2(4\pi)^{4}\Gamma(1-2\varepsilon)} \nonumber \\
&& \int_0^{\pi}{\rm d}\theta_1 \left(\sin\theta_1\right)^{1-2\varepsilon}\int_0^{\pi}{\rm d}\theta_2 \left(\sin\theta_2\right)^{-2\varepsilon}\langle|\mathcal{A}_{{\rm LP+NLP},q\bar{q}\rightarrow \gamma g g}|^2\rangle.
\end{eqnarray}
We can now use our NLP result for the amplitude (Eq.~(\ref{eq:NLPqq})). Inserting the form of the momentum shifts of Eq.~(\ref{eq:deltamom}) and integrating, one finds
\begin{eqnarray}
vw(1-v)s\frac{{\rm d}\sigma^{{\rm LP+NLP}}_{q\bar{q}\rightarrow \gamma g g }}{{\rm d}{v}{\rm d}w} &=& Q_q^2\alpha\alpha_s^2 \frac{C_F}{C_A}\Bigg[-\frac{1}{\varepsilon}\frac{4C_F T_{q\bar{q}}}{(1-w)_+} -  \frac{1}{\varepsilon}\left\{2C_F\frac{4v(v-1)^2-1}{1-v}\right\} \nonumber \\
&& +\left(\frac{\ln(1-w)}{1-w}\right)_+ 2(4C_F-C_A)T_{q\bar{q}} \nonumber\\
&& +\frac{1}{(1-w)_+}\Bigg\{T_{q\bar{q}}\left(2C_A\ln(1-v) +8C_F\ln(v)-2 C_A - 8C_F\ln\frac{\bar{\mu}^2}{s}\right) \nonumber\\
&& \hspace{2.2cm} +8C_F((v-1)v+1)\Bigg\} \nonumber \\
&& +\ln(1-w)\left\{(4C_F-C_A)\frac{4v(v-1)^2-1}{1-v}\right\}\nonumber \\
&& + \mathcal{O}\left(\delta(1-w)\right)+\mathcal{O}(1) \Bigg]\label{eq:qqbarNLP},
\end{eqnarray}
where $T_{q\bar{q}} = 2v(v-1)+1$. We may compare this result with the full NLO calculation of Ref. \cite{Gordon:1993qc}, collected  for convenience up to NLP order in the threshold expansion in appendix~\ref{app:NLOcalc}. Upon doing so we observe that, as before, the next-to-soft formalism correctly reproduces LL terms at both LP and NLP orders. Subleading terms would again require the addition of hard (next-to-)collinear information, but it is reassuring that LL information is correctly reproduced even in a less inclusive situation.
Similary to the DIS and $e^+e^-$ cases, the missing hard-collinear information is encoded in the jet functions. Here however, the missing information {\it cannot} be accounted for by the soft quark formalism as there are no quarks in the final state.  One might wonder why here it is not enough to add the possibility that the other gluon carrying momentum $p_R$ becomes soft, while in the case where the jet function was initiated by a fragmenting quark this was indeed enough. In the latter case, the fragmenting quark (Fig. \ref{fig:1d}) dresses the non-radiative amplitude according to
\begin{eqnarray}
\langle|\mathcal{A}_{1d}|^2 \rangle &\propto& \frac{1}{(p_i \cdot k)}{\rm Tr}\left(\slashed{k}\mathcal{M}\mathcal{M}^{\dagger}\right) + \mathcal{O}(\varepsilon).
\end{eqnarray}
Here, the $k\rightarrow 0$ limit does not contribute at LP order, as then the soft singularity will vanish. This is related to the fact that the quark propagator is suppressed by one power of soft momentum. However, the $p_i \rightarrow 0$ limit does contribute, and constitutes exactly the missing collinear information, which can be added by the soft quark formalism.\\
The fragmenting gluon case of Fig. \ref{fig:1e} yields
\begin{eqnarray}
\langle|\mathcal{A}_{1e}|^2 \rangle &\propto& \frac{1}{(p_i \cdot k)^2}\left[p_i^{\rho}k^{\rho'}+k^{\rho}p_i^{\rho'}\right]{\rm Tr}\left(\mathcal{M}_{\rho}\mathcal{M}_{\rho'}^{\dagger}\right) + \mathcal{O}(\varepsilon),
\end{eqnarray}
where we see the appearance of three separate contributions: $k\rightarrow 0$, $p_i\rightarrow 0$ and $k \cdot p_i \rightarrow 0$ with neither being soft. The latter case is, by definition, not captured by the next-to-soft formalism, and contributes a LP NLL term that is then a truly hard-collinear effect. By explicit calculation this contribution reads
\begin{eqnarray}
vw(1-v)s\frac{{\rm d}\sigma^{{\rm jet}}_{q\bar{q}\rightarrow \gamma g g }}{{\rm d}{v}{\rm d}w} &=& Q_q^2\alpha\alpha_s^2\frac{C_F}{C_A}\left[\frac{1}{(1-w)_+}\frac{T_{q\bar{q}}}{6}\right],
\end{eqnarray}
which is indeed the missing LP NLL term. 
The second contribution that is missing is one that originates from a final state gluon splitting into a quark-antiquark pair. This results in a $\frac{1}{(1-w)_+}$ contribution, which the authors of Ref. \cite{Gordon:1993qc} added to the $q\bar{q}\rightarrow g g \gamma$ subprocess. Formally it is part of the $q\bar{q}\rightarrow \gamma q\bar{q}$ and $q\bar{q}\rightarrow \gamma q'\bar{q}'$ subprocesses, therefore we will treat this contribution in the next section.

The above cross-section contains infrared poles, that must be absorbed into the parton distributions via the usual mass factorisation procedure. This leads to a novel source of NLP contributions in the fully subtracted cross-section, which is worth drawing attention to. Mass factorisation can be performed by adding a counter cross section, which is a convolution of a scaled Born cross section with the parton distribution functions. The phase space for the counter term consists of a two-body final state, and is given by 
\begin{eqnarray}
{\rm d}\Phi_2 = \left(\frac{4\pi\mu^2}{s}\right)^{\varepsilon}\frac{\left(v(1-v)\right)^{-\varepsilon}}{8\pi \Gamma(1-\varepsilon)}\int {\rm d} v\int {\rm d}w \delta(1-w).
\end{eqnarray}
There is then a difference in the $\varepsilon$-dependence with respect to the three-body phase space of Eq. (\ref{eq:3bodyphase}), such that subtracting the counterterm leads schematically to an NLP contribution in the finite part of the cross-section:
\begin{eqnarray}
&& \frac{1}{\varepsilon}\frac{1}{(1-w)_+}\left[\left(\frac{4\pi\mu^2}{s}\right)^{2\varepsilon}\frac{\left(v^2(1-v)w(1-w)\right)^{-\varepsilon}}{\Gamma(1-2\varepsilon)}-\left(\frac{4\pi\mu^2}{s}\right)^{\varepsilon}\frac{\left(v(1-v)\right)^{-\varepsilon}}{\Gamma(1-\varepsilon)}\right] \nonumber \\
&& \hspace{1.5cm} =  -\left(\frac{\ln(1-w)}{1-w}\right)_+ + \frac{\ln\left(\bar{\mu}^2/s\right)}{(1-w)_+} - \frac{\ln(v)}{(1-w)_+}- \frac{\ln(w)}{(1-w)_+}.
\end{eqnarray}
The first and second term on the second line of this equation are part of the LP LL and NLL contribution, whereas the fourth term is of NLP order.
The third term on the second line gives subleading logarithmic contributions at both LP and NLP orders. To see this, note that in the prompt photon production process the observable is $p_T$, and the threshold limit is given by $4p_T^2 \rightarrow s$ \cite{Catani:1998tm}. The Mellin moment is then taken with respect to $x_T^2 = 4p_T^2/s$ and reads \cite{Catani:1998tm}
\begin{eqnarray}
\tilde{\sigma}(N) = \int^1_0{\rm d} x_T^2 \, (x_T^2)^{N-1} \, \frac{p_T^3{\rm d}\sigma(p_T)}{{\rm d}p_T} = \frac{1}{2} \int^1_0{\rm d} v \int^1_0 {\rm d} w \,\,\,(4v(1-v)w)^{N+1}\, \frac{s{\rm d}\sigma(v,w)}{{\rm d}v {\rm d}w},
\end{eqnarray}
where 
\begin{eqnarray}
\int^1_0 {\rm d} v (4v(1-v)w)^{N+1}f(v) = f\left(\frac{1}{2}\right) + \mathcal{O}\left(\frac{1}{N}\right).
\end{eqnarray}
The LP contributions are therefore fixed at $v=1/2$, with $\mathcal{O}(1/N)$ terms appearing for $v\neq 1/2$, which in particular affect the LL NLP contribution. Thus, it is no longer true in the final result for the (subtracted) cross-section that all LL NLP behaviour can be predicted simply by classifying the structure of the squared amplitude: there are leading logarithmic terms originating from phase space effects in the mass factorisation procedure. This could be a crucial ingredient in future numerical studies of NLP effects. 

In this section we calculated the NLP contribution of the $q\bar{q}\rightarrow \gamma g g$ channel to the exclusive prompt photon production process. Similar to the DIS and $e^+ e^-$ processes, the next-to-soft gluon formalism indeed correctly reproduces the LL terms at both the LP and NLP order. We again miss a hard-collinear contribution, which in this case cannot be accounted for by considering a soft quark contribution, as there are no soft quarks in the final state. Here we explicitly need to add a hard-collinear gluon contribution to the next-to-soft gluon formalism to include the missing LP NLL information. We also saw that when carrying out mass factorisation for the full cross-section, one also has to carefully keep track of additional LL NLP contributions, which is perhaps not surprising. Having understood this particular partonic process, let us now consider a second sub-channel in the following section.

\subsection{$q \bar{q} \rightarrow q \bar{q} \gamma $ channel}
\label{sec:qqbartoqqbargamma}

\begin{figure}
\centering
\centering
\hspace{1cm}
\begin{fmffile}{kinqq1}
\begin{fmfchar*}(50,50)
  \fmfleft{i1,i2}
  \fmfright{f1,f2,f3}
  \fmfforce{(0,0)}{i2} 
  \fmfforce{(0,h)}{i1} 
  \fmfforce{(.5w,.5h)}{v1}
  \fmfforce{(w,h)}{f1} 
  \fmfforce{(w,0)}{f2} 
  \fmfforce{(.4w,.h)}{f3} 
  \fmfforce{(.2w,.8h)}{v2}
  \fmf{fermion}{i1,v2}
  \fmf{fermion}{v2,f3}
  \fmf{gluon}{v2,v1}
  \fmf{fermion}{v1,i2}
  \fmf{photon}{f1,v1}
  \fmf{fermion}{f2,v1}
  \fmflabel{$p_{1,c_i}$}{i1}
  \fmflabel{$p_{2,c_k}$}{i2}
  \fmflabel{$p_{R,c_j}$}{f3}
  \fmflabel{$k_{,c_m}$}{f2}
  \fmflabel{$p_{\gamma,\nu}$}{f1}
  \fmfv{decor.shape=circle,decor.filled=shaded, d.si=.4w}{v1}
  \fmfv{decor.shape=circle,decor.filled=full, d.si=.02w}{v2}
\end{fmfchar*}
\end{fmffile}
\hspace{1cm}
\begin{fmffile}{kinqq2}
\begin{fmfchar*}(50,50)
  \fmfleft{i1,i2}
  \fmfright{f1,f2,f3}
  \fmfforce{(0,0)}{i2} 
  \fmfforce{(0,h)}{i1} 
  \fmfforce{(.5w,.5h)}{v1}
  \fmfforce{(w,h)}{f1} 
  \fmfforce{(w,0)}{f2} 
  \fmfforce{(.4w,0)}{f3} 
  \fmfforce{(.2w,.2h)}{v2}
  \fmf{fermion}{i1,v1}
  \fmf{gluon}{v1,v2}
  \fmf{fermion}{v2,i2}
  \fmf{fermion}{f3,v2}
  \fmf{photon}{f2,v1}
  \fmflabel{$p_1$}{i1}
  \fmflabel{$p_2$}{i2}
  \fmflabel{$p_R$}{f1}
  \fmflabel{$k$}{f3}
  \fmflabel{$p_{\gamma}$}{f2}
  \fmf{fermion}{v1,f1}
  \fmfv{decor.shape=circle,decor.filled=shaded, d.si=.4w}{v1}
  \fmfv{decor.shape=circle,decor.filled=full, d.si=.02w}{v2}
\end{fmfchar*}
\end{fmffile}
\hspace{0.8cm}
\begin{fmffile}{kinqq3}
\begin{fmfchar*}(50,50)
  \fmfleft{i1,i2}
  \fmfright{f1,f2,f3}
  \fmfforce{(0,0)}{i2} 
  \fmfforce{(0,h)}{i1} 
  \fmfforce{(.5w,.5h)}{v1}
  \fmfforce{(w,h)}{f1} 
  \fmfforce{(w,0)}{f2} 
  \fmfforce{(.w,.6h)}{f3} 
  \fmfforce{(.8w,.8h)}{v2}
  \fmf{fermion}{i1,v1}
  \fmf{fermion}{v1,i2}
  \fmf{gluon}{v1,v2}
  \fmf{fermion}{f3,v2}
  \fmf{fermion}{v2,f1}
  \fmf{photon}{f2,v1}
  \fmflabel{$p_2$}{i2}
  \fmflabel{$p_1$}{i1}
  \fmflabel{$p_R$}{f1}
  \fmflabel{$k$}{f3}
  \fmflabel{$p_{\gamma}$}{f2}
  \fmfv{decor.shape=circle,decor.filled=shaded, d.si=.4w}{v1}
  \fmfv{decor.shape=circle,decor.filled=full, d.si=.02w}{v2}
\end{fmfchar*}
\end{fmffile}
\hspace{0.8cm}
\begin{fmffile}{qq4}
\begin{fmfchar*}(50,50)
  \fmfleft{i1,i2}
  \fmfright{f1,f2,f3}
  \fmfforce{(0,0)}{i2} 
  \fmfforce{(0,h)}{i1} 
  \fmfforce{(.5w,.5h)}{v1}
  \fmfforce{(.8w,.8h)}{v2}
  \fmfforce{(w,h)}{f1} 
  \fmfforce{(w,0)}{f2} 
  \fmfforce{(.w,.6h)}{f3} 
  \fmf{fermion}{i1,v1}
  \fmf{fermion}{v1,i2}
  \fmf{fermion}{v1,v2}
  \fmf{fermion}{v2,f3}
  \fmf{photon}{v2,f1}
  \fmf{fermion}{f2,v1}
  \fmflabel{$p_2$}{i1}
  \fmflabel{$p_1$}{i2}
  \fmflabel{$p_R$}{f3}
  \fmflabel{$k$}{f2}
  \fmflabel{$p_{\gamma}$}{f1}
  \fmfv{decor.shape=circle,decor.filled=shaded, d.si=.4w}{v1}
\end{fmfchar*}
\end{fmffile}
\hspace{0.8cm}
\begin{fmffile}{qq5}
\begin{fmfchar*}(50,50)
  \fmfleft{i1,i2}
  \fmfright{f1,f2,f3}
  \fmfforce{(0,0)}{i2} 
  \fmfforce{(0,h)}{i1} 
  \fmfforce{(.5w,.5h)}{v1}
  \fmfforce{(.8w,.2h)}{v2}
  \fmfforce{(w,h)}{f1} 
  \fmfforce{(w,0)}{f2} 
  \fmfforce{(.w,.4h)}{f3} 
  \fmf{fermion}{i1,v1}
  \fmf{fermion}{v1,i2}
  \fmf{fermion}{v1,f1}
  \fmf{fermion}{f3,v2}
  \fmf{photon}{v2,f2}
  \fmf{fermion}{v2,v1}
  \fmflabel{$p_2$}{i1}
  \fmflabel{$p_1$}{i2}
  \fmflabel{$p_R$}{f1}
  \fmflabel{$k$}{f3}
  \fmflabel{$p_{\gamma}$}{f2}
  \fmfv{decor.shape=circle,decor.filled=shaded, d.si=.4w}{v1}
\end{fmfchar*}
\end{fmffile}
\vspace{0.3cm}
\caption{NLO Feynman diagrams for the NLP contributions of the process $q(p_1)\bar{q}(p_2)\rightarrow q(p_R) \bar{q}(k) \gamma(p_{\gamma})$.}
\label{fig:qqbartoqqbar}
\end{figure}
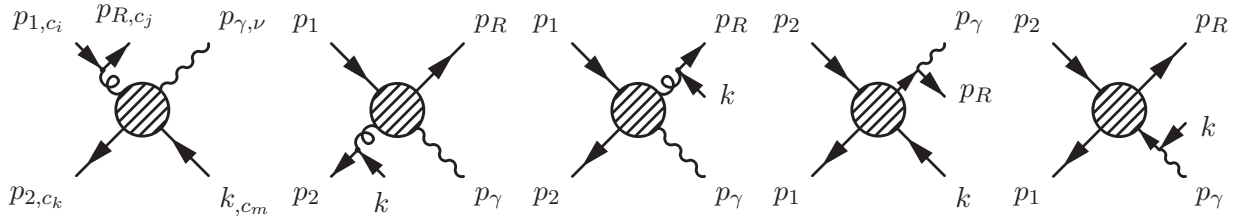

We now turn to the partonic sub-process of Fig.~\ref{fig:qqbartoqqbar}, whose final state contains only quarks in addition to the photon. To compute the NLP differential cross-section, we therefore only need to consider Eq.~(\ref{eq:NLPquark}). There are 5 types of contributions, indicated in Fig.~\ref{fig:qqbartoqqbar}, which can be split into three categories: initial state splitting ($I$), final state splitting ($F$) and final state fragmentation ($FF$). The contributions can be obtained by letting the quark emission operator $\mathcal{Q}_j$ act on the tree level processes $qg\rightarrow q\gamma$, $\bar{q}g\rightarrow \bar{q}\gamma$, $q\bar{q}\rightarrow g\gamma$ and $q\bar{q}\rightarrow q \bar{q}$. From the processes thus obtained, we then select only those with the specific partonic assignment $q\bar{q}\rightarrow q\bar{q}\gamma$. We obtain 
\begin{eqnarray}
\mathcal{A}_{{\rm NLP,quarks}} &=& \Bigg[-\frac{ig_st^{a}_{c_jc_i}}{ 2 p_1\cdot p_R} \bar{u}(p_R)\gamma_{\mu} u(p_1) \mathcal{M}^{\mu,a}_{(g)\bar{q}\rightarrow \bar{q}\gamma}(p_1,p_2,k,p_{\gamma})\nonumber \\
&& - \frac{ig_st^{a}_{c_kc_m}}{2p_2\cdot k} \bar{v}(p_2)\gamma_{\mu}v(k)\mathcal{M}^{\mu,a}_{q(g)\rightarrow q\gamma}(p_1,p_2,p_R,p_{\gamma})\Bigg] \nonumber\\
&& +\frac{ig_s t^a_{c_jc_m}}{2p_R\cdot k}\bar{u}(p_R) \gamma_{\mu}v(k) \mathcal{M}^{\mu,a}_{q\bar{q} \rightarrow (g)\gamma}(p_1,p_2,p_R,p_{\gamma})  \nonumber\\
&&+\Bigg[- \frac{ig_{\rm EM}Q_q}{2p_{\gamma}\cdot p_R} \bar{u}(p_R)\slashed{\epsilon}^*(p_{\gamma})\slashed{p}_{\gamma}\mathcal{M}_{q\bar{q}\rightarrow (q)\bar{q}}(p_1,p_2,p_{\gamma},k) \nonumber\\
&&  + \frac{ig_{\rm EM}Q_q}{2p_{\gamma}\cdot k} \mathcal{M}_{q\bar{q}\rightarrow q(\bar{q})}(p_1,p_2,p_R,p_{\gamma}) \slashed{p}_{\gamma}\slashed{\epsilon}^*(p_{\gamma})v(k)\Bigg] \nonumber \\
&\equiv& \mathcal{A}_{I} + \mathcal{A}_{FF}+ \mathcal{A}_{F},
\label{ampqq}
\end{eqnarray}
where the notation $(a)$ in each hard scattering matrix element ${\cal M}$ indicates that the latter does not include the external wavefunction for parton $a$~\footnote{In the third line of Eq.~(\ref{ampqq}), we have been careful to include only one part of the result of the $\mathcal{Q}$ operator so as to avoid double counting, as explained in appendix~\ref{app:definitions}.}. The complete NLP cross-section can be written as a sum of these contributions as 
\begin{eqnarray}
vw(1-v)s\frac{{\rm d}\sigma^{{\rm NLP}}_{q\bar{q}\rightarrow q\bar{q}\gamma}}{{\rm d}{v}{\rm d}w} =
Q_q^2\alpha\alpha_s^2\left[\Sigma_{I,I}+\Sigma_{I,F}+\Sigma_{I,FF}+\Sigma_{F,F}+\Sigma_{F,FF}+\Sigma_{FF,FF}\right],
\end{eqnarray}
where $\Sigma_{I,J}$ ($I, J\in\{I,F,FF\}$) denotes the contribution from the integrated, summed and averaged soft quark squared amplitude $\langle \mathcal{A}_I \mathcal{A}_J^{\dagger}\rangle$ (plus the complex conjugate if $I\neq J$). The individual contributions are found to be
\begin{eqnarray}
\Sigma_{I,I} &=& -\frac{C_F}{2C_A}\frac{1}{\varepsilon}\frac{2v^4-4v^3+4v^2-2v+1}{1-v}+\frac{C_F}{2C_A}\ln(1-w)\frac{2v^4-4v^3+4v^2-2v+1}{1-v}+\mathcal{O}(1) \nonumber\\
\Sigma_{I,F} &=& \ln(1-w)\left\{\frac{C_F}{C_A^2}v^2(1-v)-\frac{C_F}{C_A}v(3-2v(1-v))\right\} + \mathcal{O}(1) \nonumber\\
\Sigma_{I,FF} &=& \mathcal{O}(1)\nonumber\\
\Sigma_{F,F} &=& -\frac{1}{\varepsilon}\left\{\frac{C_F}{C_A^2}\frac{v(3v^3-6v^2+4v-1)}{1-v}+\frac{C_F}{C_A}\frac{2v^6-6v^5+8v^4-6v^3+5v^2-3v+1}{1-v}\right\} \nonumber\\
&& + \ln(1-w)\Bigg\{\frac{C_F}{C_A^2}\frac{v(3v^3-6v^2+4v-1)}{1-v}\nonumber\\
&&\hspace{2.5cm}+\frac{C_F}{C_A}\frac{2v^6-6v^5+8v^4-6v^3+5v^2-3v+1}{1-v}\Bigg\}+\mathcal{O}(1)\nonumber\\
\Sigma_{F,FF} &=& \mathcal{O}(1)\nonumber\\
\Sigma_{FF,FF} &=& \frac{1}{(1-w)_+}\frac{C_F}{3C_A}T_{q\bar{q}} + \mathcal{O}(1).
\end{eqnarray}
By separating them, we can explicitly identify the jet contribution $\Sigma_{FF,FF}$, which contributes at LP NLL order. As we have already seen in the DIS and $e^+e^-$ cases, also here it is actually the collinear information carried by the soft quark that creates the LP term.  Furthermore, there are two types of soft-collinear contributions ($\Sigma_{I,I}$ and $\Sigma_{F,F}$) whose NLP log can easily be guessed from the collinear pole. There is also, however, an interference term that contributes at NLP level and does not come with a collinear pole: $\Sigma_{I,F}$. This term can be regarded as arising from the wide angle emission of a soft quark. The contributions where a soft quark emission from an observed final state parton interferes with a similar emission from an unobserved parton vanish at $\mathcal{O}(1)$ and will therefore only contribute beyond NLP LL order. %\MB{can we understand why?} \\

Putting everything together, the NLP differential cross-section for this subprocess is 
\begin{eqnarray}
vw(1-v)s\frac{{\rm d}\sigma^{{\rm NLP}}_{q\bar{q}\rightarrow q\bar{q}\gamma}}{{\rm d}{v}{\rm d}w} &=&  \\
&& \hspace{-3.0cm} Q_q^2\alpha\alpha_s^2 \Bigg[\frac{1}{\varepsilon}\Bigg\{-\frac{C_F}{C_A^2}(v(3(v-1)v+1)) -\frac{C_F}{C_A}\frac{T_{q\bar{q}}(2(v-1)v((v-1)v+1)+3)}{2(1-v)}\Bigg\} \nonumber \\
&& \hspace{-1cm} + \ln(1-w)\left\{\frac{C_F}{C_A^2}v(1-2v)^2+\frac{C_F}{C_A}\frac{2(v-1)v((v-1)v(2(v-1)v+5)+7)+3}{2(1-v)}\right\} \nonumber \\
&&  \hspace{-1cm} + \frac{1}{(1-w)_+}\frac{C_F}{3C_A}T_{q\bar{q}} + \mathcal{O}(1)
\Bigg].\nonumber
\label{eq:qqNLPq}
\end{eqnarray}
This result is remarkable, in that it demonstrates that the quark emission operator that we have introduced in section \ref{sec:NLPquark} can be used to correctly obtain the NLP contribution to the NLO cross-section. Since the emission of a quark is already at NLP order due to the momentum information that is carried by the spinor, we do not need the momentum shift of the LO matrix elements. 
%==========================

\subsection{$q g \rightarrow q g \gamma $}
\label{sec:qgfinal}
\begin{figure}
\centering
\centering
\begin{fmffile}{kin3a}
\begin{fmfchar*}(60,60)
  \fmfleft{i1,i2}
  \fmfright{f1,f2,f3}
  \fmfforce{(0,0)}{i2} 
  \fmfforce{(0,h)}{i1} 
  \fmfforce{(.5w,.5h)}{v1}
  \fmfforce{(w,h)}{f1} 
  \fmfforce{(w,0)}{f2} 
  \fmfforce{(w,.7h)}{f3} 
  \fmfforce{(.8w,.8h)}{v2}
  \fmf{fermion}{i1,v1}
  \fmf{gluon}{v1,i2}
  \fmf{photon}{f2,v1}
  \fmflabel{$p_{2,c_i}$}{i1}
  \fmflabel{$p_{1,\mu,a}$}{i2}
  \fmflabel{$p_{R,c_j}$}{f1}
  \fmflabel{$k_{\sigma,b}$}{f3}
  \fmflabel{$p_{\gamma,\nu}$}{f2}
  \fmf{fermion}{v2,f1}
  \fmf{fermion}{v1,v2}
  \fmf{gluon}{v2,f3}
  \fmfv{decor.shape=circle,decor.filled=shaded, d.si=.4w}{v1}
  \fmfv{decor.shape=circle,decor.filled=full, d.si=.02w}{v2}
\end{fmfchar*}
\end{fmffile}
\hspace{1.5cm}
\begin{fmffile}{kin4a}
\begin{fmfchar*}(60,60)
  \fmfleft{i1,i2}
  \fmfright{f1,f2,f3}
  \fmfforce{(0,0)}{i2} 
  \fmfforce{(0,h)}{i1} 
  \fmfforce{(.5w,.5h)}{v1}
  \fmfforce{(w,h)}{f1} 
  \fmfforce{(w,0)}{f2} 
  \fmfforce{(.5w,0)}{f3} 
  \fmfforce{(.2w,.2h)}{v2}
  \fmf{fermion}{i1,v1}
  \fmf{gluon}{v1,v2}
  \fmf{gluon}{v2,i2}
  \fmf{gluon}{v2,f3}
  \fmf{photon}{f2,v1}
  \fmflabel{$p_2$}{i1}
  \fmflabel{$p_1$}{i2}
  \fmflabel{$p_R$}{f1}
  \fmflabel{$k$}{f3}
  \fmflabel{$p_{\gamma}$}{f2}
  \fmf{fermion}{v1,f1}
  \fmfv{decor.shape=circle,decor.filled=shaded, d.si=.4w}{v1}
  \fmfv{decor.shape=circle,decor.filled=full, d.si=.02w}{v2}
\end{fmfchar*}
\end{fmffile}
\hspace{1cm}
\begin{fmffile}{kin5a}
\begin{fmfchar*}(60,60)
  \fmfleft{i1,i2}
  \fmfright{f1,f2,f3}
  \fmfforce{(0,0)}{i2} 
  \fmfforce{(0,h)}{i1} 
  \fmfforce{(.5w,.5h)}{v1}
  \fmfforce{(w,h)}{f1} 
  \fmfforce{(w,0)}{f2} 
  \fmfforce{(.5w,h)}{f3} 
  \fmfforce{(.2w,.8h)}{v2}
  \fmf{fermion}{i1,v2}
  \fmf{fermion}{v2,v1}
  \fmf{gluon}{v1,i2}
  \fmf{gluon}{v2,f3}
  \fmf{photon}{f2,v1}
  \fmflabel{$p_2$}{i1}
  \fmflabel{$p_1$}{i2}
  \fmflabel{$p_R$}{f1}
  \fmflabel{$k$}{f3}
  \fmflabel{$p_{\gamma}$}{f2}
  \fmf{fermion}{v1,f1}
  \fmfv{decor.shape=circle,decor.filled=shaded, d.si=.4w}{v1}
  \fmfv{decor.shape=circle,decor.filled=full, d.si=.02w}{v2}
\end{fmfchar*}
\end{fmffile}
\hspace{1cm}
\begin{fmffile}{kin6a}
\begin{fmfchar*}(60,60)
  \fmfleft{i1,i2}
  \fmfright{f1,f2,f3}
  \fmfforce{(0,0)}{i2} 
  \fmfforce{(0,h)}{i1} 
  \fmfforce{(.5w,.5h)}{v1}
  \fmfforce{(w,h)}{f1} 
  \fmfforce{(w,0)}{f2} 
  \fmfforce{(.w,.5h)}{f3} 
  \fmf{fermion}{i1,v1}
  \fmf{gluon}{v1,i2}
  \fmf{gluon}{v1,f3}
  \fmf{photon}{f2,v1}
  \fmflabel{$p_2$}{i1}
  \fmflabel{$p_1$}{i2}
  \fmflabel{$p_R$}{f1}
  \fmflabel{$k$}{f3}
  \fmflabel{$p_{\gamma}$}{f2}
  \fmf{fermion}{v1,f1}
  \fmfv{decor.shape=circle,decor.filled=shaded, d.si=.4w}{v1}
\end{fmfchar*}
\end{fmffile}
\vspace{0.5cm}
\caption{NLO Feynman diagrams for the NLP contributions of the process $g(p_1)q(p_2)\rightarrow q(p_R) g(k) \gamma(p_{\gamma})$.}
\label{fig:qgtoqg}
\end{figure}
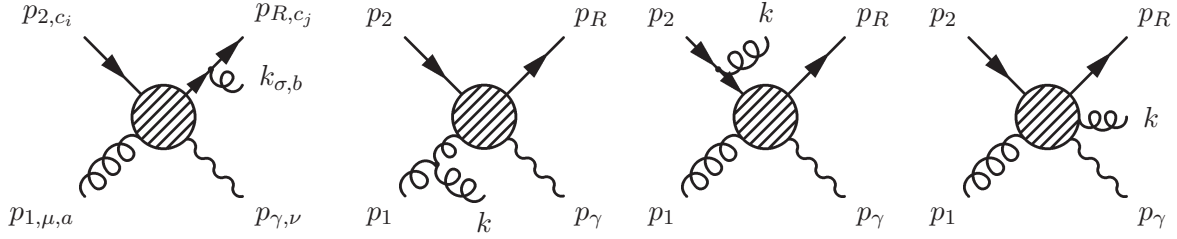

This is the only subprocess for NLO prompt photon production that contains NLP contributions due to both quark and gluon emission. Let us first consider the radiation of a gluon, where the diagrams that we need are shown in Fig.~\ref{fig:qgtoqg}.
The derivation of the (next-to-)soft gluon contribution is directly analogous to the $q\bar{q}\rightarrow g g \gamma $ (next-to-)soft gluon amplitude analysed previously, and hence will not be presented in full detail here. The result is 
\begin{eqnarray}
\langle |\mathcal{A}_{{\rm LP+NLP},qg\rightarrow \gamma qg}|^2 \rangle &=& \frac{Q_q^2 g_s^4g_{\rm EM}^2}{8C_A(1-\varepsilon)}\Bigg[\left(C_F-\frac{1}{2}C_A\right)\frac{2p_1\cdot p_R}{(p_1\cdot k)(p_R\cdot k)} \\
&&\hspace{4cm} \times H_{gq\rightarrow q\gamma}(p_1+\delta p_{1;R},p_2,p_R-\delta p_{R;1},p_{\gamma}) \nonumber \\
&&\hspace{0.8cm} + \frac{1}{2}C_A\frac{2p_1\cdot p_2}{(p_1\cdot k)(p_2\cdot k)}H_{gq\rightarrow q\gamma}(p_1+\delta p_{1;2},p_2+\delta p_{2;1},p_R,p_{\gamma}) \nonumber \\
&& \hspace{0.5cm} + \frac{1}{2}C_A\frac{2p_2\cdot p_R}{(p_2\cdot k)(p_R\cdot k)}H_{gq\rightarrow q\gamma}(p_1,p_2+\delta p_{2;R},p_R-\delta p_{R;2},p_{\gamma}) \Bigg],\nonumber
\end{eqnarray}
where the factor of $1-\varepsilon$ in the common denominator stems from the fact that gluons can take $2(1-\varepsilon)$ different spin orientations in $d=4-2\varepsilon$ dimensions. 
The expression for the LP+NLP soft gluon amplitude leads to the following differential cross-section, valid up to NLP order:
\begin{eqnarray}
\label{eq:qggluon}
vw(1-v)s\frac{{\rm d}\sigma^{{\rm g, LP+NLP}}_{qg\rightarrow \gamma q g }}{{\rm d}{v}{\rm d}w} &=&  \frac{Q_q^2\alpha\alpha_s^2}{C_A(1-\varepsilon)}\Bigg[-\frac{1}{\varepsilon}\frac{T_{qg}}{(1-w)_+}(C_A+C_F)\\
&&+\frac{1}{\varepsilon}\left\{C_F T_{qg} -C_A\frac{v(v(v(2v-5)+4)-2)}{1-v}\right\} \nonumber\\
&& + \left(\frac{\ln(1-w)}{1-w}\right)_+(2C_A+C_F)T_{qg}\nonumber \\
&&+ \frac{1}{(1-w)_+}\Bigg\{T_{qg}\left(
C_F\ln\left(v^2(1-v)\right)+2C_A\ln v -2(C_F+C_A)\ln\frac{\bar{\mu}^2}{s}\right)\nonumber\\
&& \hspace{2cm}+2C_Av((v-3)v+3)+C_Fv(v-2)^2\Bigg\}\nonumber\\
&& -\ln(1-w)\left\{C_F T_{qg}+C_A\frac{v(v((18-7v)v-16)+8)}{2(1-v)}\right\}\nonumber\\
&&+\mathcal{O}(\delta(1-w))+\mathcal{O}(1)\Bigg]\nonumber,
\end{eqnarray}
where $T_{qg} = v(1+(1-v)^2)$. 

\begin{figure}
\centering
\begin{fmffile}{kin3b}
\begin{fmfchar*}(60,60)
  \fmfleft{i1,i2}
  \fmfright{f1,f2,f3}
  \fmfforce{(0,0)}{i2} 
  \fmfforce{(0,h)}{i1} 
  \fmfforce{(.5w,.5h)}{v1}
  \fmfforce{(w,h)}{f1} 
  \fmfforce{(w,0)}{f2} 
  \fmfforce{(w,.7h)}{f3} 
  \fmfforce{(.8w,.8h)}{v2}
  \fmf{gluon}{i2,v1}
  \fmf{fermion}{i1,v1}
  \fmf{photon}{f2,v1}
  \fmflabel{$p_{2,c_i}$}{i1}
  \fmflabel{$p_{1,\mu,a}$}{i2}
  \fmflabel{$p_{R,c_j}$}{f3}
  \fmflabel{$k_{\sigma,b}$}{f1}
  \fmflabel{$p_{\gamma,\nu}$}{f2}
  \fmf{gluon}{f1,v2}
  \fmf{fermion}{v1,v2}
  \fmf{fermion}{v2,f3}
  \fmfv{decor.shape=circle,decor.filled=shaded, d.si=.4w}{v1}
  \fmfv{decor.shape=circle,decor.filled=full, d.si=.02w}{v2}
\end{fmfchar*}
\end{fmffile}
\hspace{1.5cm}
\begin{fmffile}{kin4b}
\begin{fmfchar*}(60,60)
  \fmfleft{i1,i2}
  \fmfright{f1,f2,f3}
  \fmfforce{(0,0)}{i2} 
  \fmfforce{(0,h)}{i1} 
  \fmfforce{(.5w,.5h)}{v1}
  \fmfforce{(w,h)}{f1} 
  \fmfforce{(w,0)}{f2} 
  \fmfforce{(.5w,0)}{f3} 
  \fmfforce{(.2w,.2h)}{v2}
  \fmf{fermion}{i1,v1}
  \fmf{fermion}{v1,v2}
  \fmf{gluon}{v2,i2}
  \fmf{fermion}{v2,f3}
  \fmf{photon}{f2,v1}
  \fmflabel{$p_2$}{i1}
  \fmflabel{$p_1$}{i2}
  \fmflabel{$k$}{f1}
  \fmflabel{$p_R$}{f3}
  \fmflabel{$p_{\gamma}$}{f2}
  \fmf{gluon}{f1,v1}
  \fmfv{decor.shape=circle,decor.filled=shaded, d.si=.4w}{v1}
  \fmfv{decor.shape=circle,decor.filled=full, d.si=.02w}{v2}
\end{fmfchar*}
\end{fmffile}
\hspace{1cm}
\begin{fmffile}{kin5b}
\begin{fmfchar*}(60,60)
  \fmfleft{i1,i2}
  \fmfright{f1,f2,f3}
  \fmfforce{(0,0)}{i2} 
  \fmfforce{(0,h)}{i1} 
  \fmfforce{(.5w,.5h)}{v1}
  \fmfforce{(w,h)}{f1} 
  \fmfforce{(w,0)}{f2} 
  \fmfforce{(w,.7h)}{f3} 
  \fmfforce{(.8w,.8h)}{v2}
  \fmf{fermion}{i1,v1}
  \fmf{gluon}{v1,i2}
  \fmf{gluon}{f2,v1}
  \fmflabel{$p_{2}$}{i1}
  \fmflabel{$p_{1}$}{i2}
  \fmflabel{$p_{R}$}{f3}
  \fmflabel{$k_{\sigma}$}{f2}
  \fmflabel{$p_{\gamma}$}{f1}
  \fmf{photon}{v2,f1}
  \fmf{fermion}{v1,v2}
  \fmf{fermion}{v2,f3}
  \fmfv{decor.shape=circle,decor.filled=shaded, d.si=.4w}{v1}
  \fmfv{decor.shape=circle,decor.filled=full, d.si=.02w}{v2}
\end{fmfchar*}
\end{fmffile}
\hspace{1cm}
\vspace{0.5cm}
\caption{Feynman diagrams for the NLP contributions of the process $g(p_1)q(p_2)\rightarrow q(p_R) g(k) \gamma(p_{\gamma})$. }
\label{fig:NLPqg}
\end{figure}
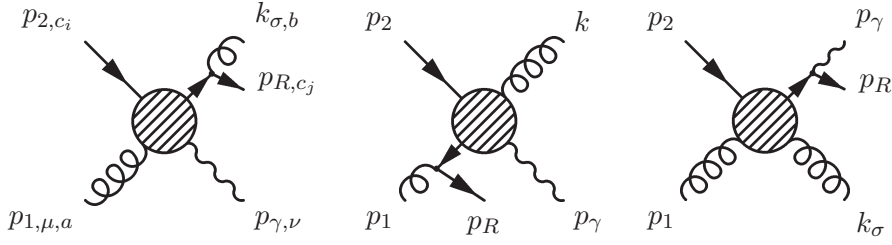
Next, we need the soft quark radiative contribution, and there are three $2\rightarrow 2$ hard scattering diagrams on which we can use the quark emission operator $\mathcal{Q}_j$ to turn it into the process $q g \rightarrow q g \gamma $. These processes are $qg\rightarrow q\gamma$, $q\bar{q}\rightarrow g\gamma$ and $qg\rightarrow q g$.  As in section \ref{sec:qqbartoqqbargamma}, we will only select the resulting Feynman diagrams that describe the $qg\rightarrow qg\gamma$ sub-process. The generated NLP Feynman diagrams are given in Fig.~\ref{fig:NLPqg}, and the soft quark amplitude then consists of three pieces:
\begin{eqnarray}
\mathcal{A}_{{\rm NLP,quarks}} &=& -\frac{ig_st^{b}_{c_jc_k}}{ 2 k\cdot p_R} \epsilon^*_{\sigma}(k)\bar{u}(p_R)\gamma^{\sigma}\slashed{k}\mathcal{M}^{a}_{c_kc_i,gq\rightarrow \gamma (q)}(p_1,p_2,p_{\gamma},k) \nonumber\\
&& - \frac{ig_st^{a}_{c_jc_k}}{2p_1\cdot p_R} \epsilon_{\mu}(p_1)\bar{u}(p_R)\gamma^{\mu}\slashed{p_1}\mathcal{M}^{b}_{c_kc_i,(\bar{q})q\rightarrow \gamma g}(p_1,p_2,p_{\gamma},k) \nonumber\\
&& - \frac{ig_{\rm EM}Q_q}{2p_{\gamma}\cdot p_R} \epsilon^*_{\nu}(p_{\gamma})\bar{u}(p_R)\gamma^{\nu}\slashed{p}_{\gamma}\mathcal{M}^{ab}_{c_jc_i,gq\rightarrow (q)g}(p_1,p_2,p_{\gamma},k) \nonumber\\
&\equiv& \mathcal{A}_{FF} + \mathcal{A}_{I}+ \mathcal{A}_{F}.
\end{eqnarray}
In section~\ref{sec:qqbargamma}, we discussed the need to potentially include external ghost contributions when summing over all gluon polarisations in the final state. In that previous case, these contributions were absent at NLP. Here they will contribute owing to the presence of two hard gluons, as the quark is already soft and in order to observe the photon it needs to recoil against at least one other hard particle in the final state. The complete quark NLP cross-section then turns out to be:
\begin{eqnarray}
\label{eq:qgquark}
vw(1-v)s\frac{{\rm d}\sigma^{{\rm q,NLP}}_{qg\rightarrow qg\gamma}}{{\rm d}{v}{\rm d}w} \equiv
 \frac{Q_q^2\alpha\alpha_s^2}{C_A(1-\varepsilon)}\left[\Sigma_{I,I}+\Sigma_{I,F}+\Sigma_{I,FF}+\Sigma_{F,F}+\Sigma_{F,FF}+\Sigma_{FF,FF}\right],
\end{eqnarray}
where the various contributions are as follows:
\begin{eqnarray}
\Sigma_{I,I} &=& -\frac{1}{\varepsilon}\frac{C_F}{C_A}\frac{v(2v(v-1)+1)}{2(1-v)}+\ln(1-w)\frac{C_F}{C_A}\frac{v(2v(v-1)+1)}{2(1-v)}+\mathcal{O}(1) \nonumber\\
\Sigma_{I,F} &=& \ln(1-w)\left\{\frac{C_F}{C_A}v^3+\frac{v^4}{2(1-v)}\right\}+\mathcal{O}(1)\nonumber \\
\Sigma_{I,FF} &=& \mathcal{O}(1)\nonumber \\
\Sigma_{F,F} &=& -\frac{1}{\varepsilon}\left\{\frac{1}{2}\frac{C_F}{C_A}v(1-v)(v^2+1)+\frac{1}{2}\frac{v^2(v^2+1)}{(1-v)}\right\}\nonumber\\
&&+\ln(1-w)\left\{\frac{1}{2}\frac{C_F}{C_A}v(1-v)(v^2+1)+\frac{1}{2}\frac{v^2(v^2+1)}{(1-v)}\right\}+\mathcal{O}(1)\nonumber \\
\Sigma_{F,FF} &=& \mathcal{O}(1) \nonumber\\
\Sigma_{FF,FF} &=& \frac{C_F}{4C_A}\frac{T_{qg}}{(1-w)_+}+\mathcal{O}(1). 
\end{eqnarray}
The complete NLP differential cross-section for the $qg \rightarrow  q g \gamma$ subprocess, where one additional quark is radiated, is
\begin{eqnarray}
\label{eq:quark}
vw(1-v)s\frac{{\rm d}\sigma^{{\rm q,NLP}}_{qg\rightarrow qg\gamma}}{{\rm d}{v}{\rm d}w} &=&  \frac{Q_q^2\alpha\alpha_s^2}{C_A(1-\varepsilon)}\Bigg[-\frac{1}{\varepsilon}\left\{C_F\frac{v(v(v((v-2)v+4)-4)+2)}{2(1-v)}+C_A\frac{v^2(v^2+1)}{2(1-v)}\right\} \nonumber\\
&&+\ln(1-w)\left\{C_F\frac{v((v-2)v((v-2)v+2)+2)}{2(1-v)}+C_A\frac{v(2v^3+v)}{2(1-v)}\right\} \nonumber\\
&&+ \frac{1}{(1-w)_+}\frac{C_F}{4}T_{qg} + \mathcal{O}(1)
\Bigg].
\end{eqnarray}
We may now combine the gluon and quark radiative contributions, where the sum of Eqs.~(\ref{eq:qggluon}) and~(\ref{eq:quark}) yields
\begin{eqnarray}
vw(1-v)s\frac{{\rm d}\sigma^{{\rm LP+NLP}}_{qg\rightarrow qg\gamma}}{{\rm d}{v}{\rm d}w} &=& \frac{Q_q^2\alpha\alpha_s^2}{C_A(1-\varepsilon)}\Bigg[-\frac{1}{\varepsilon}\frac{1}{(1-w)_+}(C_F+C_A)T_{qg}\\
&&+\frac{1}{\varepsilon}\left\{C_A\frac{v}{2}(5(v-1)v+4)-C_F\frac{v(v(v^3-2v+4)-2)}{2(1-v)}\right\}\nonumber\\
&&+ \left(\frac{\ln(1-w)}{1-w}\right)_+(2C_A+C_F)T_{qg} \nonumber\\
&& + \frac{1}{(1-w)_+}\Bigg\{T_{qg}\left(C_F\ln((1-v)v^2)+2C_A\ln v-2(C_F+C_A)\ln\frac{\bar{\mu}^2}{s}\right)\nonumber\\
&& \hspace{2cm} + 2C_Av(v(v-3)+3) -\frac{C_F}{4}v((18-5v)v-18)\Bigg\}\nonumber\\
&& +\ln(1-w)\left\{-C_A\frac{v}{2}(9(v-1)v+8)+C_F\frac{v\left(v\left((v-2)v^2+4\right)-2\right)}{2(1-v)}\right\}\Bigg].\nonumber
\end{eqnarray}
This agrees with the full NLO calculation truncated to NLP order, as presented here in appendix~\ref{app:NLOcalc}. Our result shows that we can separately treat the (next-to-)soft gluon and fermion radiation, as is implied by our general analysis in section~\ref{sec:NLP}. Note also in this case that we do not have to add an additional hard collinear piece to obtain all LP NLL effects. They are fully generated by taking the effect of a soft quark emission into account. 

We have here presented three of a total of seven different partonic sub-channels for prompt photon production at NLO. The remaining channels work analogously to the ones already presented, and are listed in appendix \ref{app:remain}.  We also present results for the unsubtracted NLO cross-sections (up to NLP order) in appendix \ref{app:NLOcalc}, given that these have not previously been presented in the literature.

\section{Discussion}
\label{sec:discuss}

In this paper, we have examined the role of next-to-soft effects in processes containing one or more coloured particles in the final state. A complication of such processes is that hard collinear real radiation  produces terms that are enhanced near threshold, making recently derived next-to-soft theorems potentially incomplete. It is then necessary to examine what the domain of applicability of such theorems is, or, in other words, to which logarithmic order these theorems are valid. Another complication is the presence of soft quarks.

We have examined DIS, hadroproduction in electron-positron annihilation and prompt photon production, and find a number of interesting results. More specifically, the NLO cross-section up to next-to-soft order in the emitted {\it gluon} momentum can be expressed in terms of the non-radiative amplitude with shifted momenta for all cases. This is directly analogous to a similar conclusion reached for colour singlet production in Ref.~\cite{DelDuca:2017twk}. Due to the presence of unobserved quarks in the final state for all considered processes, we need to complement the next-to-soft gluon amplitude with a {\it soft quark} amplitude. This can be treated completely independently from the next-to-soft gluon amplitude, and itself factorises in terms of a universal quark emission operator, that we have defined. Adding this soft (and potentially wide-angle) quark contribution to the next-to-soft gluon amplitude shows that we are able to capture all LL NLP behaviour, and also fill in missing LP NLL information. This is due to the fact that the soft expansion is systematically building up the effects of collinear quark emission, where the latter contributes at LP NLL and beyond. Corrections to the next-to-soft formalism (starting at NLL) arise from collinear emissions that are next-to-next-to soft and beyond. These corrections can include an NLP component, suggesting that a systematic study of {\it next-to-collinear} effects would be useful (see Ref.~\cite{Nandan:2016ohb} for work in a more formal context).

When both the emitting and the emitted particles are gluons, one needs to include the known gluon jet functions of e.g. Ref.~\cite{Sterman:1987aj} to capture the missing LP NLL hard-collinear information. That the LL contributions are correctly captured by the next-to-soft formalism is not surprising: na\"{i}vely, one expects that the leading singular behaviour should come from where all emitted radiation is maximally soft and/or collinear. However, that this is true even for {\it next-to-soft} emissions is itself a non-trivial result. Furthermore, it is a highly useful piece of information for considering the resummation of NLP effects, given that one would start by proving that LL contributions can be resummed. At fixed order, we have also clarified that leading logarithmic terms at (N)LP order may arise from the mass factorisation procedure, which would need to be kept track of in potential numerical studies.

\section*{Note added}

In the final stages of preparing this paper, Ref.~\cite{Moult:2019mog} appeared, which addresses the emission of soft and collinear radiation (including quarks) up to next-to-leading power, within the framework of soft collinear effective theory.

\section*{Acknowledgments}

This article is based upon work from COST Action CA16201 PARTICLEFACE
supported by COST (European Cooperation in Science and Technology). CDW was supported by the Science and Technology Facilities Council (STFC) Consolidated Grant ST/P000754/1 \textit{``String theory, gauge theory \& duality"}, and by the European Union's Horizon 2020 research and innovation programme under the Marie Sk\l{}odowska-Curie grant agreement 
No.~764850 {\it ``\href{https://sagex.ph.qmul.ac.uk}{SAGEX}"}. WB, EL and MvB acknowledge support from the Dutch NWO-I program 156, "Higgs as Probe and Portal". MvB also acknowledges 
support from the Christine Mohrmann Stipendium. We wish to thank Lorenzo Magnea, Jort Sinninghe Damst\'{e}, Leonardo Vernazza and the participants of the 2018 Nikhef workshop "Next-to-leading power corrections in particle physics" for useful discussions. 

\appendix

\section{Useful definitions}
\label{app:definitions}

In this appendix, we collect useful formulae relating to the action of various operators appearing in Eqs.~(\ref{eq:NLPgluon}, \ref{eq:NLPquark}), as well as Fig.~\ref{fig:Soperator}. First, in Eq.~(\ref{eq:NLPgluon}), repeated here for convenience
\begin{eqnarray}
\mathcal{A}_{\rm NLP} &=&\sum_{j=1}^{n+2} \frac{g_s \mathbf{T}_j}{2p_j \cdot k}\left(\mathcal{O}_{{\rm scal},j}^{\sigma}+\mathcal{O}_{{\rm spin},j}^{\sigma}+\mathcal{O}_{{\rm orb},j}^{\sigma}\right)\otimes i\mathcal{M}_{\rm H}(p_1,\dots,p_i,
\dots,p_{n+2})\epsilon^*_{\sigma}(k),
\end{eqnarray} 
we must consider a general colour generator ${\bf T}_j$ acting on an external parton line $j$. This leads to a colour factor dressing the nonradiative amplitude
\begin{equation*}
\bold{T}_j \equiv 
\left\{
\begin{array}{rl}
t^{c}_{c_jc_i} &\text{ for an incoming quark or outgoing anti-quark with colour label $c_i$; }\\
-t^{c}_{c_ic_j} &\text{ for an outgoing quark or incoming anti-quark with colour label $c_i$; }\\
if^{cab}& \text{ for an external gluon with colour label $a$,}
\end{array}
\right.
\end{equation*}
where $\{t^a_{ij}\}$ are components of a generator in the fundamental representation. Next, we collect results for the numerator of the scalar contribution appearing in Eq.~(\ref{eq:NLPgluon}). This can be written as
\begin{equation}
\mathcal{O}_{{\rm scal},j}^{\sigma}\equiv (2p_j^{\sigma}+\eta k^{\sigma}),
\end{equation}
where $\eta=+1$ (-1) for a hard emitting particle in the final (initial) state respectively. The spin contribution in Eq.~(\ref{eq:NLPgluon}) can be written in the generic form
\begin{equation}
\mathcal{O}_{{\rm spin},j}^{\sigma}\equiv 2ik_{\alpha}\Sigma^{\sigma\alpha}_j .
\end{equation}
Here, $\Sigma^{\sigma\alpha}_j$ is a Lorentz generator in the appropriate representation of parton $j$, and given in specific cases by
\begin{eqnarray*}
\Sigma^{\sigma\alpha}_j \equiv \left\{
\begin{array}{cl}
S^{\alpha\sigma}& \text{ for an incoming or outgoing quark;}\\
S^{\sigma\alpha}& \text{ for an incoming or outgoing anti-quark; }\\
M^{\sigma\alpha,\mu\rho} & \text{ for an incoming gluon carrying Lorentz index $\mu$; }\\
M^{\sigma\alpha,\rho\mu} & \text{ for an outgoing gluon carrying Lorentz index $\mu$, }
\end{array}
\right.
\end{eqnarray*}
where the relevant generators are defined in Eqs.~(\ref{Sspindef}) and~(\ref{Mspindef}). The third term in Eq.~(\ref{eq:NLPgluon}) can be written as
\begin{equation}
\mathcal{O}_{{\rm orb},j}^{\sigma} \equiv 2 k_{\alpha} i L^{\sigma \alpha}_j ,
\end{equation}
where $L^{\sigma\alpha}_j$ is the orbital angular momentum operator of parton $j$, defined by
\begin{eqnarray*}
L^{\sigma \alpha}_j \equiv \left\{
\begin{array}{cl}
i\left(p_j^{\sigma}\frac{\partial}{\partial p_{j \alpha}}-p_j^{\alpha}\frac{\partial}{\partial p_{j \sigma}}\right)  & \text{ for an initial state emission; }\\
i\left(p_j^{\alpha}\frac{\partial}{\partial p_{j \sigma}}-p_j^{\sigma}\frac{\partial}{\partial p_{j \alpha}}\right) & \text{ for a final state emission. }
\end{array}
\right.
\end{eqnarray*}
Considering now the emission of soft quarks, the next-to-soft amplitude of Eq.~(\ref{eq:NLPquark}), repeated for convenience here
\begin{eqnarray}
\mathcal{A}_{\rm NLP,\mathcal{Q}} = \sum^{n+2}_{j=1}\frac{g_s}{2p_j\cdot k}\mathcal{Q}_j\otimes i\mathcal{M}_j(p_1,p_2,\dots,p_j,\dots,p_{n+2}),
\end{eqnarray}
contains the quark emission operator ${\cal Q}_j$, whose action on all possible species of incoming/outgoing parton legs is depicted in Fig.~\ref{fig:Soperator}. In terms of the amplitude, we may think of this operator as acting on the wavefunction for leg $j$, as follows: 
\begin{eqnarray*}
\mathcal{Q}_j\left(u(p_j)\right) &=& t^{a}_{c_jc_m} \epsilon_{\mu}(p_j)\slashed{p}_j\gamma^{\mu}v(k) \\
\mathcal{Q}_j\left(\bar{u}(p_j)\right) &=& -t^a_{c_m c_j}\epsilon^*_{\mu}(p_j)\bar{u}(k)\gamma^{\mu}\slashed{p}_j \\
\mathcal{Q}_j\left(v(p_j)\right)& =& t^a_{c_j c_m}\epsilon_{\mu}^*(p_j)\slashed{p}_j\gamma^{\mu}v(k) \\
\mathcal{Q}_j\left(\bar{v}(p_j)\right) &=& -t^a_{c_m c_j}\epsilon_{\mu}(p_j)\bar{u}(k)\gamma^{\mu}\slashed{p}_j \\
\mathcal{Q}_j\left(\epsilon_{\mu}(p_j)\right)& =& -\left(t^a_{c_m c_j}\bar{u}(k)\gamma_{\mu}u(p_j) + t^a_{c_j c_m}\bar{v}(p_j)\gamma_{\mu}v(k)\right) \\
\mathcal{Q}_j\left(\epsilon^*_{\mu}(p_j)\right) &=& t^a_{c_jc_m}\bar{u}(p_j)\gamma_{\mu}v(k) + t^a_{c_mc_j}\bar{u}(k)\gamma_{\mu}v(p_j).
\end{eqnarray*}
There are a couple of further subtleties regarding how to apply this operator in practice. Firstly, in cases where $p_j$ is an initial state particle, or is explicitly observed in the final state (i.e. in an observable that is defined in a way that is not fully inclusive), one must only include those contributions arising from the ${\cal Q}_j$ operator such that the (observed) parton appearing in the LO process has the hard momentum.  Secondly, in the final line of Fig.~\ref{fig:Soperator}, one includes the possibility that either the quark or antiquark is soft. If neither of the decay products of the gluon are explicitly observed, but instead summed over inclusively, then one double counts the quark/antiquark emission contribution due to the integration over all possible momenta $k$. This double counting must then be corrected for by a factor of $1/2$. Thirdly, the polarisation vector may also belong to a photon. In this case, the coupling that appears in Eq.~\eqref{eq:NLPquark} should be modified to $g_{\rm EM}$ and the generator becomes $Q_q$. We see an explicit example of these subtleties in the prompt photon analysis of section~\ref{sec:promptphoton}.

\section{Results for the remaining channels}
\label{app:remain}
Here we report the results of the prompt photon channels that are not discussed in section \ref{sec:promptphoton}. All of the remaining channels only have quarks in the final state, hence only the soft quark formalism is needed to derive these results. The general form of the obtained NLP cross-section is
\begin{eqnarray}
vw(1-v)s\frac{{\rm d}\sigma^{{\rm NLP}}_{{\rm q}}}{{\rm d}{v}{\rm d}w} \equiv
 \alpha\alpha_s^2 \left[\Sigma_{I,I}+\Sigma_{I,F}+\Sigma_{I,FF}+\Sigma_{F,F}+\Sigma_{F,FF}+\Sigma_{FF,FF}\right],
\end{eqnarray}
where $I$ indicates that the contribution stems from initial state radiation, and $F$ ($FF$) indicates that the contribution stems from final state radiation where the other particle is (un)observed.

The separate contributions for the $gg\rightarrow q\bar{q}\gamma$ sub-process are
\begin{eqnarray}
\Sigma_{I,I} &=& \frac{vQ_q^2}{2 C_AC_F}\left[-\frac{1}{\varepsilon} C_F  (2 (v-1) v+3) -\ln(1-w)  \left(C_A -C_F(2 (v-1) v+5)\right) +\mathcal{O}\left(1\right)\right] \nonumber\\
\Sigma_{F,F}&=& \frac{v\left(C_A(v-1)v+C_F\right)Q_q^2}{2C_AC_F}\left[-\frac{1}{\varepsilon} (2 (v-1) v+1) +\log(1-w) \left(2 (v-1) v+1\right)+\mathcal{O}\left(1\right)\right]\nonumber \\
\Sigma_{I,F} &=& \frac{Q_q^2 v}{2C_A C_F}\left[\log (1-w)\left(-C_A(3
   (v-1) v+1)+2 C_F(2 (v-1) v+1)\right)+\mathcal{O}\left(1\right)\right]\nonumber \\
\Sigma_{I,FF} &=& \Sigma_{FF,FF} = \Sigma_{F,FF} = 0,
\end{eqnarray}
and those from the $qq\rightarrow qq\gamma$ sub-process read
\begin{eqnarray}
\Sigma_{I,I} &=& \frac{Q_q^2 C_F}{2C_A^2} \Bigg[-\frac{1}{\varepsilon}\left\{\frac{C_A(2 (v-1) v ((v-1) v+1)+1)}{1-v}\right\}\nonumber \\
&& \hspace{1cm} +\ln(1-w)\left\{\frac{C_A(2 (v-1) v ((v-1) v+1)+1)}{1-v}+2v\right\}+\mathcal{O}\left(1\right)\Bigg]\nonumber \\
\Sigma_{F,F}&=& \frac{Q_q^2 C_F}{C_A^2}\Bigg[-\frac{1}{\varepsilon}\left\{\frac{C_A((v-1)v((v-1)v+3)+1)}{1-v}-v\right\}\nonumber \\
&&\hspace{1.2cm} +\ln(1-w)\left\{\frac{C_A((v-1) v ((v-1)
   v+3)+1)}{1-v}-v\right\}+\mathcal{O}\left(1\right)\Bigg]\nonumber \\
\Sigma_{I,F} &=& \frac{Q_q^2 C_F}{C_A^2}\left[\ln(1-w)\left\{v
   \left((2 (v-1) v+3) C_A+2\right)-\frac{1}{1-v}\right\}+\mathcal{O}\left(1\right)\right]\nonumber  \\
\Sigma_{I,FF} &=& \Sigma_{FF,FF} = \Sigma_{F,FF} = 0.
\end{eqnarray}

For the $qq'\rightarrow qq'\gamma$ sub-process we find
\begin{eqnarray}
\Sigma_{I,I} &=& \frac{C_F}{2C_A} \Bigg[\frac{1}{\varepsilon}\left\{Q_q^2\left(v^2+1\right) (v-1) - Q_{q'}^2 \frac{v^2((v-2)v+2)}{1-v} \right\}\nonumber \\
&&\hspace{1cm}-\ln(1-w)\left\{Q_q^2 \left(v^2+1\right) (v-1) - Q_{q'}^2\frac{v^2((v-2) v+2)}{1-v}\right\}+\mathcal{O}\left(1\right)\Bigg]\nonumber \\
\Sigma_{F,F}&=& \frac{C_F}{2C_A} \Bigg[ \frac{1}{\varepsilon}\left\{Q_q^2 ((v-2) v+2) (v-1)  - Q_{q'}^2 \frac{v^2(v^2+1)}{1-v}\right \} \nonumber \\
&&\hspace{1cm} -\ln(1-w)\left\{Q_q^2((v-2)v+2)(v-1) - Q_{q'}^2\frac{v^2(v^2+1)}{1-v}\right\} +\mathcal{O}\left(1\right) \Bigg]\nonumber \\
\Sigma_{I,F} &=&  \frac{Q_q Q_{q'}C_Fv}{C_A}\left[\ln(1-w)(2(v-1)v+3) + \mathcal{O}\left(1\right)\right] \nonumber \\
\Sigma_{I,FF} &=& \Sigma_{FF,FF} = \Sigma_{F,FF} = 0.
\end{eqnarray}

Finally, the separate contributions for the $q\bar{q}\rightarrow q'\bar{q}'\gamma$ sub-process are
\begin{eqnarray}
\Sigma_{F,F} &=& \frac{Q_{q'}^2C_F}{C_A} \Bigg[-\frac{1}{\varepsilon}(1-v)v^2(2v^2-2v+1) \nonumber \\
&&\hspace{1.5cm} + \ln(1-w)(1-v)v^2(2v^2-2v+1)+ \mathcal{O}(1) \Bigg]\nonumber \\
\Sigma_{FF,FF} &=& \frac{Q_q^2C_F}{C_A} \left[\frac{1}{(1-w)_+}\frac{T_{q\bar{q}}}{3} + \mathcal{O}(1) \right] \nonumber \\
\Sigma_{F,FF} &=& \mathcal{O}(1) \nonumber \\
\Sigma_{I,F} &=& \Sigma_{I,I} = \Sigma_{I,FF} = 0.
\end{eqnarray}
These results are in full agreement with the exact NLO results, which are presented in appendix \ref{app:NLOcalc}.
\section{NLO cross-section for prompt photon production}
\label{app:NLOcalc}

In this appendix we write down the NLO cross-sections for the $q\bar{q}\rightarrow g g \gamma$, $q\bar{q}\rightarrow q\bar{q} \gamma$ and $qg \rightarrow q g \gamma$ processes, expanded up to NLP order and before mass factorisation. The expressions for the cross-sections after mass factorisation can be found in Ref. \cite{Gordon:1993qc}. We will cast the cross-section for all sub-processes in the form
\begin{eqnarray}
vw(1-v)s\frac{{\rm d}\sigma^{{\rm NLP}}}{{\rm d}{v}{\rm d}w} &=& \alpha\alpha_s^2 \Big[c_1\frac{1}{\varepsilon}\frac{1}{(1-w)_+} + c_2\frac{1}{\varepsilon} + c_3\frac{1}{(1-w)_+} +  c_3'\frac{1}{(1-w)_+}\ln\frac{\bar{\mu}^2}{s}  \\
&& \hspace{2cm}+ c_4\left(\frac{\ln(1-w)}{1-w}\right)_+ + c_5\ln(1-w)+\mathcal{O}(\delta(1-w))+\mathcal{O}(1)\Big].   \nonumber
\end{eqnarray}
The coefficients for the $q\bar{q}\rightarrow gg \gamma$ sub-process (section \ref{sec:qqbargamma}) read
\begin{eqnarray}
c_1 &=& -4\frac{Q_q^2C_F^2}{C_A} T_{q\bar{q}}\nonumber \\
c_2 &=& -2\frac{Q_q^2C_F^2}{C_A} \frac{4(v-1)^2v-1}{1-v}\nonumber \\
c_3 &=& \frac{Q_q^2C_F}{C_A}\left(8C_F\left(((v-1)v+1)+T_{q\bar{q}}\ln v\right)+C_A\left(T_{q\bar{q}}\left(-\frac{11}{6}+2\ln(1-v)\right)\right) \right)\nonumber\\
c_3' &=& -8\frac{Q_q^2C_F^2}{C_A} T_{q\bar{q}}\nonumber \\
c_4 &=& 2 \frac{Q_q^2C_F}{C_A}T_{q\bar{q}}(4C_F-C_A)\nonumber\\
c_5 &=& \frac{Q_q^2C_F}{C_A}\left(4C_F-C_A\right)\frac{4(v-1)^2v-1}{1-v},
\end{eqnarray}
and those for the $q\bar{q}\rightarrow q\bar{q}\gamma$ sub-process (section \ref{sec:qqbartoqqbargamma}):
\begin{eqnarray}
c_1 &=& c_3' = c_4 = 0\nonumber \\
c_2 &=& -\frac{Q_q^2C_F}{C_A^2}(v(3(v-1)v+1))-\frac{Q_q^2C_F}{C_A}\frac{T_{q\bar{q}}(2(v-1)v((v-1)v+1)+3)}{2(1-v)} \nonumber\\
c_3 &=& \frac{Q_q^2C_F}{C_A}\frac{T_{q\bar{q}}}{3} \nonumber\\
c_5 &=& \frac{Q_q^2C_F}{C_A^2}v(1-2v)^2+\frac{Q_q^2C_F}{C_A}\frac{2(v-1)v((v-1)v(2(v-1)v+5)+7)+3}{2(1-v)}.
\end{eqnarray}
Next, we have the coefficients for the $qg\rightarrow qg\gamma$ sub-process  (section \ref{sec:qgfinal}),
\begin{eqnarray}
c_1 &=& -\frac{Q_q^2}{C_A} T_{qg}(C_A+C_F) \nonumber \\
c_2 &=& \frac{vQ_q^2}{2C_A}\left(C_A (5(v-1)v+4)-C_F \frac{v(v^3-2v+4)-2}{1-v}\right)\nonumber  \\
c_3 &=& \frac{Q_q^2}{C_A}\left(C_A\left(v(v-2)^2+2T_{qg}\ln v \right)+C_F\left(T_{qg}\ln\left((1-v)v^2\right) +\frac{v}{4}((v-10)v+10)\right)\right)\nonumber \\
c_3' &=& -\frac{2Q_q^2}{C_A} T_{qg}(C_A+C_F) \nonumber \\
c_4 &=&\frac{Q_q^2}{C_A}(2 C_A+C_F)T_{qg}\nonumber \\
c_5 &=& \frac{vQ_q^2}{2C_A(1-v)}\left(C_A\left(v-1\right)\left(9(v-1)v+8\right)+C_F\left(v((v-2)v^2+4)-2\right)\right)
\end{eqnarray}
followed by the $gg\rightarrow q\bar{q}\gamma$ sub-process:
\begin{eqnarray}
c_1 &=& c_3' = c_3 = c_4 = 0 \nonumber \\
c_2 &=& \frac{vQ_q^2}{2C_AC_F}\left(C_A v(v(-2v(v-2)-3)+1)-4C_F(v(v-1)+1)\right)\nonumber \\
c_5 &=& \frac{vQ_q^2}{C_A C_F}\left(C_A(v^4-2v^3+v-1)+4C_F((v-1)v+1)\right).
\end{eqnarray}
For the $qq\rightarrow qq\gamma$ sub-process, we find
\begin{eqnarray}
c_1 &=& c_3' = c_3 = c_4 = 0 \nonumber \\
c_2 &=& \frac{Q_q^2C_F}{C_A^2}\left(v-C_A\frac{(2(v-1)v+1)(2(v-1)v+3)}{2(1-v)}\right)\nonumber \\
c_5 &=& \frac{Q_q^2C_F}{C_A^2}\frac{C_A(2v^2-2v+3)-4v^2+4v-2}{2(1-v)},
\end{eqnarray}
and for the $qq'\rightarrow qq'\gamma$ sub-process (note that Ref. \cite{Gordon:1993qc} has $Q_q \leftrightarrow Q_{q'}$ as a result of an interchange in the assigned initial state momenta)
\begin{eqnarray}
c_1 &=& c_3' = c_3 = c_4 = 0 \nonumber \\
c_2 &=& -\frac{C_F}{C_A}\frac{2(v-1)v+3}{2(1-v)}\left(Q_q^2(1-v)^2+Q_{q'}^2v^2\right)\nonumber \\
c_5 &=& \frac{C_F}{C_A}\frac{2(v-1)v+3}{2(1-v)}\left(Q_q^2(1-v)^2+Q_qQ_{q'}2v(1-v)+Q_{q'}^2 v^2\right).
\end{eqnarray}
Finally, the coefficients for the $q\bar{q}\rightarrow q'\bar{q}'\gamma$ sub-process are given by
\begin{eqnarray} 
c_1 &=& c_3' = c_4 = 0 \nonumber \\
c_3 &=& \frac{Q_{q}^2 C_F}{C_A}\frac{T_{q\bar{q}}}{3}\nonumber  \\
c_2 &=& -\frac{C_F}{C_A}Q_{q'}^2(1-v)v^2(2v^2-2v+1)\nonumber \\
c_5 &=& \frac{C_F}{C_A}Q_{q'}^2(1-v)v^2(2v^2-2v+1).
\end{eqnarray}

\bibliographystyle{JHEP}
\bibliography{spires}

\end{document}